\documentclass{aa}  

\usepackage{graphicx}
\usepackage{placeins}
\usepackage{txfonts}
\usepackage[dvipsnames]{xcolor}

\usepackage[capitalise]{cleveref} 
\usepackage{appendix}

\begin{document} 
 
\newcommand{\kms}{km\,s$^{-1}\,$}
\newcommand{\um}{$\mu$m}
\newcommand{\lam}{$\lambda$}
\newcommand{\cms}{cm$^{-2}$}
\newcommand{\cmc}{cm$^{-3}$}
\newcommand{\ergs}{erg\,s$^{-1}$}
\newcommand{\ergscm}{erg\,s$^{-1}$\,cm$^{-2}$}
\newcommand{\wm}{W\,m$^{-2}$}
\newcommand{\lsol}{L$_{\odot}$}
\newcommand{\msol}{M$_{\odot}$}
\newcommand{\vLSR}{V$_{\rm LSR}$}
\newcommand{\vsys}{V$_{\rm sys}$}
\newcommand{\denshh}{n$_{\rm H_{2}}$}
\newcommand{\nhh}{\N$_{\rm H_{2}}$}
\newcommand{\pvpar}{PV$_\parallel$}
\newcommand{\pvperp}{PV$_{\perp}$}
\newcommand{\pvperpy}{PV$_{\perp,y}$}
\newcommand{\lamphi}{$\lambda_{\phi}$}

\newcommand{\cd}[1]{{\color{cyan}[CD:~#1]}} 

\title{ALMA chemical survey of disk-outflow sources in Taurus (ALMA-DOT)}
   \subtitle{VII: The layered molecular outflow from HL Tau
   and its relationship with the ringed disk}

   \author{F. Bacciotti \inst{1}
          \and
          T. Nony \inst{1}
          \and
          L. Podio \inst{1}
        \and
        C. Dougados \inst{2}
        \and
        A. Garufi \inst{3}
        \and
         S. Cabrit \inst{2,4}
        \and
        C. Codella \inst{1}
        \and
        N. Zimniak \inst{2}
        \and
        J. Ferreira \inst{2}
          }

   \institute{INAF - Osservatorio Astrofisico di Arcetri, Largo E. Fermi 5, 50125 Firenze, Italy\\
              \email{francesca.bacciotti@inaf.it}
           \and 
Univ. Grenoble Alpes, CNRS, IPAG, 38000 Grenoble, France
\and
INAF – Istituto di Radioastronomia, Via P. Gobetti 101, 40129 Bologna, Italy
\and
PSL University, Sorbonne Université, Observatoire de Paris, LERMA, CNRS UMR 8112, 75014 Paris, France
}

   \date{Received -; accepted -}


  \abstract
   {
   The ALMA image of the ringed disk around HL Tau stands out as the iconic signature of planet formation, but the origin of the observed substructures is still debated. The HL Tau system also drives a powerful bipolar wind, detected in atomic and molecular lines, that may have important feedback on the process.
   }
   {The outermost component of the wind traced by CO emission was analyzed in detail to determine its relationship with the disk and its substructures.}
   {A spectro-imaging investigation was conducted using ALMA observations of the ${}^{12}$CO (2-1) line at 1.3\,mm, with 0.2\,\kms and $\sim0\farcs28$ spectral and angular resolution, in the framework of the ALMA-DOT project. 
The relevant wind parameters were derived,  
allowing a tomographic reconstruction of the morphology and kinematics of the redshifted lobe of the outflow to be compared with theoretical models.
}
{The data channel maps and position-velocity diagrams show a rich substructure of
concatenated bubble- and arc-shaped features, whose size and distance from the source continuously increase with velocity. The superposition of such features generates the apparent conical shape. 
The spatial-kinematic properties suggest that the flow presents distinct nested shells with a
progressively higher propagation speed and a steeper speed gradient with distance from the source
for the shells closest to the axis, and rotating in the same direction as the disk. 
The wind parameters were compared with the predictions of magnetohydrodynamic (MHD) disk winds. Under this hypothesis, the launch radii of the three outermost shells are found to correspond to the positions of three rings in the dust emission distribution of the disk located at 58, 72, and 86 astronomical units from the star. We derive a magnetic lever arm $\lambda \sim 4 - 5$, higher than that commonly adopted in models of MHD winds from the outer disk. Interpretations are discussed.
}
{
The properties of the CO outflow from HL Tau appear to be compatible with magnetized disk winds with launch radii in the region at 50 to 90 au from the source. As such, the wind may be capable of removing angular momentum also from the outer disk.
The arrangement of the wind in
nested shells with brighter emission rooted at the location of ring substructures could support the results of recent non-ideal MHD simulations according to which magnetic instabilities can spontaneously generate both the ring-gap system and a connected inhomogeneous layered wind, alternatively to the action of protoplanets. Further observational analyses and comparisons with other classes of models will help establish the role of magnetic effects in the process of planet formation.
}

  \keywords{Protoplanetary disks -- Jets and outflows -- ISM: molecules -- Stars: individual: HL Tau}

\authorrunning{Bacciotti et al.}

\titlerunning{The layered molecular outflow from HL Tau}

   \maketitle

\section{Introduction} \label{intro}

One of the most spectacular discoveries of the last decade has been the detection in the continuum emission at millimeter wavelengths of concentric gaps and rings in disks around young stars. The first example found was the disk around HL Tau. Observed with the Atacama Large Millimeter Array (ALMA),\footnote{https://www.eso.org/sci/facilities/alma.html} it showed in exquisite detail a sequence of concentric rings and gaps \citep{ALMA15}, now commonly seen in many other disks (e.g., \citealt{Andrews18}).

The first interpretation offered was that these gaps are carved by young massive planets embedded in the disk (e.g., \citealt{Lodato19}). However, this scenario raises open issues, such as the amount of mass necessary to produce massive planets at such large distances from the star (up to 100 au in HL Tau) and on such short timescales (less than 1 Myr). In addition, the predicted embedded planets inside the gaps remain elusive, except for the cases of PDS 70 \citep{Keppler18} and WISPIT 2 \citep{vanCapelleveen25}, 
where the gap actually is a large inner cavity.

Several other disk processes are able to produce ring--gap structures. Perhaps the most generic are magnetic instabilities in the low-ionization dead zone \citep[e.g., ][]{Suriano18, Riols19}. This process could be widespread in young disks that harbor a residual primordial magnetic field.  In this case, an inhomogeneous magnetohydrodynamic (MHD) disk wind (DW) is created, and the disk self-organizes in a system of rings and gaps, with denser wind layers launched from the rings and less dense ones from the gaps (e.g., \citealt{Suriano19, Riols20}). An asymmetry also develops between the red and blue lobes \citep{Bethune17}, as observed in almost all cases (e.g., \citealt{Podio11, Podio21}). Finding firm evidence of such layered MHD disk winds is thus an essential step in a clarification of the origin of disk substructures and of the role of magnetic fields in planet formation.

Signatures of layered outflows have indeed been found with ALMA in a growing number of young
systems, in the form of rotating molecular winds flowing in a cavity of apparent
conical shape (e.g., TMC1A, \citep{Bjerkeli16}; HH 46/47, \citep{Zhang16, Zhang19};
HH212 \citep{Lee17}; DO Tau, \citep{Fernandez20}; HH 30, \citep{Louvet18, Lopez24, Ai24}; DG Tau B \citep{Zapata15, DeValon20, DeValon22}; HH270mms1-A, \citep{Omura24}; L1448-mm,  \citep{Nazari24}).
The CO outflow is associated in some cases with a coaxial molecular wind that emits in $H_2$ lines, with a fast collimated atomic jet flowing along the axis of the system (e.g., HH 46/47, \citep{Nisini24}, DG Tau B \citep{Delabrosse24}). 
The origin of the layered structure in these CO winds is debated. In addition to the magnetic instabilities mentioned above, other studies invoke the formation of swept-up wind shells \citep{Zhang19}, peculiarities of the MHD disk-wind accretion--ejection engine \citep{Louvet18,DeValon20}, flow wiggling \citep{DeValon22}, the interaction of a pulsed jet with its environment \citep{Tabone18}, jet-driven wide bow shocks expanding in a stratified medium \citep{Rabenanahary22}, 
composite scenarios including entrainment of the envelope material, rotating winds and inner shocked winds \citep{Lopez24b},
or the evolution of a unified wide-angle magnetized flow with a jet-bearing X-wind  \citep{Shang20,Shang23,Ai24}.

Among the observable sources, HL Tau stands out as the first and only
object where a possible coincidence between the wind structure and the distribution of disk rings and gaps can be studied. 
In  the other cases the flow is associated with a disk where no gaps and rings are detected. 
In some cases, this can be due to the geometry of the observation (for example, the edge-on disks of HH~30). In general, however, 
the accretion--ejection activity associated with young stellar objects is expected to decrease with the source evolutionary stage; hence, powerful and bright molecular winds and jets are associated with Class 0 protostars
(e.g., \citealt{Lee18, Tabone17, Podio21}),
while only faint molecular emission is associated with evolved Class II stars.
On the other hand, substructures such as rings and gaps are commonly detected in pre-main sequence disks (e.g., SHARP LP, \citealt{Andrews18}), while they are rarely detected in younger disks
(see, e.g., \citealt{Segura-Cox20, sheehan17, sheehan18, DeValon20}).
In this context, HL Tau is intermediate between an evolved accreting protostar (Class I) and a classical T Tauri star (Class II),  and is associated with both a disk with clearly detected rings and gaps and a bright extended and structured outflow. HL Tau is therefore an ideal target for investigating the possible relationship between the disk substructures and the wind properties and for testing the proposed scenarios for their formation.

\begin{figure*}
 \centering
\includegraphics[width=0.485\linewidth]{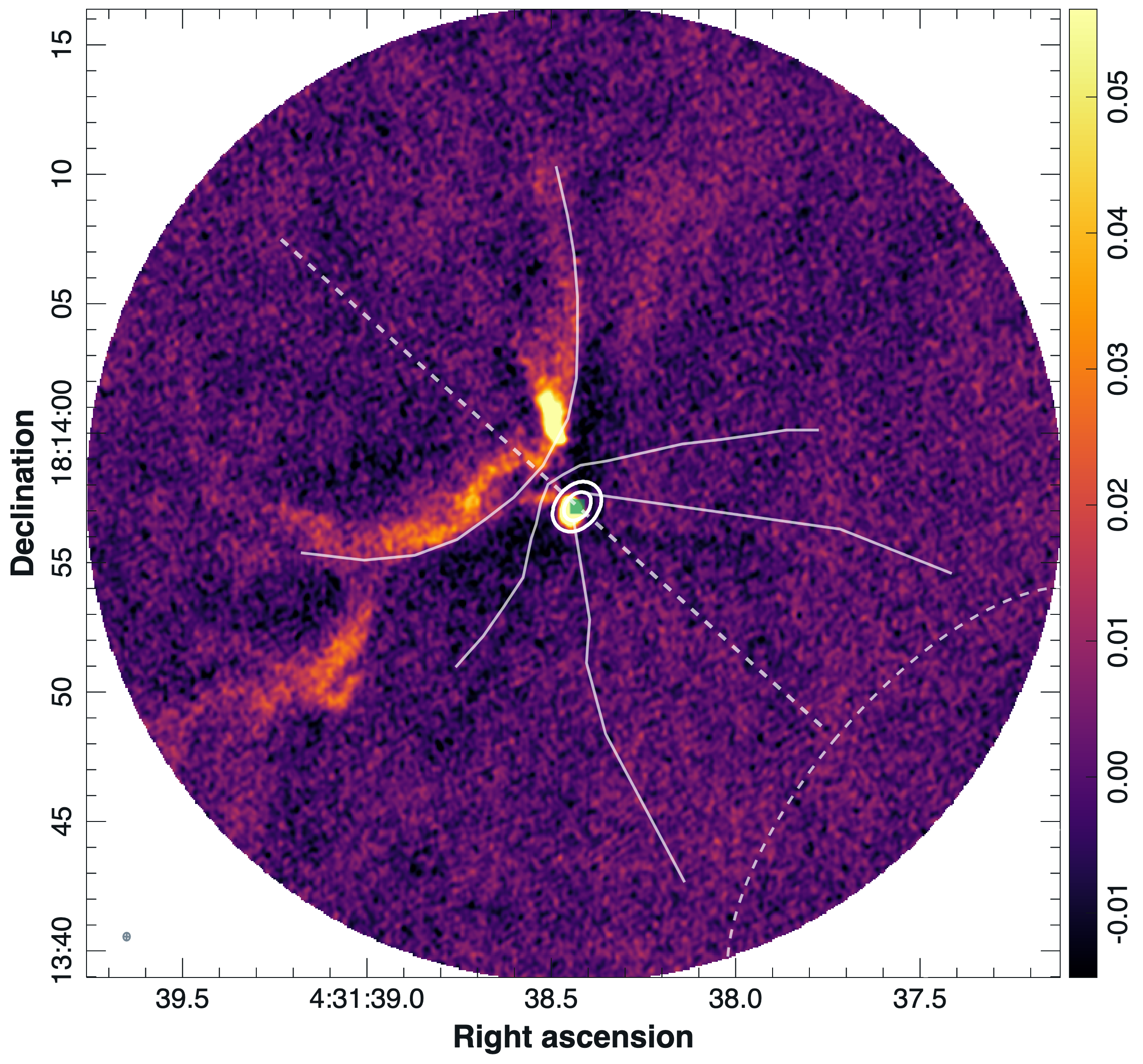}
\includegraphics[width=0.50\linewidth]{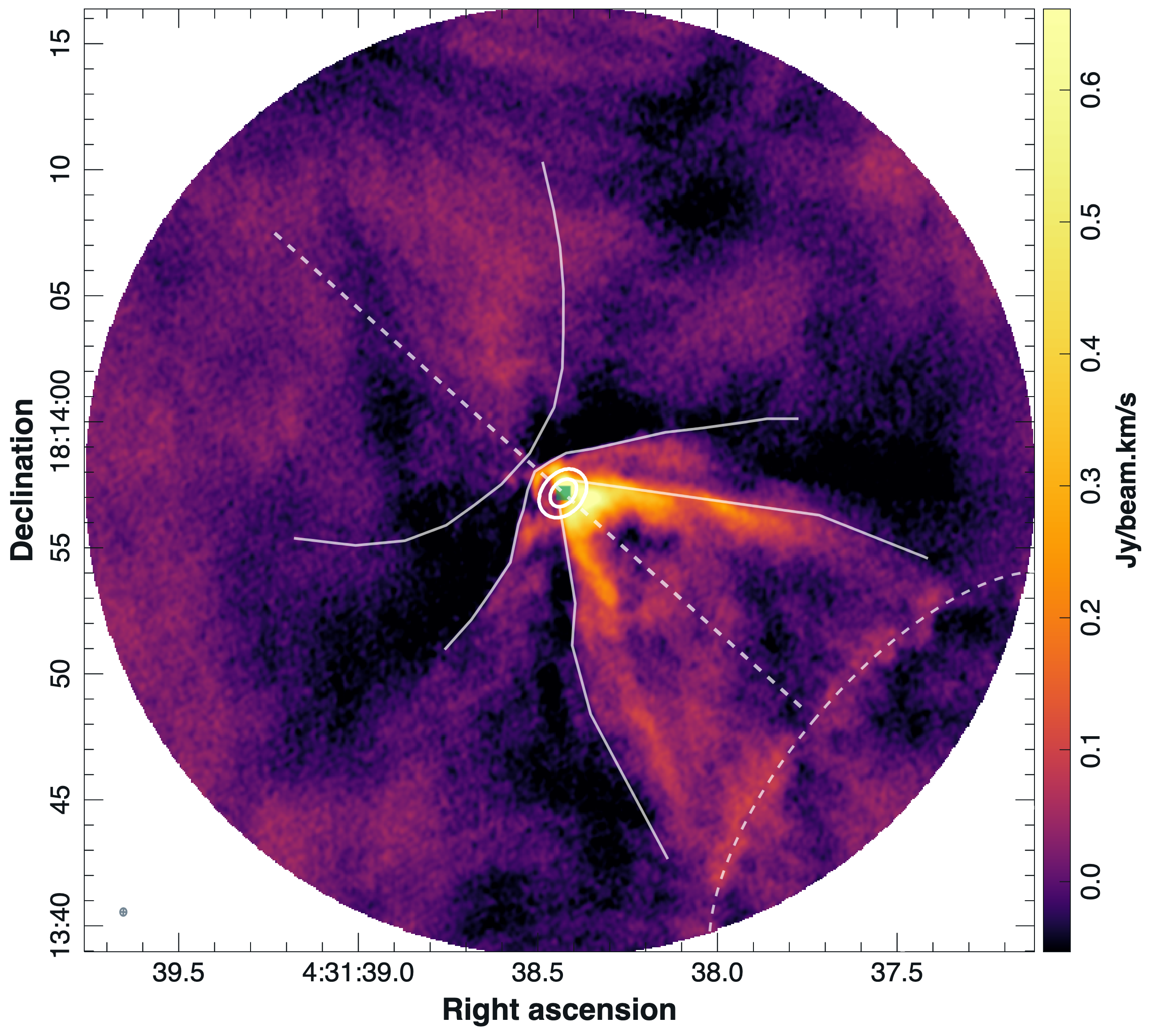}
   \caption{{\em Left:} Moment 0 maps of the CO (2$-1)$ emission integrated over the velocity interval 
    V$_{\rm LSR} = (-10, +3)$ \kms, highlighting the morphology of the blueshifted outflow lobe.
   The red contours at the center of the panel corresponds to the disk continuum emission at 1.3 mm drawn at  [10, 200]$\sigma_{\rm c}$, with $\sigma_{\rm c}$=1.2$\times$10$^{-4}$ Jy beam$^{-1}$. 
   The green square in the center of the contours marks the position of the continuum peak. 
   The white dashed line at PA=48$^{\circ}$ indicates the orientation of the disk minor axis, and the  white solid curves outline the spatial limits of the main wind components described in the text. {\em Right:} Same as left panel, but with velocity integration range  V$_{\rm LSR} = (+8, +30)$ \kms, illustrating the structure of the redshifted lobe. The beam is drawn in the bottom left corner. We note the difference in brightness between the two lobes:  the blueshifted lobe is about ten times fainter. 
   }
   \label{fig:moment0maps}
    \end{figure*}

\begin{figure}
 \centering
 \includegraphics[width=0.98\linewidth]{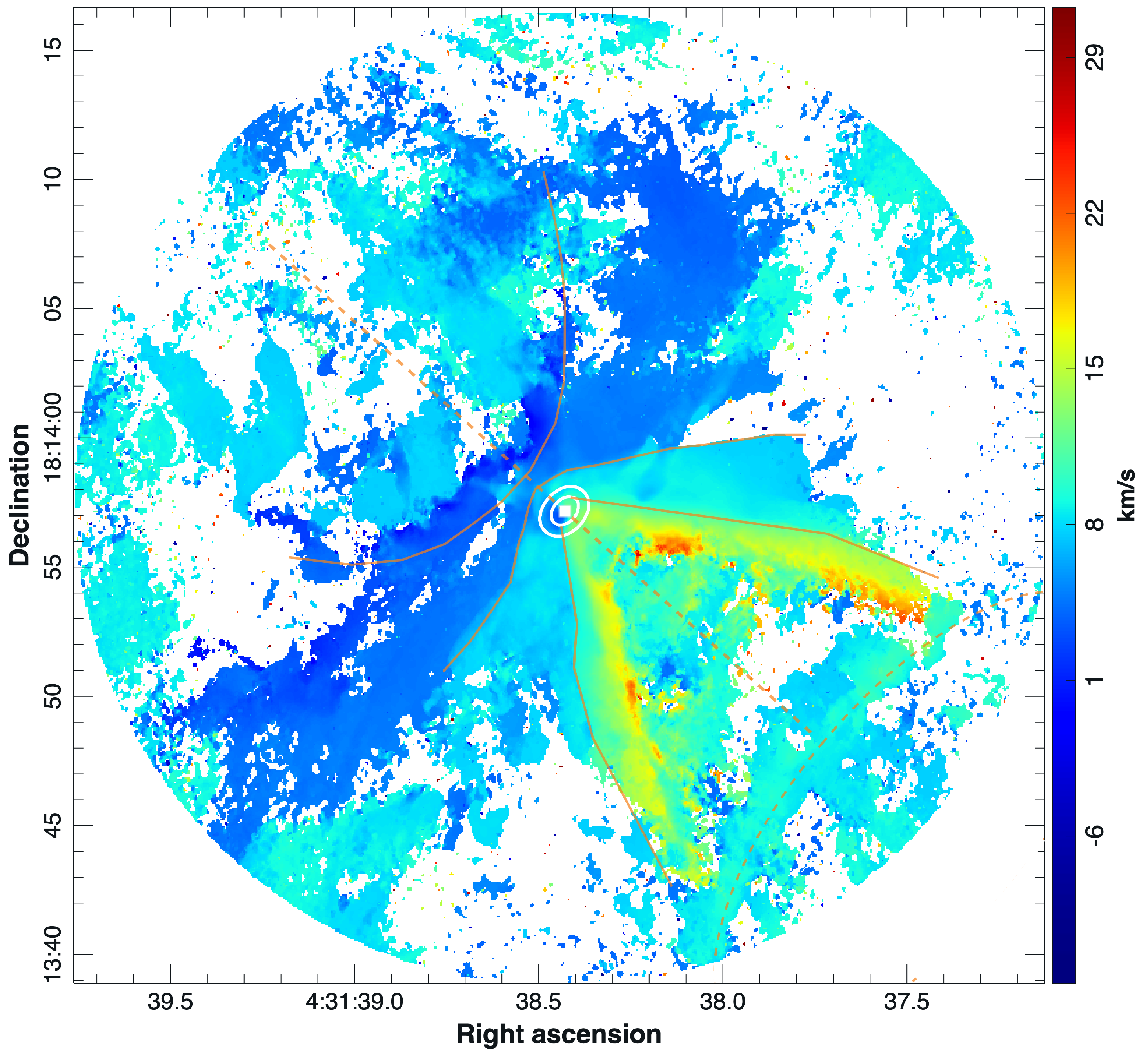}
   \caption{Moment 1  map of the CO (2$-1)$ emission over the velocity interval V$_{\rm LSR} = (-10, +32)$ \kms. 
   The solid curves, disk contours,  dashed line, and central square
   are as in Fig.\ref{fig:moment0maps}. 
   The map was produced using a 4$\sigma$ clipping. 
   }
   \label{fig:moment1map}
    \end{figure}

The central object, a young star of mass 2.1$\pm 0.2 $\,\msol\, \citep{Yen19a}, 
is  the source of a composite outflow seen at optical, infrared, and millimeter wavelengths.
The collimated atomic jet that shines in shock-excited lines such as [S\,II]\lam\lam6716,6731, H$\alpha$, and [O\,I]\lam6300, stretches out of the reflection nebula surrounding the star for about 1.$\arcmin$5 
along the direction PA $\sim$ 50\degr, 
with an inclination angle with respect to the line of sight of $40^{\circ} - 50^{\circ}$
\citep{Mundt90, Anglada07, Pyo06, Movsessian12}.
The jet is compatible with being perpendicular to the disk, which has an inclination angle with respect to the line of sight $i_{\rm disk}$ = 46$^{\circ}.2 \pm 0^{\circ}$.2 (defined as $0^{\circ}$ for face-on), as determined in \cite{ALMA15} from the data at 0\farcs2 resolution. 
The same study reports that the major axis of the disk lies along PA$_{\rm disk}$ = +138$^{\circ}.2 \pm 0^{\circ}$.2 measured from north to east.

The system was imaged at 0\farcs2 resolution with adaptive optics in the near-infrared, showing a collimated jet  in [FeII]\,1.64\,\um\, and a surrounding wider-angle warm wind emitting in H$_2$\,2.12\,\um\, that flows in a conical cavity detected in scattered light \citep{Takami07, Beck08}. 
A coaxial molecular outflow has been observed in the CO(3$-2$) and CO(1$-0$) lines \citep{Lumbreras14, ALMA15,  Klaassen16}. The data show a prominent conical redshifted flow to the southwest of HL Tau, with the axis
almost aligned with the atomic jet, and with
a full opening angle starting at 90\degr and then narrowing at 60\degr farther from the source \citep{Klaassen16}. The inclination with respect to the line of sight is assumed to be along the perpendicular to the disk (i.e. $i$=$i_{\rm disk}$). 
The northeastern lobe of the molecular outflow is much fainter and appears as a wide cavity spread within the ambient.

For this paper we investigated the structure of the CO outflow in greater detail by analyzing a dataset obtained within the  ALMA chemical survey of Disk-Outflow sources in Taurus (ALMA-DOT) program, an observational campaign aimed at characterizing the gas in disks and outflows of embedded sources in the Taurus star-forming region. The high angular and spectral resolution
of ALMA allows us to disentangle and characterize the emission from the various elements of the systems (disk, outflows, streamers, envelope). Previous papers in the ALMA-DOT series focus on the chemical properties of the disks
\citep{Podio19,Podio20a,Podio20b,Codella20} and of the associated streamers and outflows \citep{Garufi20,Garufi22}, while \citet{Garufi21} describes the collective properties of the targets examined as a whole. The present work focuses on the observations of the molecular outflow around HL Tau, as imaged in the line of ${}^{12}$CO (2$-1$) at 1.3 mm with a resolution of 0\farcs3 and 0.2 \kms, and on the implications of these findings for the structure of the disk.

\begin{figure*}[htp!]
  \centering
\includegraphics[width=\textwidth]{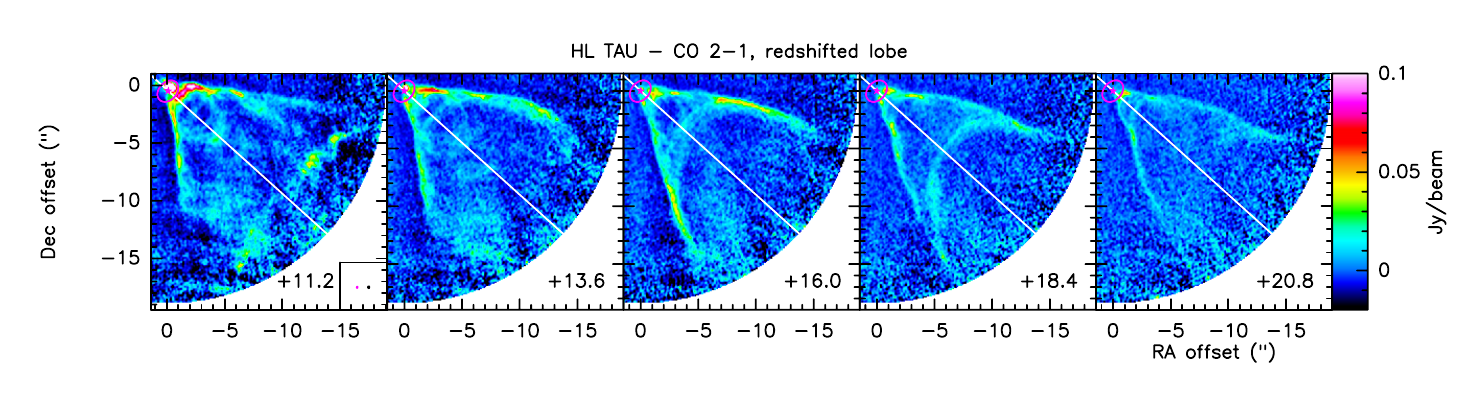}
   \caption{Channel maps at selected \vLSR\, velocities of the redshifted emission in the CO (2-1) line  in the SW outflow lobe. 
   The star symbol and the contour indicate the position of the source and the disk emission at 1.3 mm drawn at 10$\sigma_{\rm c}$, respectively. 
   The white solid line 
   lying at PA=228$^{\circ}$ coincides with the direction of the disk minor axis.
   The magenta and black dots in the lower right corner of the leftmost panel report the beam size in the continuum and line observations, respectively.  
} 
        \label{fig:ellipses1}
    \end{figure*}

\begin{figure*}[htp!]
\includegraphics[width=\textwidth]{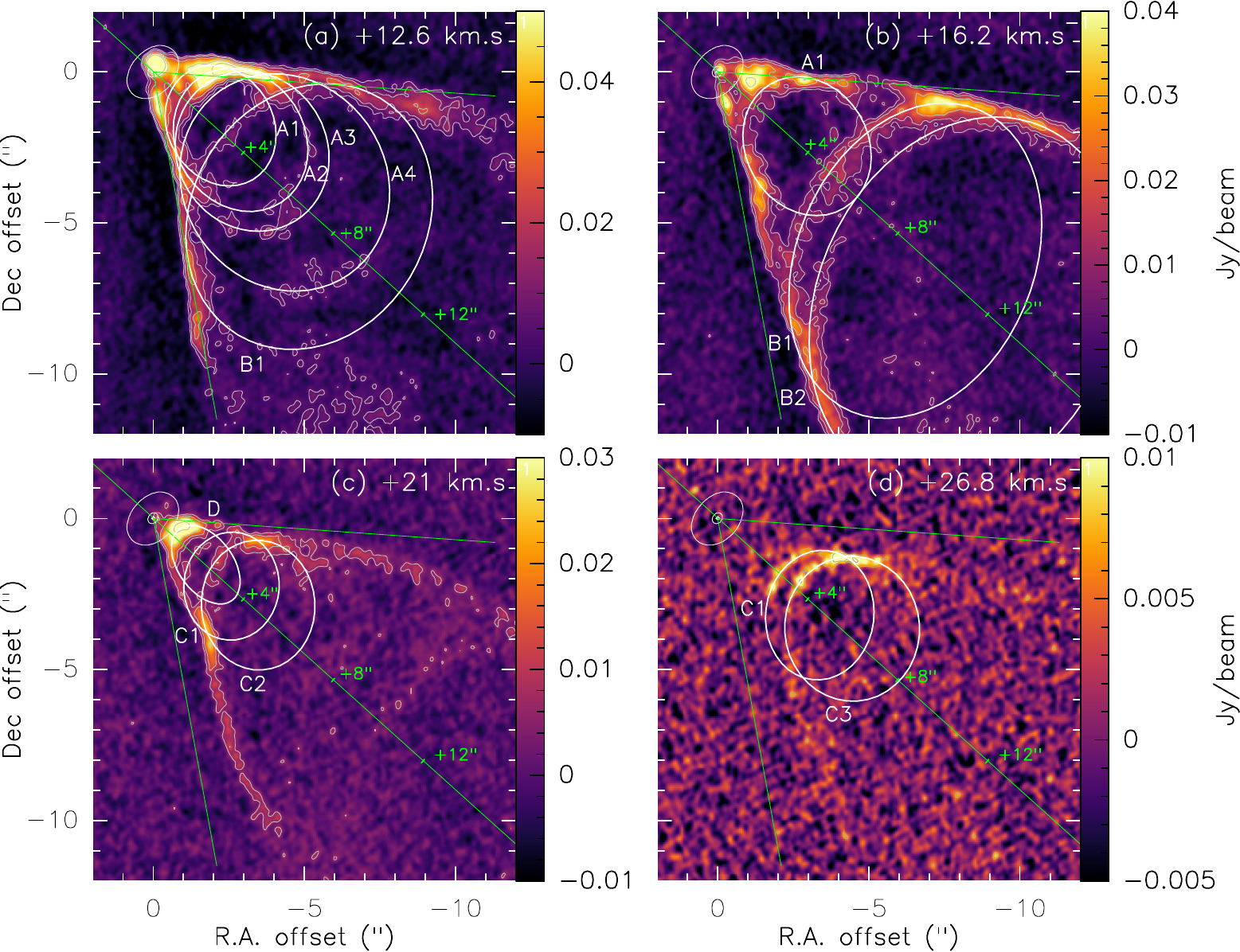}
   \caption{Zoomed-in images of channel maps at selected \vLSR velocities, 
   illustrating the system of arcs and bubble-like substructures visible in the redshifted outflow.
   The CO emission contours are drawn at [3, 6, 12, 24, 48]$\sigma_\mathrm{CO}$. 
   The two white contours
   of the 1.3 continuum at [30, 700]$\sigma_\mathrm{c}$ indicate the disk position.
   The green line oriented at PA=228$\degr$ coincides with the direction of the disk minor axis, and  defines the reference axis of the offset $y$ from the source in the plane of the sky, with positive offsets toward SW.   
   The other green lines mark  the $\pm38\degr$ average opening angle of the apparent conical cavity at 12.6 \kms, to illustrate the narrowing of the flow with increasing velocity. 
   In the vicinity of the source, a number of arcs are seen to combine to form  closed curves.  Farther from the source only arcs opened toward the SW are visible. The superposed white ellipses are examples of visual fits to the identified features, labeled in families sharing common properties. 
   }
          \label{fig:ellipses2}
    \end{figure*}

The paper is organized as follows. 
Section~\ref{obs} details the observations and the reduction of the data. Section~\ref{resu} describes the results obtained, while the derived hypothesis for the structure of the wind is described in \cref{su:shells}. In \cref{disc} we discuss the comparison of the results with theoretical models in a MHD disk wind scenario. Our conclusions of this study and the perspectives for future analyses are outlined in Section~\ref{concl}.

\section{Observations and data reduction} \label{obs}

Cycle 6 observations of the ALMA-DOT campaign \citep{Podio19} were taken with ALMA-Band 6 on October 28, 2018 in an extended configuration with baselines ranging from 15~m to 1.4 km (project 2018.1.01037.S, PI: L. Podio). The field of view was a circular region around HL Tau of radius $\sim$ 18\farcs5, corresponding to about 2.7$\times$10$^{3}$ au at the distance from the source, 147.3 pc \citep{Galli18}. The integration time summed up to $\sim$ 110 minutes \citep{Garufi21}. The bandpass and phase calibrators were J0423-0120 and J0510+1800, respectively. 
The correlator setup consisted of high-resolution (0.141 MHz) spectral windows (SPWs) covering ${}^{12}$CO (2$-1$),  CN (2$-1$), o-H$_2$CO (3$_{1,2}-2_{1,1})$ CS (5$-4$),
and CH$_3$OH ($5_{0,5}-4_{0,4}$) (A). The molecular parameters were taken from the Cologne Database of Molecular Spectroscopy (CDMS) \citep{muller01}. In this paper, we focus on the ${}^{12}$CO (2$-1$) emission (rest frequency 230.538 GHz), 
while the description in other molecular lines can be found in \citet{Garufi20}.

Data were reduced using the standard procedure using the
Common Astronomy Software Applications package (CASA,
\citealt{McMullin07}) version 4.7.2.
Self-calibration was performed on the continuum emission and applied on the line-free continuum, and the continuum subtracted line emission. The procedure improved the continuum S/N by a factor of 3.3. The final continuum maps and spectral cubes were produced with \textsc{tclean} by applying a manually selected mask to the signal and iterating until the residuals revealed no significant source emission.
We adopted Briggs weighting with robust=0.0 for the bright CO line, to maximize angular resolution, and set a channel width of 0.2 \kms.
The clean beam of the self-calibrated maps in CO is  0.26\arcsec$\times$0.31\arcsec\ with a PA of -3.12\degr. The r.m.s. noise per channel is $\sigma_{\rm CO} = 2$ mJy beam$^{-1}$.  
The beam in the continuum images is   0.23\arcsec$\times$0.26\arcsec\ with  PA= -6.77\degr. The r.m.s. for the continuum image is  $\sigma_{\rm c} = $0.12 mJy beam$^{-1}$.

\section{Results} \label{resu}

\subsection{Continuum emission}
\label{su:cont}

The continuum at 1.3 mm has an integrated flux of 844.9 $\pm$ 1.9 mJy,                      
with peak intensity of 95.48 $\pm$ 0.19 mJy beam$^{-1}$ located at R.A.  04$^{\rm h}$ 31$^{\rm m}$ 38$^{\rm s}$.43, and Dec. 18$^{\circ}$ 13$^{\prime}$ 57$\farcs$17. 
At our angular resolution the intensity distribution is smooth, and the sequence of rings and gaps is not visible.
The FWHM along the major and minor axes, determined with a 2D Gaussian fit
deconvolved from the beam, is $812\pm 2$ and $559\pm 1$ mas, respectively. 
From these values, we estimated an inclination of the disk $i_{\rm disk}\sim 46^{\circ}.5 \pm 0^{\circ}.3$  (with 0\degr\, corresponding to face-on). The fit gives a disk position angle, PA$_{\rm disk}$ = +138$^{\circ}.06 \pm 0^{\circ}$.25. Both values agree with
those reported in \cite{ALMA15}.

\subsection{CO (2$-1$) emission}
\label{su:moments}

The ALMA datacube in the CO (2$-1)$ line consists of 290 velocity channels with \vLSR\, ranging from -26 to +32 \kms, in steps of 0.2 \kms. The emission of the outflow is detected from -10 \kms to +31.8 \kms. The systemic velocity is \vsys=+7.1 \kms \citep{Garufi21}. 

Figure \ref{fig:moment0maps} presents the intensity integrated over the velocity (moment-0 map) in the full field of view. 
The left panel shows the emission integrated over the blueshifted velocity range V$_{\rm LSR} = (-10, +3)$, while the integration in the right panel is for V$_{\rm LSR} $ velocities between +8 and +30 \kms, illustrating the redshifted emission.
The faint blueshifted emission is distributed along a wide-angle cavity of semi-spherical shape pointing toward the northeast (NE), with a distinct bright feature on the northern side.
The emission in the redshifted southwest (SW) lobe presents
two main components, an inner flow of apparent conical shape around an axis oriented at PA = 228\degr (coinciding with the minor axis of the disk)
with a full opening angle starting at 90\degr and then narrowing at 60\degr farther from the source \citep{Klaassen16}
and an outer faint conical wide wind, in projection surrounding both the disk and the base of the inner outflow, 
with vertex at $\sim$ 1\farcs5 NE of the source
and opening angle of $\sim$ 120\degr\,.
To guide the eye, the approximate spatial limits
of these components are indicated by solid white curves in the figure.\footnote{
The opening angle was determined with a linear fit to the wind edges, identified by the position of the  zeroes of the second-order derivative of the emission along cuts parallel to axis, in the first 2\arcsec of propagation.
}

A molecular emission ridge is evident at 14" from the source. This is part of a wider elliptical feature, as shown in \citet{ALMA15}, Fig. 1, and reported here with a dashed white elliptical curve.

The intensity-weighted mean velocity (moment-1 map) of the emission is presented in \cref{fig:moment1map}. 
The NE sector hosts blueshifted material that is not collimated, while the redshifted SW lobe is structured. The transition in velocity between the two wind components of the red lobe identified in the moment-0 map shows that the outer wide wind is slower than 10 \kms, while the inner flow has increasing velocities from the border toward the axis, up to about 30 \kms.

The asymmetry between the blue- and redshifted lobes may be an effect of the inclination angle with respect to the plane of the sky. For example,  \citet{Klaassen16} illustrates how the large aperture of the initial portion of the HL Tau flow causes the far side of the approaching lobe to be imaged at redshifted velocities close to the systemic. However, while this effect may explain the loss of the upper part of the large arc at blueshifted velocities, it would require the approaching lobe to have an opening angle larger than the receding one to explain the absence of a well-developed conical structure. A trace of faint conical arms of aperture angle similar to that in the red lobe is barely seen at the base of the blueshifted lobe, but it is not seen farther out, and it is covered by the signal of the large semicircular arc. 
This structure of the blue lobe may be due to
HL Tau being located on the wall of a large-scale molecular bubble inflated by nearby XZ Tau, at 23\arcsec east of the source \citep{Welch00, Yen19b}. The expanding bubble may have compressed the outer
envelope of HL Tau on its NE side, smashing the CO cavity, while the SW lobe
remained unaffected. Signatures of such a large-scale bubble are visible at low blueshifted velocities in \cref{fig:moment1map}.

In the following, we concentrate the analysis on the redshifted lobe, which presents a well-defined and relatively bright structure over a wide range of velocities. This allows a detailed inspection of the spatio-kinematic properties, for a meaningful comparison with proposed acceleration mechanisms. We defer the study of the fainter blueshifted lobe to a forthcoming investigation, adding that
the squared 10\arcsec at the center of the region were discussed in \citet{Garufi22}, where the kinematics of the rotating disk is analyzed from the same dataset, as well as possible accretion streamers. 

\subsection{Velocity channel maps}
\label{su:ch-maps}

A comprehensive atlas of the channel maps of CO (2-1) emission
is given in Appendix \ref{app:chmaps}.  
The analysis of the individual channels reveals a faint but very rich flow substructure, not visible in the moment-0 and moment-1 maps.
The emission in the blue lobe is barely detectable, but the signal is strong in the redshifted lobe, thus allowing a detailed inspection.
A selection of channel maps illustrating the configuration of the inner flow in the redshifted SW lobe is presented in \cref{fig:ellipses1}. The maps show an apparent conical structure, 
with aperture decreasing with velocity, 
from about 80$\degr$ at 10 \kms to 55$\degr$ at 26 \kms (see \cref{app:chmaps}). 

Inscribed in the conical apparent shape, one finds a series of ‘bubbles’ and ‘arcs’ of different sizes, positions and brightness. The arcs appear to be part of concatenated closed curves that can be visually fitted by ellipses, as illustrated in \cref{fig:ellipses2}. About ten such structures can be identified, grouped in four families that we label  A, B, C, and D.
Each feature continuously changes size and position with channel maps and acquires progressively larger extension and separation from the star as \vLSR\, velocity increases, but maintains its shape unchanged. 

The arcs of family A are observed at low velocity (10-16 \kms, see \cref{fig:ellipses2}a-b) and at small separation. Despite the blending of the emission on the source side and the fragmented appearance at larger distances, these structures clearly form closed bubble-like curves expanding with increasing velocities.
Arcs B1 and B2 have a large extension and only show the convex branch on the side of the source. They can be followed over a wide range of velocities, from 11 to 19 \kms, in which they continuously change position from 3$\arcsec$ to 10$\arcsec$ from the source, as illustrated in \cref{fig:ell_move}.
Arcs C are visible only at high velocity (20-29 \kms, see \cref{fig:ellipses2}c-d). They are relatively faint, but their positional shift with velocity can be clearly recognized. 
Finally, the bright arc D is visible at a small distance from the source (about 0\farcs9) and at intermediate velocities (20 \kms, see \cref{fig:ellipses2}c). 

In the following, we define as the axis of the system the symmetry axis of the disk. Its projection on the plane of the sky is shown as a solid green line in \cref{fig:ellipses2}, and is taken to be the reference axis of the positional offset $y$ from the source, with positive offsets toward SW (PA=228$\degr$ north to east). 
The orientation of the features deviates slightly from this direction, as is evident, for example, for the features of the B family
in panel \textit{b} of Fig. \ref{fig:ellipses2}.

In general, inspection of channel maps makes clear that the flow cannot be defined to have a simple and continuous conical shape. In contrast, the apparent conical border is revealed to arise from the superposition of the discrete limb-brightened arc-shaped substructures.

\begin{figure*}
  \centering
  \sidecaption
  \includegraphics[width=12cm]{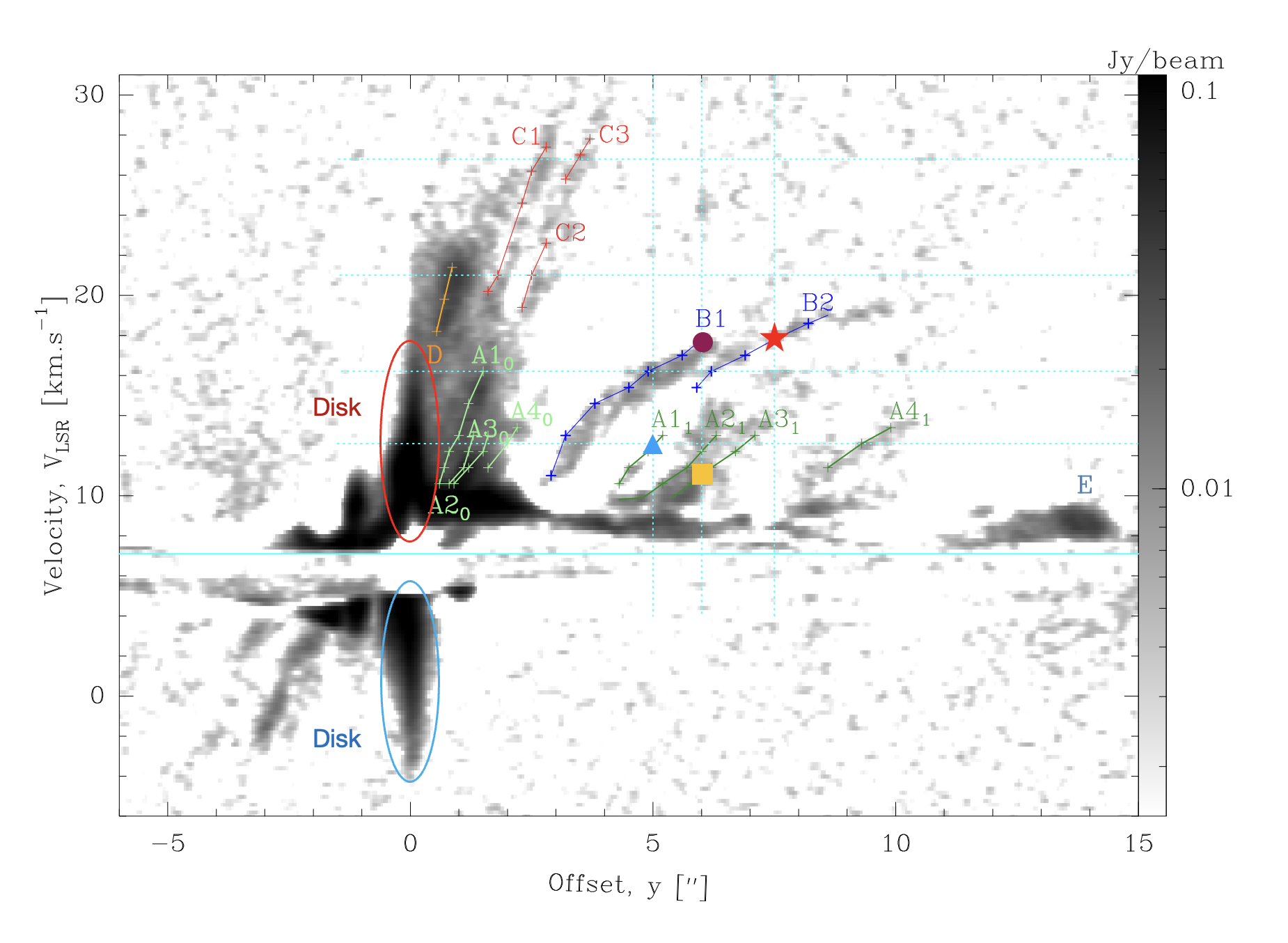}
   \caption{Longitudinal position-velocity diagram (\pvpar) formed along the axis (PA=228$^{\circ}$) with the source in $y=0$ and positive offsets toward the SW.
   For the redshifted lobe, and for each branch, the connected crosses correspond to the intersections of the ellipse traces in the channel maps with the axis, each family with a different color. For the A features, the subscript $0$ ($1$) refers to the intersections of the side of the ellipses closer to (farther from) the source. The horizontal dotted lines mark the \vLSR\, of the channel maps in Fig.~\ref{fig:ellipses2}; the vertical dotted lines are the offsets $y$ of the transverse PV diagrams shown in \cref{fig:PVperp-selec}. The solid cyan line is at \vsys. The symbols in color indicate corresponding emission points in the diagrams of \cref{fig:PVperp-selec}.}
          \label{fig:PVparB&W}
    \end{figure*}

\begin{figure*}[h]
 \centering
\includegraphics[width=0.9\linewidth]{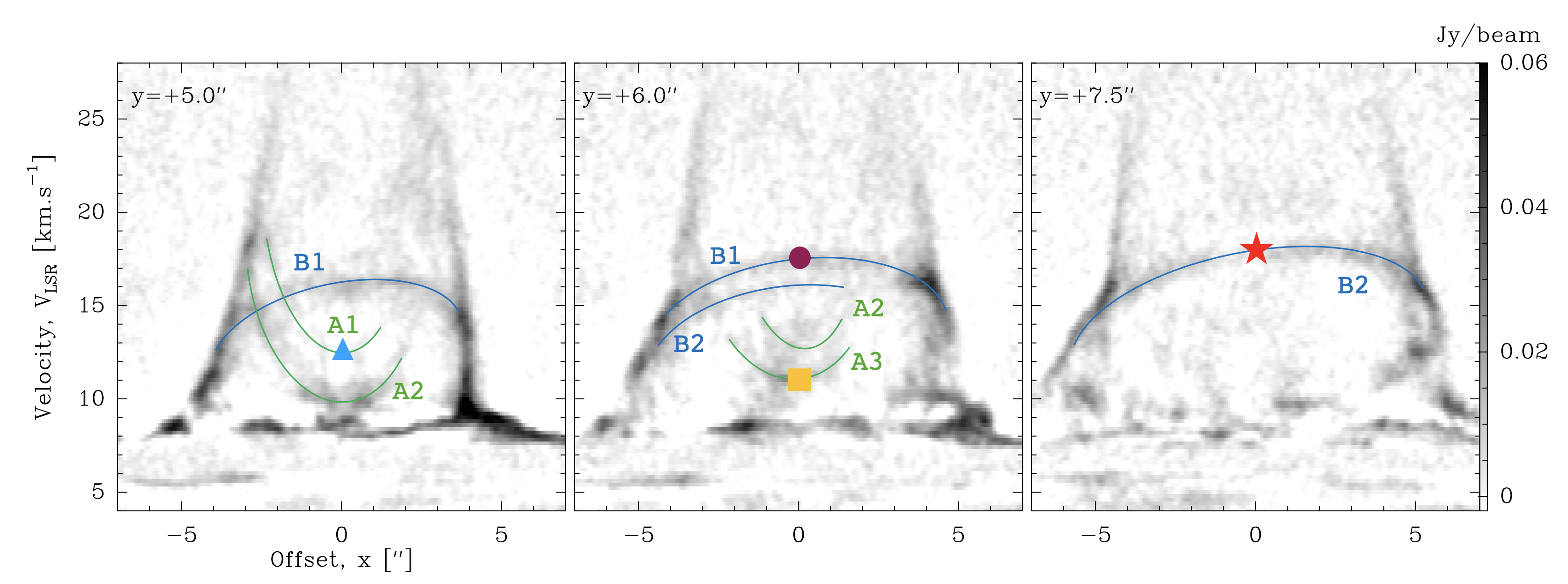}
   \caption{Selection of transverse position-velocity diagrams (\pvperp) formed perpendicularly to the axis at separation $y=5\farcs0,\, 6\farcs0$, and $7\farcs5$ from the source, and with orientation PA=318\degr\, (i.e., positive offsets toward the NW). The curves indicate the corresponding substructures in  \cref{fig:ellipses2} and \ref{fig:PVparB&W}, 
   labeled with the same name. The points of intersection of a given curve with the $x=0$ axis can easily be found in the other orthogonal projections of the datacube.  For example, the point in the \pvperp\, at $y=$7\farcs5 indicated by the red star, lying on substructure B2 at ($x=$0, \vLSR = 18 \kms), is found
   in the \pvpar\, of Fig. \ref{fig:PVparB&W} along the trace B2 at ($y=$7\farcs5, \vLSR = 18 \kms ), and in the channel map for \vLSR = 18 \kms, at the intersection of the arc B2 with the axis, located at $y=$7\farcs5 from the source.}
          \label{fig:PVperp-selec}
\end{figure*}

\begin{figure*}
 \centering
\includegraphics[width=15cm]{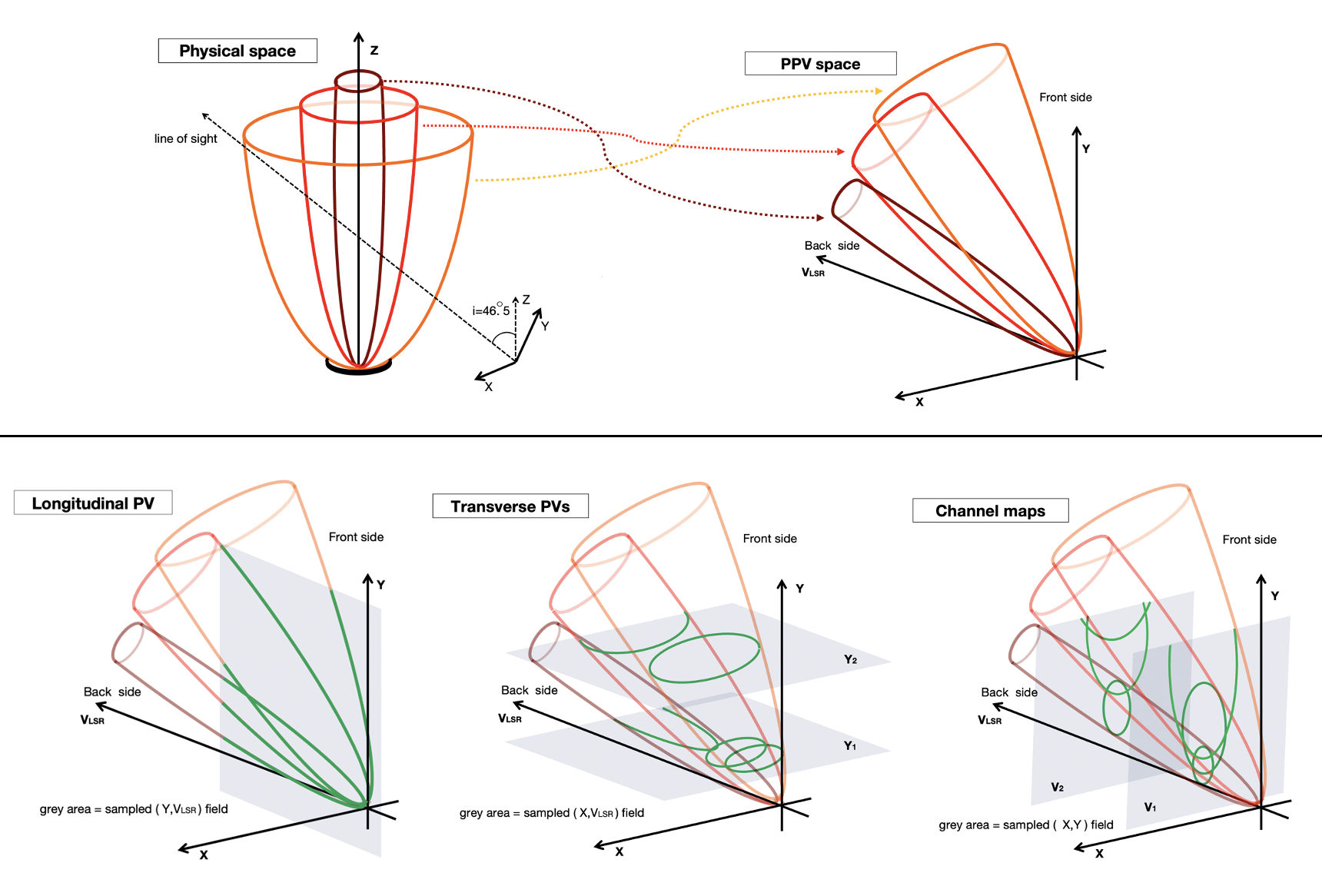}
   \caption{
 Proposed structure of the wind derived from the data analysis in the simplified case of  no rotation of the flow. {\em Top:} Nested shells in the physical  space (left) have progressively smaller radii, higher poloidal velocity, and faster acceleration for shells closer the axis, so that the corresponding curved surfaces
 in the  position-position-velocity (PPV) space are progressively closer to and more inclined toward the velocity axis 
 (right). {\em Bottom:} Examples illustrating the origin of the structures seen in the orthogonal projections of the datacube. 
    }
          \label{fig:sketch}
    \end{figure*}

\subsection{Parallel position-velocity diagram}
\label{su:PVpar}

Figure \ref{fig:PVparB&W} illustrates the position-velocity diagram  obtained with a virtual slit 0\farcs3 wide (comparable to the beam size) positioned along the axis (PA=228${^\circ}$).  
The diagram (hereafter referred to as PV$_\parallel$) collects emission from both circumstellar material and extended wind.
 The emission within offset $y = \pm$ 2 arcseconds from the source is due partly to the disk and partly from the base of the wind, whose large opening produces contributions in both redshifted and blueshifted velocities on each side (e.g., \citealt{Klaassen16}).
 The emission at larger offsets and velocities comes only from the extended wind. 
 
The striking characteristic of the diagram is the fact that the emission in the wind region is not continuous, but it is distributed in a fan of separated traces on both sides of the source (superposed to a low intensity more homogeneous emission in the region $y <2\arcsec $ and \vLSR $< 23 $ \kms). 
In the blue lobe, at least three linear traces are seen to propagate toward larger blueshifted \vLSR\, velocity.
On the SW redshifted side, more than ten traces can be recognized.
Within the first 4\arcsec a fan of almost linear traces opens from the source reaching high velocities, up to at least \vLSR $\sim$ 28 \kms.
For each of them, the velocity appears to increase linearly with offset, in a way reminiscent of a Hubble law trend. 
This group of traces displays the steepest slope, that is, the highest apparent acceleration, among the features in the diagram.
A second fainter fan of almost linear segments is observed between 4 and 11\arcsec, with lower \vLSR\, velocities up to $\sim$ 15 \kms.
Interposed between the two groups, two extended traces of curved
shape reach velocities as high as 20 \kms.

The substructures detected in the diagram suggest a common origin with the arc- and bubble-shaped features identified in the channel maps. 
To illustrate this hypothesis, for each of the ellipses described in Section \ref{su:ch-maps}, and for each velocity channel, we report on the redshifted portion of the \pvpar\, diagram the $y$ offset of the intersections of the fitted ellipses with the flow axis (see \cref{fig:ellipses2}). The subscripts $0$ and $1$ indicate the intersections of the side of the ellipse closer and farther from the source, respectively. The broken lines obtained by connecting the points relative to the same side of each ellipse are strikingly coincident with sections of the emission branches in the diagram. 

The first group of traces reaching high velocities corresponds to arcs D (orange), C (red) and to the intersections of type $0$ of features A (green). 
The second group corresponds to the SW sides of the A ellipses, marked as ``$1$'' (green). 
Given the inclination angle of the wind in space ($i$=226$^\circ$.5), the A$_0$ traces appear to correspond to the rear side of the flow and the A$_1$ traces to the front side. 
We note that the A$_0$ and A$_1$ traces appear in the same order from the source, in a way that is consistent with the fact that in the channel maps the ellipses of the A family are concatenated rather than nested.
The long curved traces correspond to the features of the B family (in blue). These filaments well reproduce the consistent shift in position with velocity illustrated in \cref{fig:ell_move}. 
The emission peak at offset $y=$ 14$\arcsec$ at a constant low velocity denoted E corresponds to the ridge of emission southwest of the source and almost perpendicular to the flow direction seen in \cref{fig:moment0maps}.

\subsection{Transverse position-velocity diagrams}
\label{su:PVperp}

Appendix \ref{app:PVperp} presents an atlas of the transverse position-velocity diagrams (hereafter PV$_{\perp}$) obtained for the red lobe with a 0\farcs3 wide virtual slit centered on the axis and oriented at PA=318\degr\, (i.e., perpendicular to the axis and with positive offsets toward NW). The diagrams are formed at the separation $y$ of
3\farcs5 to 10\arcsec\, from the source, in steps 0\farcs5 (see \cref{fig:ell_chmap16.2}).  
Illustrative examples are given in Fig. \ref{fig:PVperp-selec}, for selected separations from the star.

In all panels, the emission presents a bell shape around the $x=0$ axis, narrower toward high velocities. Progressing with separation from the source, the bell becomes wider, but the lateral walls maintain coherence.  
Outside of the bell, only low-velocity  material (7.5 - 9.5 \kms) is present. 

Internal structures in the form of arcs and branches of ellipse-shaped curves are found in all the panels, with a monotonic increase in average velocity and size as the separation from the source increases. This behavior is reminiscent of the variation with velocity observed for the substructures in the channel maps and in the \pvpar\, diagram. Despite the additional difficulty due to the signal confusion created by the overlap of the emission traces,
Figure \ref{fig:PVperp-selec} proposes an identification of visible curves with features belonging to the A and B family, with the same labels as in \cref{fig:PVparB&W}.

An evident characteristic of the structures identified in the transverse diagrams is the skew with respect to the horizontal. 
This property is related to the rotation of the flow.
The two opposite borders of the wind, in fact, present a velocity shift due to the projection of the rotation (i.e., toroidal) velocity along the line of sight.  An illustration of this effect is offered in \citet{DeValon22}, Fig. 2. The observed features display higher redshifted velocities on the NW side (seen more clearly for the curves of the B family)
consistent with a clockwise rotation observed from the tip of the red lobe toward the source. This direction of rotation matches that of the disk \citep{Garufi21}. 
\vspace{0.5cm}

In summary, the ensemble of maps presented in this Section indicates that the orthogonal projections of the 
position-position-velocity (PPV) datacube all show well-defined substructures, consisting of
concatenated curves in the velocity channel maps and in the transverse \pvperp\, diagrams, and of a fan of traces opening from the source region in the parallel \pvpar. The features can be connected from one projection to the other, and this leads to inferring the existence of coherent structures in the datacube.

\begin{figure*}
  \centering
    \includegraphics[width=0.95\linewidth]{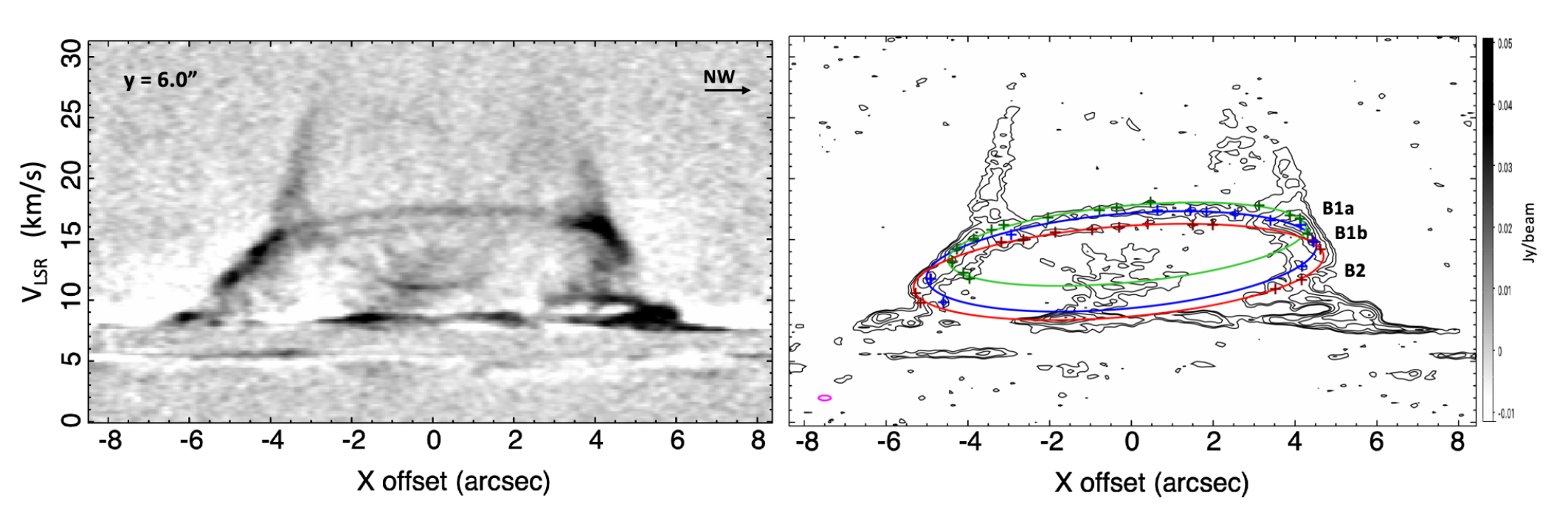}    \includegraphics[width=0.95\linewidth]{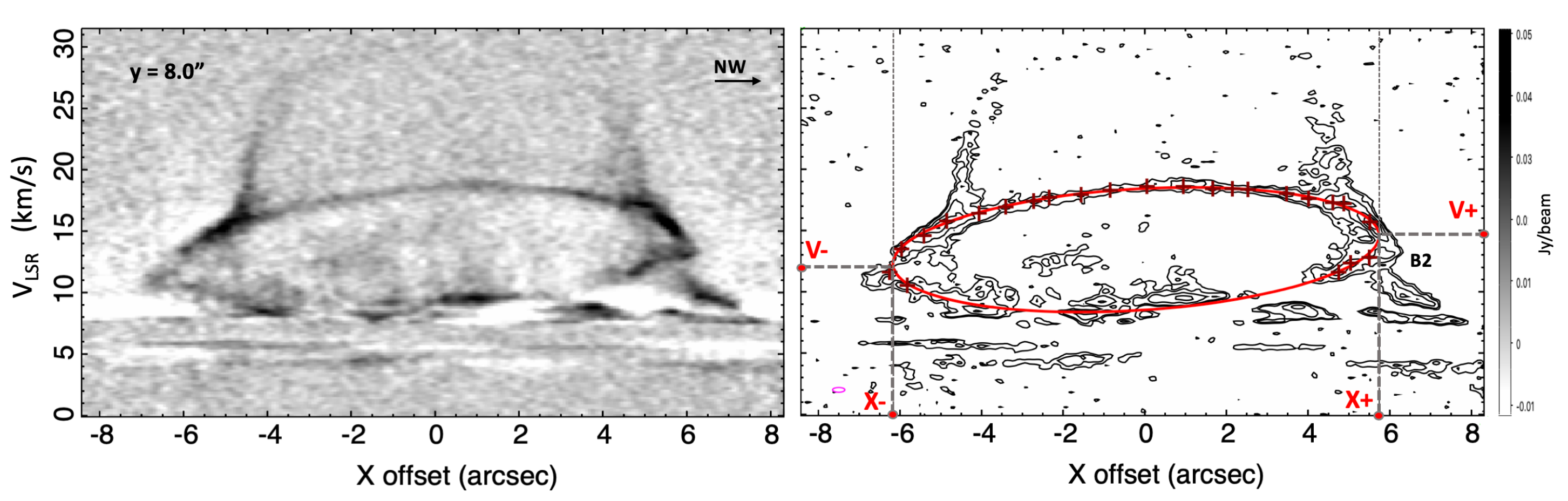}
    \caption{
    Examples of ellipse fit for the structures of family B in two \pvperp\, diagrams formed at offsets of 6 and 8 arcseconds from the source. Feature B1 visible at 6\arcsec turns out to be composed of two concatenated substructures (B1a, B1b). In the \pvperp\, at 8\arcsec only feature B2 is clearly identified. Details of the fitting procedure are described in \cref{app:PVperp}.
    The small magenta ellipse indicates the instrumental uncertainty. The bottom right panel illustrates the identification of the coordinates of the apexes ($x_{-},V_{-}$), ($x_{+},V_{+}$) reported in \cref{tab:PVperp_ell}. 
    }
    \label{fig:PVperp-fit}
\end{figure*}

\section{Interpretation of the substructures in terms of nested gas shells}
\label{su:shells}

In this section, we analyze what type of wind could determine
the emission properties in the PPV datacube. The description
above suggests a wind configuration in physical space
consisting of a series of nested rotating shells, which gradually open up
with distance from the source, with a progressively higher propagation speed
and a steeper speed gradient with distance
from the source for the shells closest to the axis. An illustration of this
hypothesis is provided
in the sketch of \cref{fig:sketch}, based on an axisymmetric flow composed of three such layers, which is assumed not to rotate for simplicity. 

In the real case, rotation and inclination introduce deformations of the traces in the projected maps.
The skew of the features due to flow rotation has been described in the previous Section. 
On the other hand, inclination with respect to the line of sight produces a deformation of the closed curves with respect to the pure ellipse shape, as described  in \cite{DeValon21}, among others. 
We adopt for the inclination angle $i$ of the redshifted flow  with respect to the line of sight a value $i = i_{\rm disk} = $ 46.5\degr\,. 
This compares well with the inclination angle of the redshifted axial atomic jet, $i_{\rm jet} = 45^{\circ} \pm 8^{\circ}$, determined by combining the tangential velocity measured for the most coherent [SII] emission knots on the SW side of the source in \cite{Anglada07} with the average radial velocity measured for the same jet lobe by \cite{Mundt90}. 
Due to this inclination,
the low-velocity part of the ellipses is expected to be flatter than the high-velocity part. However, this deformation is not easily recognizable in our maps, since the structures centered on moderate velocities are weak and/or do not have a visible upper part, while those centered on low velocities have their lower part possibly hidden by the absorption of CO gas with velocities close to \vsys. 

In any case, the overall appearance of the data supports an interpretation of the observed substructures in terms of distinct nested rotating shells in physical space. 
As mentioned in the Introduction, the same hypothesis has been put forward for other molecular winds that exhibit similar spectro-imaging properties in high-resolution data.

However, it is important to note that HL Tau is the only object among the known cases with a ring--gap configuration detected in the disk. In this case, therefore, it is possible to study the relationship between the substructures observed in the wind and those visible in the disk.

\subsection{Tomographic reconstruction of the wind shells}
\label{su:tomography}

\begin{table*}[!ht]
    \centering
    \caption{Properties of the features of family B measured in the transverse \pvperp\, diagrams and derived outflow parameters.}
    \begin{tabular}{cc|cccc|cccc|ccc|cc}
    \hline
    \hline
     &$y$ &$x_{-}$ &$x_{+}$ &$V_{-}$ &$V_{+}$ &$z$ &$r$ &$V_z$ &$V_{\phi}$  &${\alpha}$ &$V_p$ & $J = r V_{\phi}$ &$r_0$ & $\lambda_{\phi}$\\ 
   & \scriptsize{[arcsec}]  & \scriptsize{[arcsec]} & \scriptsize{[arcsec]} & \scriptsize{[\kms]} &\scriptsize{[\kms]} & \scriptsize{[10$^3$\,au]} & \scriptsize{[10$^2$\,au]} &\scriptsize{[\kms]} &\scriptsize{[\kms]} & \scriptsize{[deg]} &\scriptsize{[\kms]} & \scriptsize{[10$^2$\,au\,\kms]} & \scriptsize{[au]} & \\ 
       \hline
        
           & ~ & ~ & ~ & ~ & ~ ~~&& ~ & ~ & ~ & ~& ~& ~&\\
    B1a & 4.5 & -3.50 & 3.54 & 11.95 & 14.68 & 0.9 & 5.2 & 9.0 & 1.88 & 21.1 & 9.7 & 9.8 & 57 & 3.0 \\  
        & 5.0 & -3.88 & 3.84 & 12.40 & 15.10 & 1.0 & 5.7 & 9.7 & 1.86 & 21.1 & 10.4 & 10.6 & 58 & 3.2 \\  
        & 5.5 & -4.06 & 4.06 & 12.82 & 15.47 & 1.1 & 6.0 & 10.2 & 1.83 & 21.1 & 11.0 & 10.9 & 56 & 3.4 \\  
        & 6.0 & -4.35 & 4.27 & 13.25 & 15.95 & 1.2 & 6.3 & 10.9 & 1.86 & 21.1 & 11.7 & 11.8 & 57 & 3.6 \\  
        & 6.5 & -4.59 & 4.57 & 13.41 & 16.08 & 1.3 & 6.7 & 11.1 & 1.84 & 21.1 & 11.9 & 12.4 & 58 & 3.8 \\  
        & 7.0 & -4.84 & 4.95 & 13.46 & 16.11 & 1.4 & 7.2 & 11.2 & 1.83 & 21.1 & 12.0 & 13.2 & 61 & 3.9 \\  
      ~ &    ~ &     ~ &    ~ &      ~&     ~ &~&  ~ &    ~ &     ~ &     ~&      ~&  ~\\ 
    B1b & 4.5 & -3.89 & 3.67 & 10.80 & 13.40 & 0.9 & 5.6 & 7.3 & 1.79 & 23.2 & 7.9 & 10.0 & 71 & 2.8 \\  
        & 5.0 & -4.24 & 4.02 & 11.34 & 13.89 & 1.0 & 6.1 & 8.0 & 1.76 & 23.2 & 8.7 & 10.7 & 69 & 3.0 \\  
        & 5.5 & -4.50 & 4.41 & 11.67 & 14.18 & 1.1 & 6.6 & 8.5 & 1.73 & 23.2 & 9.2 & 11.4 & 70 & 3.2 \\  
        & 6.0 & -5.02 & 4.53 & 11.90 & 14.50 & 1.2 & 7.0 & 8.9 & 1.79 & 23.2 & 9.6 & 12.6 & 74 & 3.4 \\  
        & 6.5 & -5.15 & 4.76 & 12.09 & 14.65 & 1.3 & 7.3 & 9.1 & 1.76 & 23.2 & 9.9 & 12.9 & 73 & 3.5 \\  
        & 7.0 & -5.43 & 5.15 & 12.32 & 14.90 & 1.4 & 7.8 & 9.5 & 1.78 & 23.2 & 10.3 & 13.9 & 75 & 3.7 \\  
     ~ &    ~ &    ~ &   ~ &   ~&  ~ &  ~&  ~ &   ~ & ~ & ~ & ~& ~&  ~\\ 
    B2  & 6.0 & -5.34 & 4.70 & 11.10 & 13.60 & 1.2 & 7.4 & 7.6 & 1.72 & 17.5 & 8.0 & 12.7 & 90 & 3.1 \\  
        & 6.5 & -5.52 & 5.03 & 11.50 & 13.90 & 1.3 & 7.8 & 8.1 & 1.65 & 17.5 & 8.5 & 12.9 & 85 & 3.2 \\  
        & 7.0 & -5.70 & 5.52 & 11.60 & 14.10 & 1.4 & 8.3 & 8.4 & 1.72 & 17.5 & 8.8 & 14.2 & 89 & 3.5 \\  
        & 7.5 & -5.89 & 5.65 & 12.08 & 14.43 & 1.5 & 8.5 & 8.9 & 1.62 & 17.5 & 9.4 & 13.8 & 83 & 3.5 \\  
        & 8.0 & -6.18 & 5.73 & 12.29 & 14.57 & 1.6 & 8.8 & 9.2 & 1.57 & 17.5 & 9.6 & 13.8 & 80 & 3.6 \\  
        & 8.5 & -6.47 & 5.84 & 12.33 & 14.67 & 1.7 & 9.1 & 9.3 & 1.61 & 17.5 & 9.8 & 14.6 & 84 & 3.7 \\  
        & 9.0 & -6.56 & 6.09 & 12.41 & 14.70 & 1.8 & 9.3 & 9.4 & 1.58 & 17.5 & 9.8 & 14.7 & 84 & 3.7 \\  
        & 9.5 & -7.10 & 6.39 & 12.40 & 14.80 & 1.9 & 9.9 & 9.4 & 1.65 & 17.5 & 9.9 & 16.4 & 92 & 4.0 \\  
        & 10 & -7.16 & 6.53 & 12.45 & 14.87 & 2.0 & 10.1 & 9.5 & 1.67 & 17.5 & 10.0 & 16.8 & 93 & 4.1 \\  
       ~& ~ & ~ & ~ & ~ & ~ & ~ & ~& ~ & ~ & ~& ~&  ~\\
        \hline
     \end{tabular}
    \tablefoot{Listed parameters, from left to right: (1) Feature identification as in the bottom right panel of \cref{fig:PVperp-fit}; (2) Offset from the source of the considered \pvperp\, diagram; (3,4) $x$ coordinates and (5,6) \vLSR velocity of the SE and NW apexes of the fitted ellipse; (7) Offset from the source $z=y/\sin{i}$ deprojected along the flow axis (i=46.5$\degr$); (8,9,10) Shell radius $r$, axial velocity $V_z$, and toroidal velocity $V_{\phi}$ derived from \cref{eq:rvpvphi}; (11) Tangent angle to the shell, from the fit of $r=r(z)$; (12) Poloidal velocity $V_p=V_z/\cos{\alpha}$; (13) Specific angular momentum $J = r V_{\phi}$; (14) Radius of the shell footpoint calculated from  \cref{eq:andersonlaw};
    (15) Component \lamphi=$rV_{\phi} / r_0^2 \Omega_0$ of the magnetic lever arm $\lambda$ in \cref{eq:lambda}. 
    The uncertainties introduced by the derivation procedure are estimated to be 5\% on $r$, 4\% on $V_z$, and 10\% on $V_{\phi}$ (and hence 15\% for $J$). The dependence on angle $\alpha$ limits the accuracy in $V_p$ to 5\%. The uncertainty on $r_0$ is found to be $20\%$, from the sensitivity to variations of the input values in \cref{eq:andersonlaw}. This leads to a 25\% uncertainty on $\lambda_{\phi}$. 
   }
    \label{tab:PVperp_ell}
\end{table*}

The series of \pvperp\, diagrams illustrated in Appendix \ref{app:PVperp}\, allows, in principle, a tomographic reconstruction
of the wind shells that provides information on their morphology and kinematics in the physical space.
This can be done by measuring the width, inclination, and position of each detected feature in adjacent diagrams.

In practice, for each \pvperp\, diagram one has to identify the SE and NW apexes of the identified features and measure the corresponding position and \vLSR\, velocity. These coordinates are sufficient to determine the local shell radius $r$, the toroidal velocity $V_{\phi} \hat{e}_{\phi}$ and the axial velocity $ V_z \hat{e}_z$ (referring to a co-moving cylindrical coordinate system centered on the star, with $\hat{e}_z$ directed toward PA=228\degr with an inclination angle $i$=46.5\degr\,).
However, the task is obstructed by the confusion generated by the faintness of the signal and the overlap of structures at low \vLSR. The main difficulty lies in identifying a coherent feature that can be followed in its evolution from a diagram to the next. 
However, we illustrate the measurement by focusing on the features of family B, which are more clearly identified and have a regular shape. We note that a close inspection of the datacube projections
revealed that the B1 feature is actually composed of
two intersecting curves that we label B1a and B1b.
The analysis of the other fainter arcs, requiring a dedicated procedure, will be presented in the next paper of the series (Nony et al., in prep).

To ensure identification of the apexes in a way that is as reliable as possible, a cross-check was first performed with the positions of the
emission peaks in common with the other projections of the datacube. Secondly, 
a mathematical fit of the sequence of peaks along the identified traces was performed, assuming an ellipse as the geometrical curve closest to the shape of the features. 
The SE and NW extremes of the fitted ellipses, that is, the points on the curve furthest from the $x=0$ axis, were considered the vertices of each feature at the edges of the shell, and their coordinates were recorded. 
The procedure is illustrated in \cref{fig:PVperp-fit}, and details on sampling criteria and fitting routine, as well as a discussion of inherent limitations, are available in \cref{app:PVperp}.

Defining as ($x_{-,y},V_{-,y}$) and ($x_{+,y},V_{+,y}$) the coordinates of the SE and NW apexes of a given feature in the \pvperp\,diagram formed at separation $y$ from the source,  then:
\[
\centering
r_y = {{(x_{+,y} - x_{-,y})}\over{2}},
\]
\begin{equation}
\centering
   V_{z,y} = {{(V_{-,y}\, +\, V_{+,y} ) - 2 V_{\rm sys}}\over{2 \cos{i}}}  
 \hspace{0.3cm}
 V_{\phi,y} ={{(V_{+,y}\, -\, V_{-,y})}\over{2\sin{i}}},
\label{eq:rvpvphi} 
\end{equation}
where $V_{\rm sys} =$+7.1 \kms and $i$=46.5$^{\circ}$.\footnote{
In writing these formulas we neglected flow wiggling and precession. The validity of this assumption is discussed in \cref{app:wigggle}.}
Additionally, the tangent angle $\alpha$ to each shell was found with a linear fit
to the variation of $r_y$ with $z=y/\sin(i)$.
If the wind flows along the shell, the poloidal velocity $V_p =  V_r \hat{e}_r + V_z \hat{e}_z$ is tangential to the surface, and $V_{p,y} = V_{z,y}/ \cos(\alpha)$. To validate this assumption, one should measure independently $V_{r,y}$. 
Unfortunately, this measurement is complicated by the inclination of the flow with respect to the l.o.s., which distributes the information at displaced $y$ offsets (see \cref{fig.sketchvr}), in regions where the emission traces are faint and/or confused. This makes the test conditions prohibitive in almost all cases. However, we could measure $V_{r,y}$ and confirm the validity of the hypothesis at least for a few positions, as described in \cref{app:vr}. In the following, it is assumed that the flow streams along the shell surfaces. 
A similar but not identical tomographic analysis has been performed for the study of the flow from the young star DG Tau B in \citet{DeValon22}. The comparison of the two approaches is discussed in \cref{app:devalon}.

\subsection{Derived parameters of the wind shells}
\label{su:param_tomography}

\begin{figure}
    \centering
    \includegraphics[width=\linewidth]{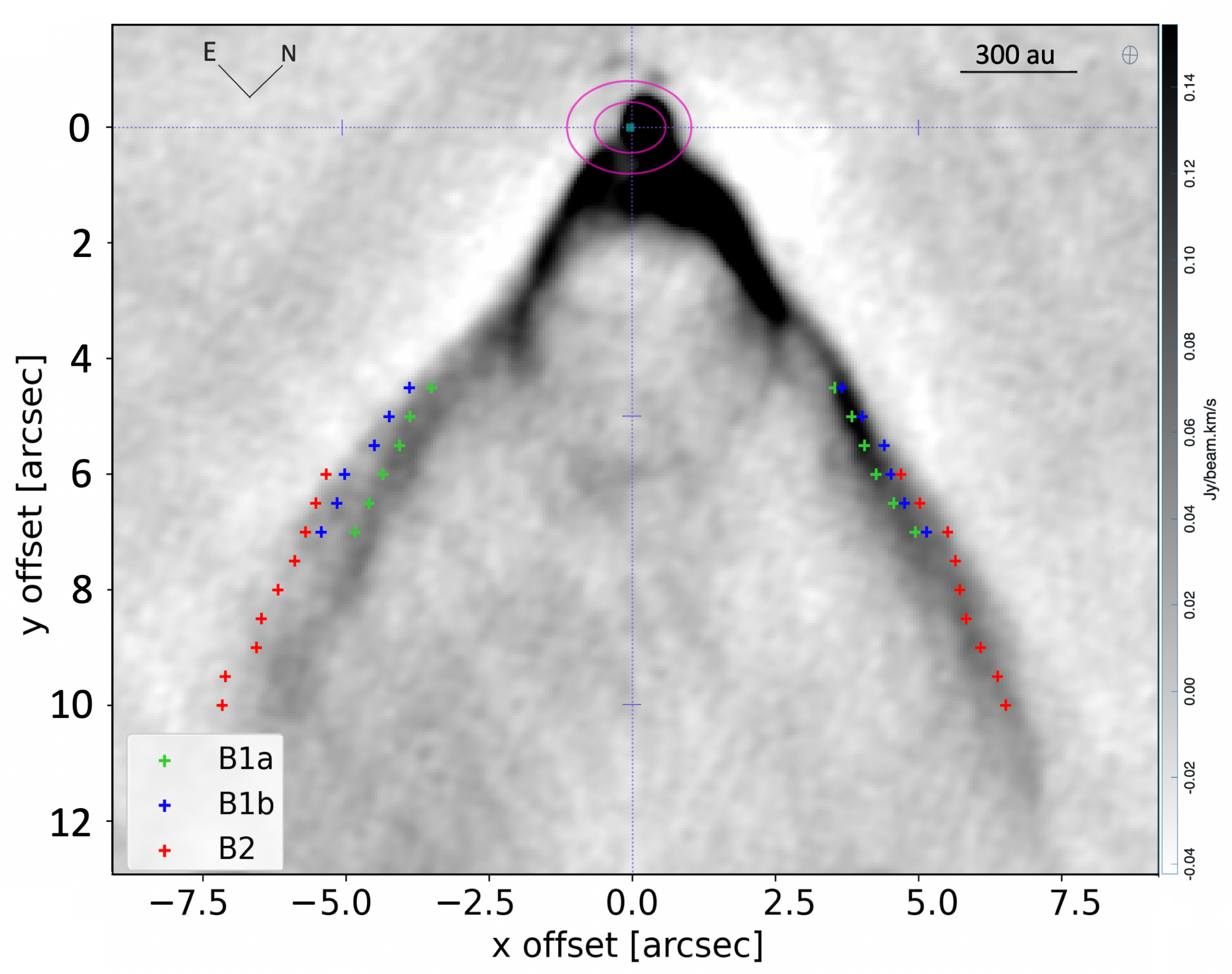}
\caption{
Position of the shells B1a, B1b, and B2 in the outflow, according to the tomographic analysis in \cref{su:tomography}. The  coordinates of the SE and NW apexes of the ellipse fit as in \cref{tab:PVperp_ell} are superposed to an intensity (moment 0) map of the redshifted lobe integrated over the velocity range 10.6 -- 16.2 \kms\, (characteristic of the considered traces), rotated 
clockwise by 48\degr to have the system axis along the vertical direction.
 The magenta ellipses are contours of the continuum emission at 10 and 200$\sigma_c$. The ellipse in the upper right 
 corner is the beam size in CO maps.
}
    \label{fig:xoffsets}
\end{figure}

\begin{figure}
    \centering
    \includegraphics[width=0.9\linewidth]{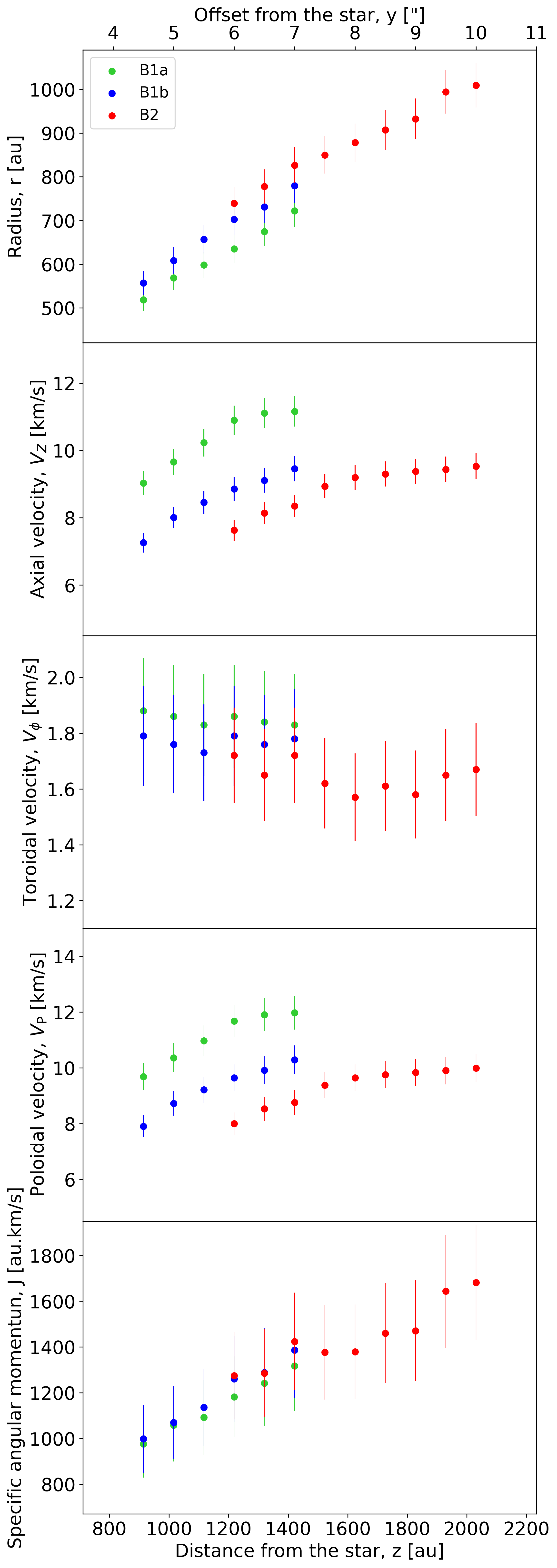}
    \caption{Variation in the physical parameters in the shells B1a (green), B1b (blue), and B2 (red) with  distance from the source  as derived from the tomographic analysis of the \pvperp\, diagrams in \cref{su:param_tomography}, and reported in \cref{tab:PVperp_ell}. 
    }
    \label{fig:rz-Vz-Jz}
\end{figure}

The results of the tomographic analysis for the offsets $y$ in which the B substructures are well identified are given in \cref{tab:PVperp_ell}. 
The same quantities are illustrated in Fig. \ref{fig:rz-Vz-Jz} as a function of both the offset $y$ and the deprojected distance from the source $z$ along the flow axis. For a visualization of the position of the three considered shells in the flow portion examined, see Fig. \ref{fig:xoffsets}, which reports the spatial offsets of the SE and NW apexes of the ellipse fit ($x_{-}$ and $x_{+}$ in \cref{tab:PVperp_ell}) superposed to a moment 0 intensity map integrated over the velocity range 10.6 - 16.2 \kms (characteristic of the studied traces), rotated to have the axis along the vertical.

The three shells turn out to have a similar trend for the radius, that increases from about
5$\times$10$^2$ to 10$^3$ au for distances from 10$^3$ to 2$\times$10$^3$ au from the star, and appear indeed to be nested, with B1a the innermost one and B2 the external layer.
The opening angle of B2, smaller than the ones of B1a and B1b, that were determined at offsets closer to the source, may indicate recollimation at large distances. 
The axial velocity $V_z$, between 7 and 11 \kms at the distances sampled, is higher for the inner shells and increases monotonically. In shell B2 the increase is less pronounced after
1.6$\times$10$^3$ au, and the same occurs for B1a and B1b, but closer to the source, after 1.3$\times$10$^3$ au. The poloidal velocity $V_p$, proportional to $V_z$, shows a similar trend.
The toroidal velocity $V_{\phi}$ is constant on average along a single shell, but is progressively higher for the shells closer to the axis, with an average along the flow of 1.85, 1.77 and 1.64 \kms, for B1a, B1b, and B2, respectively.  
The specific angular momentum $J=r V_{\phi} $ varies between 950 and 1700 au \kms. Its value increases monotonically and appears to be similar in the three shells. 
It is also interesting to estimate the dynamical time at each position along the propagation streamline based on the derived results for $V_z$. An illustration of this point is reported in \cref{app:tdyn}.

\section{Discussion} 
\label{disc}

As mentioned in the Introduction, different scenarios have been proposed for the interpretation of the recent observations of layered molecular winds. These scenarios are, in principle, also applicable to the HL Tau case. Ellipses in channel maps and transverse PVs, as well as an increase in velocity in longitudinal PVs are predicted by models of magnetized disk winds \citep{Ray21,DeValon22}, models of wide-angle winds driven swept-up shells \citep{Lee00, Zhang19}, unified models with a jet-bearing wide-angle X-wind \citep{Shang23, Ai24}. On the other hand, flow rotation is expected in both magnetically driven winds, which extract angular momentum from the disk, and thermal photoevaporative winds \citep{Alexander14}, which conserve their initial angular momentum, but do not slow down the rotation of the disk. 
The layered wind structure could also be due to the interaction with a pulsed axial jet (e.g., \citet{Rabenanahary22}). Each of these possibilities deserves a dedicated analysis.
In the following, we focus on the scenario of
extended MHD disk wind (DW), deferring the comparison of the observed trends with other classes of models to forthcoming papers.

\subsection{Basics of the MHD disk-wind theory}
\label{su:DWtheory}

According to disk-wind theory, a young star-disk system is embedded in a large-scale 
hourglass-shaped, twisted magnetic field inherited from the collapse of a magnetized rotating cloud. In this configuration, and provided sufficient ionization, 
a wind can be launched by the action of magnetic effects from the surface of the rotating disk.  The outflow carries away vertically angular momentum, and the consequent braking of the disk rotation allows accretion through the disk to occur.
Self-similar steady and axisymmetric DW solutions in ideal MHD conditions have proven to be successful in reproducing the observed properties of the axial atomic jets (see, e.g., \citealt{Frank14, Pascucci23}). 
The wind structure predicted by such models, that is, nested rotating gas shells with higher poloidal velocities along the inner shells, is actually observed also in the outer CO outflows, as shown above for the HL Tau case.
However, to properly describe global disk winds from an extended launch region, the models have to include non-ideal effects in the disk. Among them, ambipolar diffusion (AD) between neutrals and ions becomes dominant when ionization is scarce (e.g. \citet{Pinto08}). Magnetized winds also arise in this case,  as described, for instance, in the review by \citet{Lesur23}.

An important quantity of the DW theory is
the parameter $\lambda$, which represents the ratio between the extracted and the initial specific angular momentum. Following the definition of \citet{Blandford-Payne82}, $\lambda$ can be expressed as
\begin{equation}
\lambda= L / {\Omega_0 r_0^2}  = \frac{\Omega r^2}{\Omega_0 r_0^2} - \frac{r B_{\phi} B_p}{ \Omega_0 r_0^2 4\pi \rho V_p}, 
\label{eq:lambda}
\end{equation}
where $L$ is the total specific angular momentum carried away by
the MHD wind along a streamline, in the form of both matter rotation and magnetic torsion. $\Omega_0$ is the Keplerian angular rotation speed at the launch point $r_0$, $\rho$ is the mass volume density and $B_p$ and $B_{\phi}$ are the poloidal and toroidal components of the magnetic field in the wind. The higher the value of $\lambda$,  the more efficient is the extraction of angular momentum and the
acceleration of the wind.  The parameter $\lambda$ is often referred to as the 'magnetic lever arm', as it can be expressed as $(r_A / r_0)^2$, where $r_A$ is the  Alfv\'en radius, that is, the cylindrical radius at which the poloidal velocity reaches the Alfv\'en velocity $V_A = B_p /(4 \pi \rho)^{1/2}$. The locus of points in the wind where this occurs is called Alfv\'en surface, and it indicates the region above the disk where the increasing inertia of the flow starts to overcome the magnetic push. In self-similar steady DW solutions, this surface has a conical shape.

In the DW scenario $\lambda$ is constant along a streamline, which means that the angular momentum is gradually transferred from the magnetic field to the motion of the particles. Continuing along the streamline, the component $\lambda_{\phi} = \Omega r^2 / \Omega_0 r_0^2 $, representing the normalized specific angular momentum, increases and approaches $\lambda$ in the asymptotic regime, in which the magnetic contribution is sensibly reduced.  The distance from the disk at which this occurs is larger for higher values of $\lambda$, and can be several orders of magnitude higher than the altitude of the Alfv\'en surface. However, $\lambda_{\phi}$ may never reach $\lambda$, as the magnetic field maintains asymptotically a fraction of the available energy.

The importance of gas pressure versus magnetic effects in the disk can be parametrized by the plasma $\beta$ at the midplane, defined as
$\beta = 8 \pi\rho_0 c^2_s/{B_0}^2$, 
where $c_s$ is the sound speed and ${\rho}_0, B_0$ are the mean density and vertical magnetic field threading the disk.
Unfortunately, the magnetic field amplitude in the disk is not available from observations.
However, according to theoretical studies for self-similar disk wind solutions, $\lambda$ is expected to increase with smaller $\beta$ (e.g., \citealt{Jacquemin-Ide19, Lesur21}). 
It follows that the determination of $\lambda$ from the observations is crucial not only to identify the properties of the wind, but also to constrain the importance of magnetic effects in the disk.

Self-similar DW solutions from fully turbulent disks have been found for a wide range of $\lambda$  values at both high and low $\beta$ \citep{Ferreira97, CasseFer00, Jacquemin-Ide19, Zimniak24}.
On the other hand, in disk regions where the ionization is quenched in the midplane, as expected in the so-called 'dead zone' of the disk, solutions are dominated by AD, and low values of $\lambda$ are expected, up to about 3 \citep{Lesur21}. Recent results show that the dead zone could actually start closer than 1 au from the star in T Tauri disks \citep{Flock19}, and extend to the outer disk regions of interest here. 

In the next Section the principles of the DW theory will be applied to the derivation of relevant quantities in the HL Tau wind, such as the location of origin of the flow in the disk, and the magnetic lever arm. The variation with distance from the source of the wind parameters will also be tested against the prescriptions of these models.

\subsection{Derivation of the shell footpoints in the disk wind scenario}
\label{su:footpoints}

Regardless of the conditions adopted for the disk physics, the wind can be treated in ideal MHD. Assuming that the flow is a cold stationary and axisymmetric disk wind, from the measurements described in Sect. \ref{su:tomography} one can infer the so-called footpoint radius $r_0$, that is, the radius of the disk annulus from which the wind shell originates. Following the method introduced in \citet{Bacciotti02} and refined in \citet{Anderson03}, $r_0$ can be found as the real solution of the following third-order equation:
\begin{equation}
r V_{\phi} (G M_{\star})^{1/2} r_0^{-3/2} - 3/2 (GM_{\star})r_0^{-1} -  (V_p^2 + V_{\phi}^2 )/2  =  0.
\label{eq:andersonlaw}
\end{equation}
Here $G$ is the gravitation constant and $M_{\star}$ the mass of the central source. The equation is valid if the enthalpy and the gravitational force exerted by the star are negligible with respect to the kinetic terms at the height above the disk where the wind parameters are measured. These conditions are fulfilled for cold CO molecular flows with poloidal velocities around 10 \kms observed at more than 500 au from a star of solar type, conditions met in our case. 
In addition, it is assumed that the flow propagates freely. Actually, the environment around HL Tau has a very complex structure, and before reaching the observed location the outflow may have interacted with several elements disturbing the steady propagation, such as accretion streamers \citep{Garufi22}, winds inflated by nearby stars \citep{Welch00}, surrounding cavities, and with the lateral wings of the bow-shocks of the inner wind components. However, the regular geometry observed on large scales supports the assumption that the kinematical information survives the occurring perturbations. 
We note that the observed emission in the CO 2-1 line is seen just because of the density contrast, even if the flow is perfectly laminar, as the emission does not require shock heating to be produced, since the energy of the upper level is only 17 K.  However, the presence of external perturbations cannot be excluded a priori, and our simple model should be regarded as a first approximation to the real situation. The investigation of such effects has been deferred to forthcoming work.

\begin{figure}
    \centering
    \includegraphics[width=0.95\linewidth]{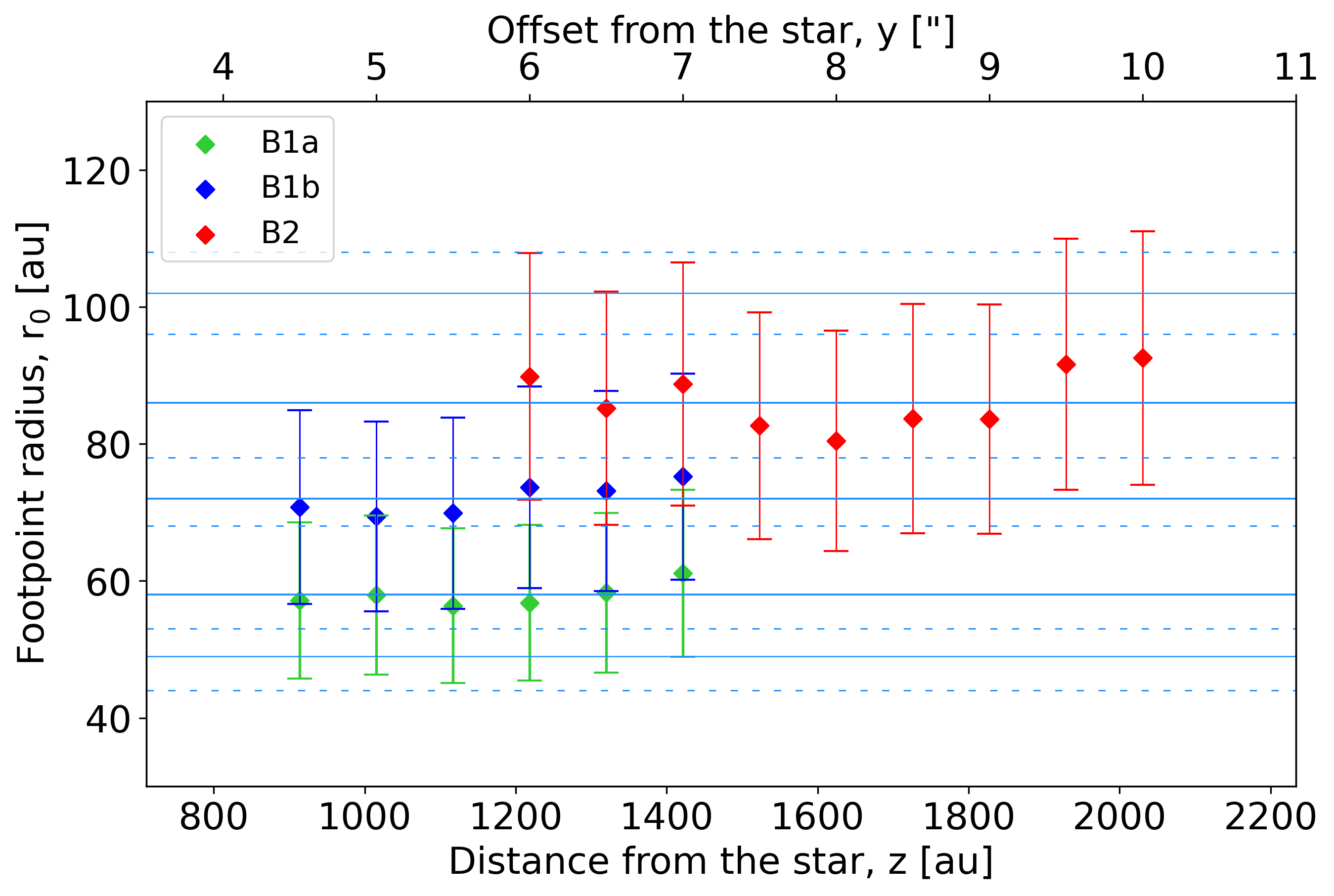}
    \caption{
    Footpoint radii $r_0$ of the B1a, B1b, and B2 shells determined from \cref{eq:andersonlaw} at increasing offset from the source (see \cref{tab:PVperp_ell}).
    The solid and dashed lines indicate the position of rings and gaps, respectively, in the continuum emission of the outer region of the disk \citep{ALMA15,Stephens23}.
    }
    \label{fig:footpoints_vs_z}
\end{figure}

The results of the application of \cref{eq:andersonlaw} to the features of family B are collected in column (14) of Table \ref{tab:PVperp_ell}, 
and are graphically illustrated in \cref{fig:footpoints_vs_z}. 
For each wind shell, the footpoint radii derived from the measurements at different separations from the star turn out to have an almost constant value, with a spread of only a few au.

Interestingly, the footpoint radii of the B1a, B1b, and B2 shells are found to correspond to the radii of three consecutive rings in the dust emission distribution in the disk, located at 58, 72, and 86 au from the source, respectively (labeled B4, B5, and B6 in \cite{ALMA15}).\footnote{The position of the rings on the disk surface given in \cite{ALMA15} were rescaled for a distance from HL Tau of 147.3 pc, following \cite{Galli18}.} 
However, we note that the uncertainty of 20\% in the determined footpoint radius (estimated by testing the sensitivity of the calculation in \cref{eq:andersonlaw} to variations in the input parameters) also covers the positions of adjacent disk substructures.

\begin{figure*}
    \centering
    \includegraphics[width=0.95\linewidth]{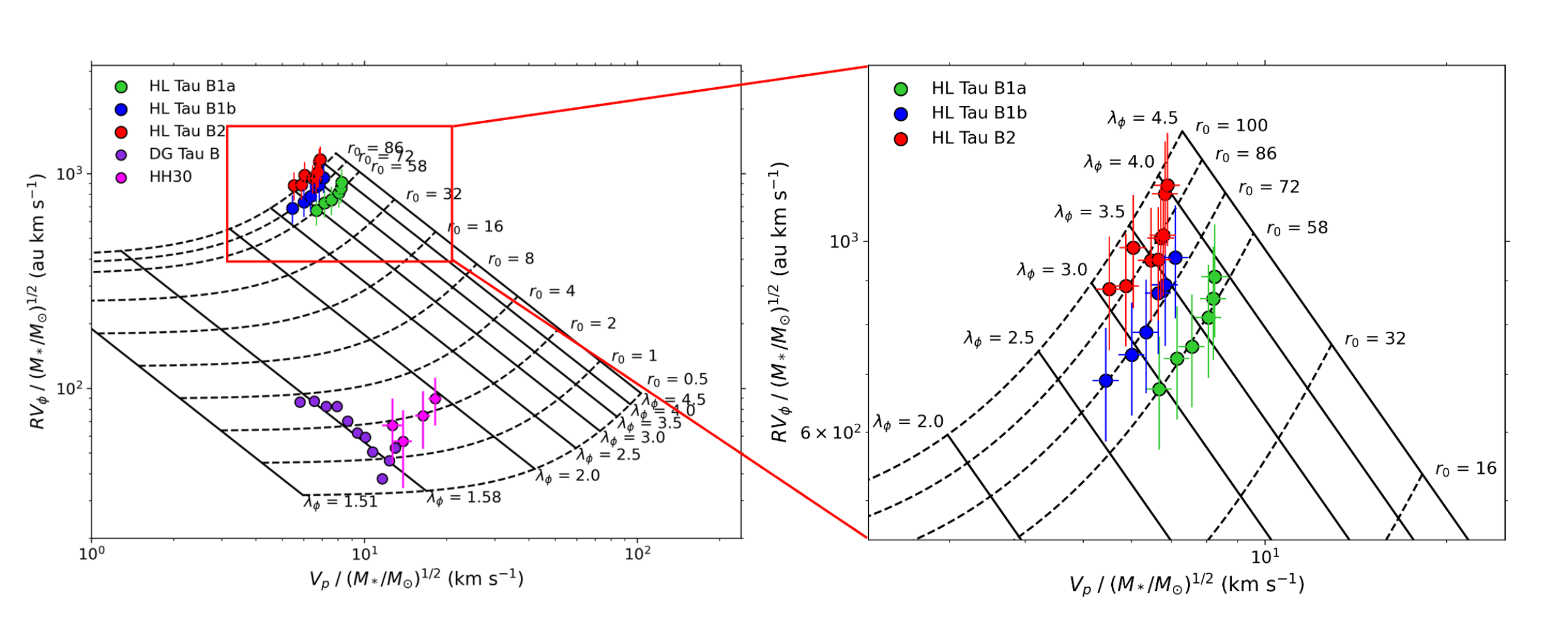}
    \caption{Footpoint radii $r_0$ and parameter $\lambda_{\phi} = \Omega r^2 / \Omega_0 r_0^2$ determined for the B1a, B1b, and B2 shells following the diagnostic procedure in \citet{Ferreira06}. The diagram also reports the results published for  the wind from DG Tau B \citep{DeValon22}, and for the wind from HH 30 \citep{Louvet18, Lopez24}. We note that the results for DG Tau B are derived with a different selection of the input data in the tomographic analysis, corresponding to different assumptions on the structure of the wind (see \cref{app:devalon}).  
    }
    \label{fig:footpoints}
\end{figure*}

An equivalent derivation of the shell footpoints is illustrated by the diagnostic diagram of \cref{fig:footpoints}, constructed following the procedure described in \cite{Ferreira06}. Under the same assumptions, in fact,  it can be shown that $J$ and $V_p$ for a given streamline depend only on the two parameters $r_0$ and $\lambda_{\phi}$. 
The location of the measured ($J,V_p$) values on the grid of theoretical curves drawn in the diagram gives estimates of the radius of the footpoint r$_0$ and of the parameter $\lambda_\phi$.
In HL Tau, for each individual shell the points lie along or near a curve
at a defined value of $r_0$, corresponding to the position of one of the disk rings B4, B5, and B6, and progress toward higher values of $\lambda_{\phi}$ as the distance from the source increases. As discussed in the next section and in \cref{app:trendsz}, this trend is consistent with the one expected in a DW scenario. 

We warn that the positions of the wind footpoints should be compared with the distribution of disk substructures in the gas, rather than in the dust, and these may not coincide. It is well known that bumps in gas pressure act as dust traps, but the average shift of the peak positions of gas pressure and dust emission depends on the model adopted (see, e.g. \citealt{Riols20, Hu22}).
\cite{Yen19a} reports profiles of gas emission in the HL Tau disk in the molecular lines of HCO$^+$(3-2) and HCO$^+$(1-0). The profiles are quite shallow, with no clear substructure detected, but it is interesting to note that the footpoint radii of the B-shells are in the vicinity of peaks in the HCO$^+$(3-2)/HCO$^+$(1-0) ratio (see their Fig. 9), which, under the assumption of constant temperature, can be considered proportional to the gas density.
According to this comparison, therefore, the wind shells appear to arise from gas rings. However, the association of the footpoint locations with rings in the gas distribution remains tentative at this stage, as a more reliable identification would require higher precision in the measurements
of the gas tracers.

The derived footpoint radii are larger than our spatial resolution limit, which is about 40 au. With reference to \cref{fig:chmap_blue}, the base of the redshifted wind appears resolved in the channel maps at low velocities between \vLSR = 7.4 and 9.4 \kms. However, the emission pattern is not clear in these maps, as contributions at low velocity and close to the star may come both from the outflow base and from parts of the envelope, possibly from accreting streamers, or from some parts of the disk.
Nevertheless, as discussed in \cref{app:chmaps}, this velocity interval is consistent with that in which, under the DW hypothesis, we expect to resolve the flow base in these observations.
In the channel maps at \vLSR$>9.5 $ \kms the wind base is not resolved on the SE side. At this velocity the flow appears to originate within a region inside the resolution limit of 40 au, it propagates initially with a large opening angle, up to a projected distance of about 0."5 from the disk, and then it appears to recollimate. 
This trend is again consistent with the DW scenario, following which the inner shells coming from an unresolved footpoint are subject to a faster acceleration, have an initial large opening angle, which then reduces to the collimated value (cf. \cref{fig:rz-Vz-Jz.vs.model}, first panel). We could not apply the same analysis to the NW side as the peak of the disk emission at these velocities covers the signal from the wind.

\subsection{Variation of parameter \lamphi\, in the wind shells}
\label{su:lamphi}

In \cref{su:footpoints}, we determined $\lambda_{\phi}$ at each sampled distance from the source either through  its direct calculation, as in \cref{tab:PVperp_ell}, or graphically as in the diagram of \cref{fig:footpoints}.  For all points relative to HL Tau $\lambda_{\phi}$ is within the range $\lambda_{\phi}$ = 2.5 - 4.0, and
for each shell, the points relative to \pvperp\, in subsequent offsets follow curves at constant $r_0$, and are found at higher values of $\lambda_{\phi}$ as the distance from the source increases. 
We note that the same behavior is shown by the points corresponding to theoretical DW models, in a zone in which
the asymptotic regime has not yet been reached (see \cref{app:trendsz}).
This appears to indicate that also in HL Tau we are probing a transition region in which the magnetic field is still transferring angular momentum to the matter. However, the increasing vicinity of the points for higher $\lambda_{\phi}$ suggests that the distances probed are not far from the asymptotic regime.
This point should be farther away for the outer streamlines than for the inner ones, due to the self-similar nature of the flow \citep{Ferreira06}.  This is in line with the fact that a similar value of $\lambda_{\phi}$ is attained closer to the star for the B1 shells than for B2 (sampled at a larger distance).

We also show in Fig. \ref{fig:footpoints} the values of $J$ and $V_p$ derived for the shells identified in the CO outflow of HH30 \citep{Louvet18, Lopez24} and in the bright conical CO outflow of DG Tau B \citep{DeValon22}. Interestingly, the $J$ and $V_p$ measurements in HH 30 are consistent with a curve at constant $r_0$, while the points in DG Tau B follow a curve at constant $\lambda_{\phi}$. 
We caution, however, that the choice of the sampling points of the tomography in DG Tau B differs from ours (see \cref{app:devalon}). 
In fact, the behavior determined in DG Tau B
is expected in the case in which the tomography is probing a surface of varying $r_0$ in a homogeneous disk wind. 
Averaging our ($J,V_p$) values along each HL Tau shell would lead to a trend similar to that observed in DG Tau B, although with higher values of $\lambda_{\phi}$ and $r_0$.

\subsection{Comparison with self-similar MHD disk-wind solutions}
\label{su:comparison}

\begin{figure}
    \centering
    \includegraphics[width=0.86\linewidth]{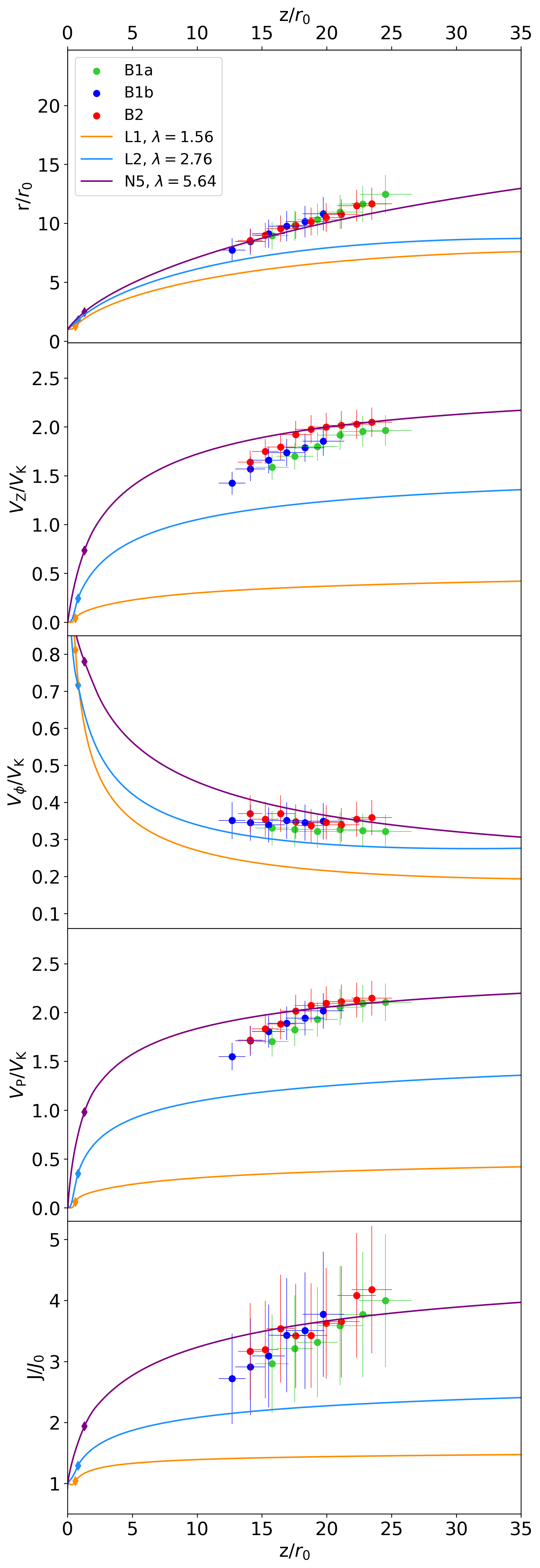}
    \caption{{\em Symbols}: Parameters for the shells B1a, B1b, and B2 as in \cref{tab:PVperp_ell}, normalized by the average $\bar{r}_0$ and $V_K (\bar{r}_0)$ of each shell. {\em Curves}: Trends predicted by the models L1 ($\lambda$=1.56, orange) and L2 ($\lambda$=2.76, blue) of winds from an AD-dominated disk region,  and by the model N5 ($\lambda$=5.64, purple) of a wind from a fully turbulent disk region (see text). The diamonds indicate the altitude at which the streamline crosses the corresponding Alfv\'en surface. 
    }
    \label{fig:rz-Vz-Jz.vs.model}
\end{figure}

The quantity \lamphi\, is only one component of the magnetic lever arm. To find observationally the value of the full $\lambda$ with the methods described above, one should examine the wind in the shell until the point at which asymptotically \lamphi\, reaches $\lambda$. However, this is hampered by the progressive weakness of the signal and by the interaction of the flow with the environment. A possible solution is to compare the variation of the wind parameters in the three shells with the trends prescribed by the DW models calculated for different values of $\lambda$. 
This also constitutes a test of the validity of the application of the DW theory to the observed outflow.

We did not attempt to explore the full range of possible self-similar solutions, but we show expectations from only three wind models that together encompass the range of $\lambda$ values compatible with the observations. Turbulence-dominated disk models should be used for MHD winds anchored below a few au, and AD-dominated models farther out. However, in both cases, the MHD wind is described by the same set of ideal MHD equations, and what differs is only the physical conditions at the jet launching point. 

We used two solutions, L1 and L2, corresponding to cold disk winds arising from an AD dominated disk \citep{Lesur21}, with, respectively, $\beta=10^5$ and $\beta=$63 and corresponding to $\lambda$= 1.56 and $\lambda$= 2.76.    
Both solutions assume an ambipolar diffusion parameter $A_m$ = 1 and no Ohmic diffusion. The disk and its atmosphere were assumed to be cylindrically isothermal (sound speed $c_s(R, z) = c_s(R, z = 0)$), with a disk aspect ratio $\epsilon = h/R$ = 0.1, where $h$ is the geometric thickness of the disk ($ h = c_s(R,z = 0)/\Omega_K(R)$, with $\Omega_K(R)$ the Keplerian angular velocity). The $\lambda$ value achieved by solution L2 is the highest published so far for AD-dominated disks.\footnote{The solutions in \citet{Lesur21} used in our work are freely available at https://github.com/glesur/PPDwind.}  

In order to investigate the comparison with a MHD wind with a higher magnetic lever arm, we computed a warm solution corresponding to the case of a fully turbulent disk, similar to the one presented in \citet{Tabone20} to reproduce the ALMA observations of the rotating SO wind in HH212. Our solution N5 describes a wind that arises from a thinner disk with $\epsilon $ = 0.01, $\beta \sim $ 10, an effective magnetic Prandtl number of 1, a turbulence parameter $\alpha_m$=2.3, a turbulence anisotropy $\chi_m $= 4.2, and no turbulent pressure ($\alpha_P$=0) \citep{Zimniak24}. Despite its large magnetic field, the ejection efficiency is high and the resulting outflow is dense. It achieves a magnetic lever arm $\lambda=$ 5.64 thanks to additional heating at the base of the wind (see the description in \citealt{CasseFer00b}). This heating is localized at the disk surface and requires only 2\% ($f$=0.02, see Eq. 25 in that work) of the dissipated power within the disk. Farther up, the wind becomes adiabatic.

The comparison with the models is illustrated in Fig. \ref{fig:rz-Vz-Jz.vs.model}. The points relative to the three B shells, with the same color code as in Fig. \ref{fig:rz-Vz-Jz}, 
describe the wind parameters normalized by their respective average footpoint radius $\bar{r}_0$ and by the corresponding Keplerian velocity $\bar{V}_K= V_K(\bar{r}_0)$.
The bottom curves are from the wind model from the AD-dominated disk region in \citet{Lesur21},  with curve L1 (orange) corresponding to the solution  for $\beta=10^5$, $\lambda$=1.56, and curve L2 (blue) for  $\beta = 63$, $\lambda$=2.76. 
The upper curve N5 (purple)  corresponds to the model of wind from a fully turbulent disk region that provides $\lambda$=5.64 (cf. \citealt{Zimniak24}).

First, it can be noted that once normalized, the points for the three observed shells in each panel tend to overlap. This is expected for self-similar solutions, in the case in which $\lambda$ is similar for all shells. 
Secondly, the values of the normalized data are in the same range as those predicted by the models. The variation of the quantities with $z/r_0$ is similar to that of the models in this range, but the gradient in the data points appears to be steeper than in the theoretical curves, for all quantities except $V_{\phi}$.

The top panel shows the comparison with the expected width of the shells. The opening angle of the wind is consistent with that predicted by the model N5 at higher $\lambda$. For the other quantities, 
the observed points lie in an intermediate position between the curves predicted by the two models L2 and N5, corresponding to an expected asymptotic value of $\lambda$ between about 4 and 5 and $\beta$ between 63 and 10.

The fact that the observational points lie between the two types of solution may have interesting implications.
For winds coming from disk regions dominated by turbulence, the usual assumptions lead to a limiting anchoring radius of a few au; so in principle our observations, indicating larger footpoint radii, should be better described by the L solutions, corresponding to winds arising from AD-dominated disk regions.  

This may indicate that the plasma $\beta$ in the outer regions of the HL Tau disk is smaller
(that is, the magnetic effects are more important) than commonly assumed for this zone in protoplanetary disks. A stronger field produces stronger vertical compression and reduced wind density, leading to a larger $\lambda$. However, other effects enter the wind acceleration mechanism, such as the aspect ratio of the disk, the existence of heating at the disk surface, the amount of effective turbulence, the degree of diffusion anisotropy, and the amount of ambipolar diffusion present in the disk.
Regarding the last point, the outer region of the disk might possibly present a level of ambipolar diffusion consistently higher than that assumed in solutions of type L, related to an unexpectedly low level of ionization. This might be due to the screening of cosmic rays (the dominant ionization agent at 50 au) by magnetic mirroring from the wind, as proposed in \citet{Cleeves15} for the TW Hya disk from chemical arguments. Reducing the irradiation of these particles could thus reduce the mass loading on the field lines and increase $\lambda$ \citep{Lesur21}.
An AD-dominated scenario, leading to a laminar-type configuration in the region where rings/gaps are observed, would also be in line with the fact that the rings appear to be remarkably axisymmetric in HL Tau, pointing toward a low level of turbulence in the plane of the disk \citep{Pinte16}.

In summary, the physical properties of the molecular CO wind from HL Tau appear to be consistent with a description in terms of a magnetized wind originating from a large region of the disk, up to 90 au and possibly beyond. The idea that not only the axial atomic jet originated in the first au from the star, but also the outer wind may be magnetic in nature implies that feedback from flows onto the disk may occur on a large scale. This may have important consequences for models of disk evolution and planet formation.
First of all, such winds may be able to sustain accretion along the disk surface at large distances from the star, in regions where the disk is internally weakly ionized and MHD turbulence, and hence viscous accretion, is suppressed. Other effects may also come into play. 
As discussed in \citet{Pascucci24}, for example, magnetically driven winds can slow the drift of solids in the direction of the star and affect the migration of planetary embryos.

\subsection{Implications of the observed shell separation}

Up to this point, the discussion regarded observational trends along the direction of the wind streamlines, indicating an overall agreement with a DW scenario. 

However, the most puzzling evidence that emerges from the data is the inhomogeneity of the wind across its width, as indicated by the presence of distinct nested shells with enhanced emission. This property, not expected in a standard self-similar DW, is actually found in many molecular outflows observed at the high spatial and spectral resolution offered by ALMA (e.g., \citealt{Zhang19, Fernandez20, Lopez24, Ai24, DeValon20, DeValon22, Omura24, Nazari24}, cf. Introduction). 

The analysis of the HL Tau wind offers the advantage of  a comparison between the structures in the wind and in the disk. The derived configuration of the wind shells, the large estimated footpoint radii, and the indication of origin of the shells from denser gas rings of the disk appear to support recent models that prescribe the spontaneous formation of a ring--gap substructure in the disk and of a connected inhomogeneous wind, as a result of magnetic instabilities \citep{Bethune17, Suriano18, Suriano19, Riols19, Riols20, Cui21, Hu22}.
In these numerical studies the secular evolution of protoplanetary disks is studied including ambipolar diffusion in the generalized Ohm's equation. As a result, one sees
the rapid development of a self-organization of the magnetic field in the disk, which accumulates in the regions of lower mass density. In this way a system of rings and gaps forms quickly in the disk, with a greater concentration of magnetic field in the gaps. The final configuration also sees the formation of a transversely inhomogeneous disk wind arranged in nested shells, with alternate layers of enhanced density originating from the gaseous disk rings and layers of faster lighter flow originating from the highly magnetized gaps. The ALMA observations like the ones described in this paper provide maps of the intensity of the line emission, and hence are prone to detect the wind layers of higher density, that is, in the above scenario, the layers arising from the disk rings.

The interest of this approach is the alternative it offers to models in which the disk substructure is due to the presence of young planets. 
At the time of writing, no planetary-mass bodies have been detected in the HL Tau disk \citep{Mullin24}, nor in other structured disks, except in the systems PDS 70 \citep{Keppler18}
and WISPIT 2 \citep{vanCapelleveen25}, in both of which, however,
the planets are located in large cavities 50--60 au wide.
This point is therefore of crucial importance, as it indicates that magnetic effects must be taken into account in researches aimed at determining the timescale for the formation of the planets themselves.

However, a number of issues need to be addressed before we can confirm the validity of the proposed association. First of all, the value of the magnetic lever arm assumed in this class of numerical models is generally $\leq$ 2. A similar simulation for $\lambda \sim 4-5$, as suggested by our analysis, would be desirable. Secondly, the vertical size of the computational domain varies among the models, but in general the simulations do not reach the region at 10$^3$ au from the star in which we observe the wind structures with ALMA.
An extension of numerical studies to large vertical scales would be needed, but it would come at the expense of a consistent computational effort. Finally, in these simulations, the disk structures are often seen to slowly move \citep{Suriano19,Martel22}, questioning the survival of their connection to the wind. However, the motion occurs on secular timescales, suggesting that the coherence of the magnetic flux tubes is probably not disrupted. 
Clearly, the applicability of the class of models described above to the observed cases needs to be tested with refined theoretical studies, which, however, are out of the scope of this work.

\section{Conclusions} \label{concl}

We investigated the properties of the molecular CO wind associated with the young HL Tau system by means of a detailed spectro-imaging analysis of the ${}^{12}$CO (2-1) line emission, observed with ALMA at $\sim$0\farcs3 angular resolution in the context of the ALMA-DOT campaign \citep{Podio20a,Garufi21}. 
Our analysis confirms previous results derived from integrated intensity maps, but
new insights on the structure of the flow come from the inspection of the individual projections of the position-position-velocity (PPV) datacube in the brighter and less perturbed redshifted lobe. 

A complex substructure is revealed in the velocity channel maps, constituted by concatenated bubble- and arc-shaped features
that progressively increase in size and distance from the source with velocity. The conical shape of the wind appears to be the result of superposition of the limb-brightened borders of these features.
The position-velocity (PV) diagram formed along the system axis presents a fan of separated almost linear traces opening from the source region, and with increasing velocity with offset from the star. The traces have strict spatial correspondence with the arcs in the channel maps.
Finally, PV diagrams taken across the flow at stepped separations from the source show a series of concatenated ellipse-shaped features that correspond directly to the traces in the longitudinal PV diagram and in the channel maps. The ellipses are tilted with respect to the system axis, indicating rotation of the flow in the clockwise direction looking from the SW lobe toward the star, which matches the direction of the disk rotation.
 
On the basis of these projections we infer a global structure of the emission in the PPV datacube consisting of a series of distinct concatenated paraboloid surfaces with the apex toward the source, with progressively larger aperture as the angle of their axis with respect to the velocity axis increases (cf. \cref{fig:sketch}). In physical space, this translates into a finite number of coaxial, distinct, rotating nested gas shells.
Ellipse fitting of the three directly identifiable structures in adjacent transverse PV diagrams allows a tomographic reconstruction of the morphology and kinematics of the outer three shells of the flow. We find smaller radii, higher poloidal and toroidal velocities, and steeper increases of the velocity with distance from the source for shells progressively closer to the system axis. 

This structure can be reproduced by a number of different wind acceleration mechanisms. In this paper, we focused on an analysis of the applicability of magnetohydrodynamic (MHD) disk winds (DW) accelerated from an extended region of the disk surface. We defer the comparison with other classes of models to future investigations.

Under the DW hypothesis, we derive from the rotation properties of the flow that each examined shell is rooted in the disk within a few au of one of three adjacent disk rings, seen in the continuum
emission of the dust, located at 58, 72, and 86 au from the star \citep{ALMA15}. However, because of the faintness of the examined flow substructures, we caution that the uncertainty on the footpoint radii is greater than the distance of the adjacent rings and gaps. 

We attempted a determination of the magnetic lever arm $\lambda$ in the wind by comparing the observed trends of the wind parameters along the shells with the predictions of models of magnetized winds from either a laminar AD-dominated disk or a fully turbulent disk. The characteristics of the HL Tau wind turn out to be intermediate between the two types of solutions, with a value of $\lambda \sim 4 - 5$. We discuss this finding in the context of the characteristics of the examined models. 
The retrieved launching region between 50 and 90 au would point to a wind from an AD-dominated outer disk, but solutions with such a high value of $\lambda$  have yet to be found (current solutions reach $\lambda$ = 2.76).
A possibility would be to consider solutions
for a higher level of
AD than is usually assumed, which may be applicable if the wind itself screens the incoming ionizing cosmic rays by magnetic mirroring \citep{Cleeves15}.
The high value of $\lambda$ could also be justified by a magnetic field amplitude in the outer disk region larger than the value commonly adopted in these models as the ejection efficiency in the wind would be higher in this case. 

A magnetically driven disk wind with the above characteristics could sustain disk accretion at large scale by removing angular momentum in regions where the effective turbulence is suppressed because of weak internal ionization. In addition, the presence of distinct denser nested wind shells rooted in disk rings appears to support the results of recent non-ideal MHD numerical simulations for AD-dominated disks, which see the spontaneous formation of a ring--gap substructure in the disk connected with an inhomogeneous layered flow (e.g., \citealt{Suriano19, Riols20}), in a way alternative to the action of the yet elusive protoplanets. 
These aspects may have profound implications for the evolution of the disk and the process of planet formation.

The inferred properties of the CO wind call for further studies of both observational and theoretical type. An analogous scrutiny of the fainter components and substructures in both the red- and blueshifted lobes in the same datacube, as well as the challenging estimate of the mass and angular momentum transported in each shell,  requires the development of dedicated procedures, which is currently underway (Nony et al., in prep.). On the theoretical side, it would be interesting to consider the extension of the numerical simulations of inhomogeneous layered disk winds to the large scales probed by our observations.
Then, a comparison of the retrieved results with other classes of wind models will be attempted, and will be the subject of future works. Finally, new observational studies for this and other systems, currently in progress, will allow us to analyze the dynamic relationship between the outer CO outflow, the coaxial H$_2$ wind, and the atomic jet. 

These investigations are expected to provide further insight into the nature of the ejection phenomenon and finally enable us to quantify the feedback of the winds on the disk evolution.

\begin{acknowledgements}

This paper uses ALMA data 2018.1.01037.S (PI L. Podio). ALMA is a partnership of ESO (representing its member states), NSF (USA) and NINS (Japan), together with NRC
(Canada), MOST and ASIAA (Taiwan), and KASI (Republic of Korea), in cooperation with the Republic of Chile. We thank the anonymous referee for precious comments that led to improved results.
The authors thank G. Lesur for fruitful discussions and for his
support in using the self-similar MHD disk wind solutions that he has made available to the public at: https://github.com/glesur/PPDwind. 
The authors thank Z.-Y. Li, Y. Bai,
T. Downes, F. Louvet, D. Fedele, and L. Testi for their helpful comments, and K. Rygl, V. Delabrosse, and U. Locatelli for their help with data processing and analysis.
This work was supported by INAF-LG 2022 "YSOs Outflows, Disks, and Accretion: towards a global framework for the evolution of planet-forming systems (YODA)"  and in part by the NextGenerationEU funds within the National Recovery and Resilience Plan (NRRP), Mission 4 - Component 2 - From Research to Business (M4C2), Investment 3.1 - Project IR0000034 – "STILES - Strengthening the Italian Leadership in ELT and SKA".
LP and CC acknowledge the projects PRIN-MUR 2020 - "BEYOND-2p: Astrochemistry beyond the second period elements" (Prot. 2020AFB3FX); ASI-Astrobiologia 2023 - "MIGLIORA-Modeling Chemical Complexity", (F83C23000800005); INAF-GO 2024 - "ICES: Tracking the history of ices from the cradles of planets to comets"; INAF-GO 2023 - "PROTO-SKA-Exploiting ALMA data to study planet-forming disks: preparing the advent of SKA" (C13C23000770005); INAF-MG 2022 - "Chemical Origins" (PI: L. Podio); financial support under the National Recovery and Resilience Plan (NRRP), Mission 4, Component 2, Investment 1.1, Call for tender No. 104 published on 2.2.2022 by the Italian Ministry of University and Research (MUR), funded by the European Union – NextGenerationEU– Project Title 2022JC2Y93 "ChemicalOrigins: linking the fossil composition of the Solar System with the chemistry of protoplanetary disks" – CUP J53D23001600006 - Grant Assignment Decree No. 962 adopted on 30.06.2023 by the Italian Ministry of Ministry of University and Research (MUR).

\end{acknowledgements}

\bibliographystyle{aa} 
\bibliography{mybib.bib} 

\begin{appendix}

\section{Atlas of velocity channel maps}
\label{app:chmaps}

\begin{figure*}[h!]
\centering
 \includegraphics[width=16cm]{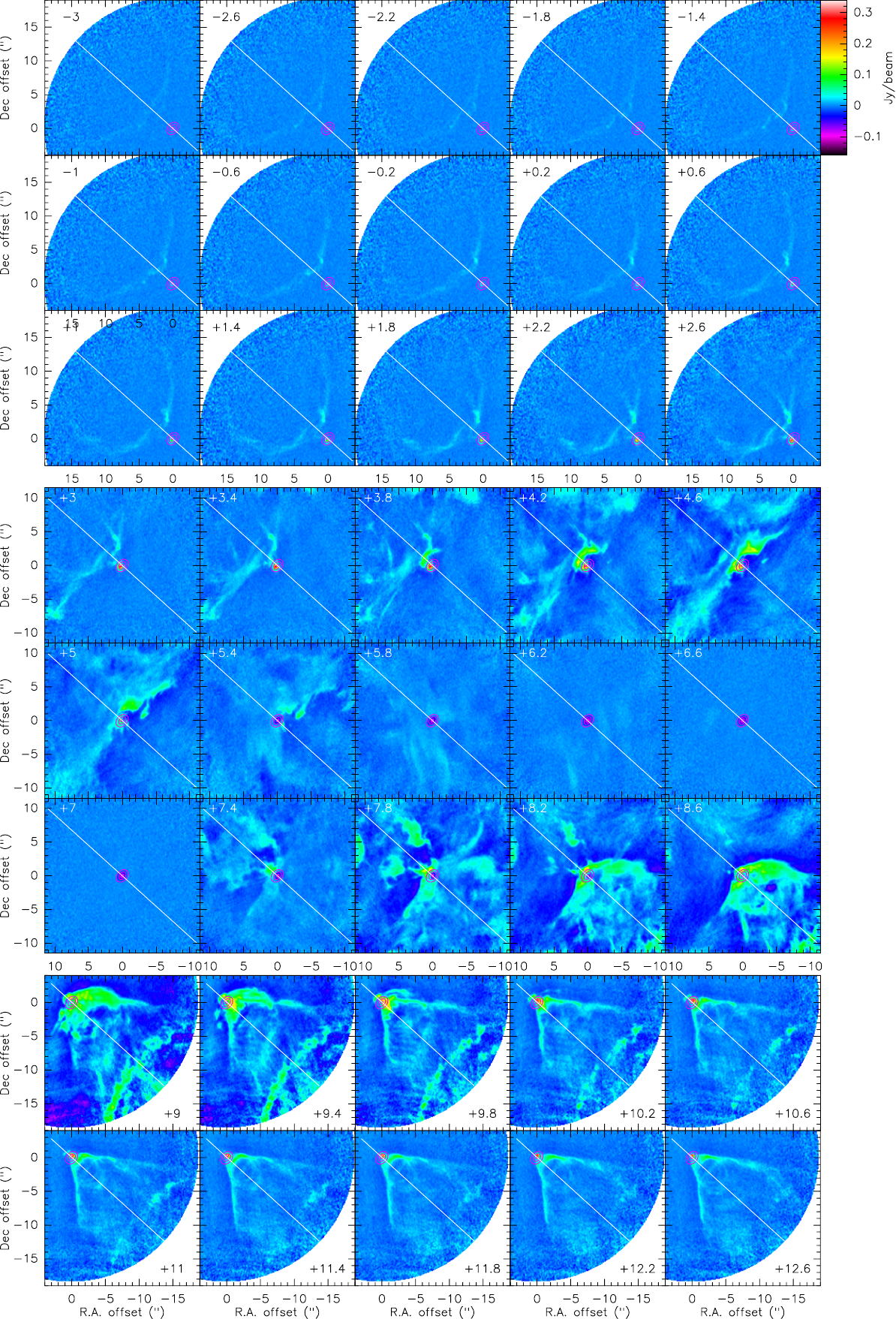}
   \caption{
   Channel maps of the CO (2-1) emission with V$_{\rm LSR}$ velocity from  -3 to +12.6 \kms. The beam size is 0\farcs31$\times$0\farcs26 (PA=-3\degr), corresponding to a spatial resolution of about 40 au at the distance of HL Tau (147.3 pc). In all the panels the disk position is indicated  by the magenta contours of the continuum emission at 1.3mm, drawn at [10, 200, 700]$\sigma_c$.
   The white line indicates the direction of the disk minor axis (PA=48$^{\circ}$). 
}
          \label{fig:chmap_blue}
    \end{figure*}

\begin{figure*}[h!]
\centering
 \includegraphics[width=16cm]{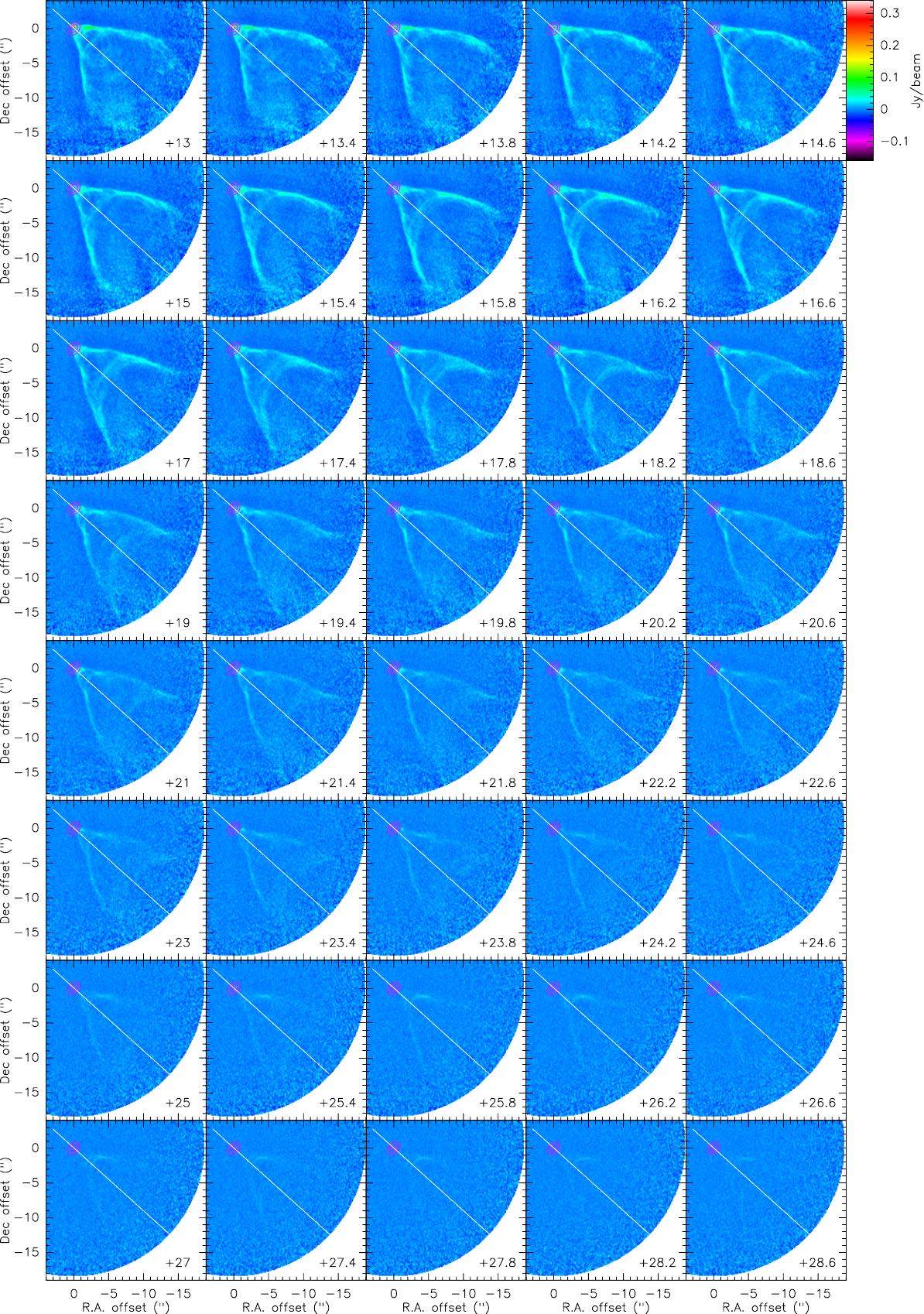}
   \caption{
   Same as Fig.\ref{fig:chmap_blue}, but for the channel maps with V$_{\rm LSR}$ from +13 to +28.6 \kms. 
}
          \label{fig:chmap_red}
    \end{figure*}

Figures \ref{fig:chmap_blue} and \ref{fig:chmap_red} present an
atlas of channel maps of CO (2-1) emission at V$_{\rm LSR}$ from -3.0 \kms to +28.6 \kms, in steps of 0.4 \kms. 
The panels at V$_{\rm LSR}$ from +3.0 to +8.6 \kms are centered on the source, to visualize the low-velocity emission in both lobes, while outside this range, emission is detected only in  the blueshifted or in the redshifted lobe, so the maps cover the relevant quadrant, with the source positioned in one corner. 

The emission in the blueshifted NE lobe is very faint. It consists of a large arc-shaped feature opening toward the NE seen at progressively larger distances from the source with increasing blueshifted velocities. A finger-like feature is identified at almost all velocities in about the same position north to the source and appears to be independent on the moving arc. 

\begin{figure}
    \centering
    \includegraphics[width=0.9\linewidth]{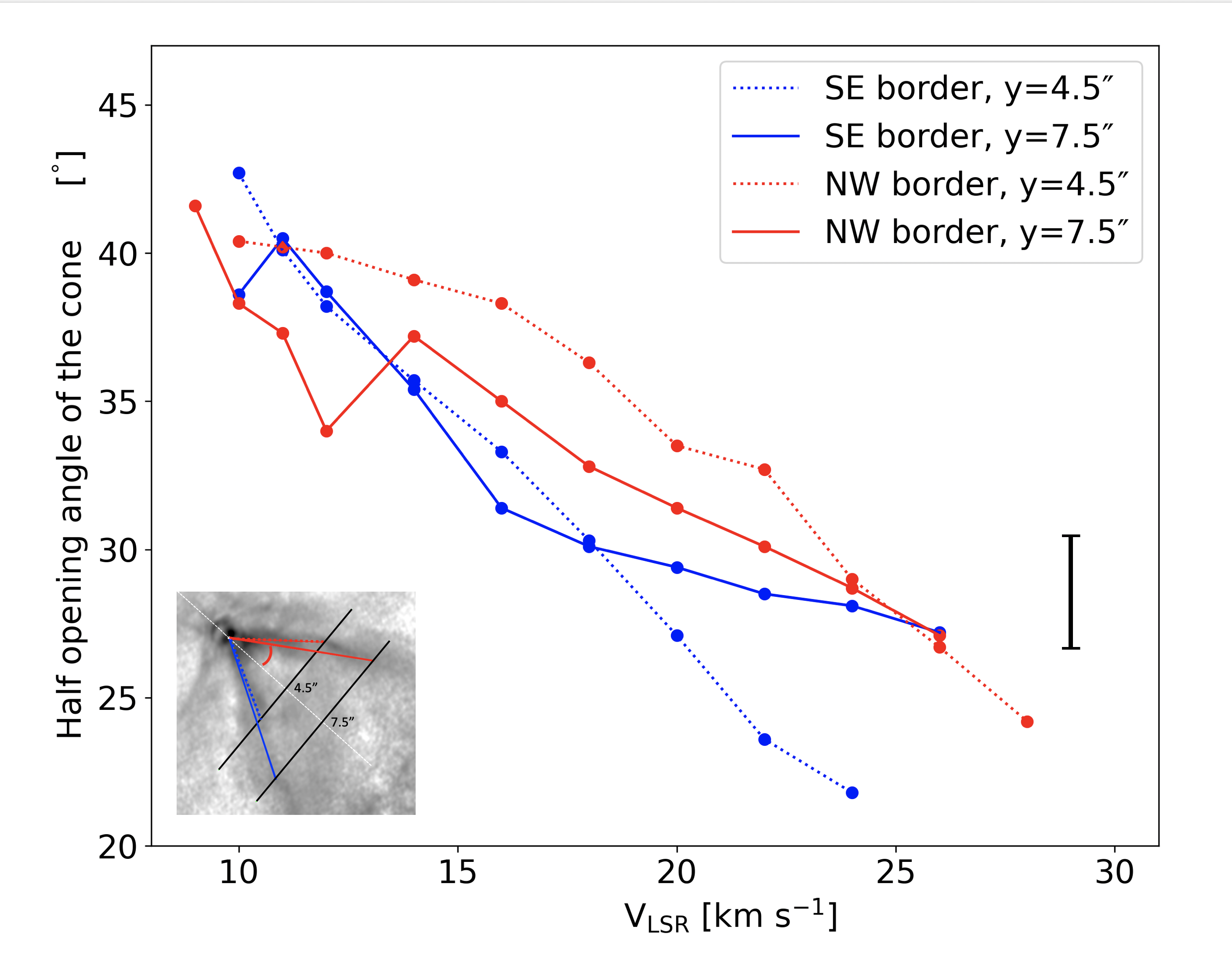}
    \caption{
    Half-opening angle of the apparent conical shape of the redshifted outflow, measured in each channel map and separately for the NW and SE sides of the flow, as the angle between the system axis and the flow border at offset 4\farcs5 and 7\farcs5 from the source (see the panel inset, drawn on a mom-0 map).  
    }
    \label{fig:opening_angle}
\end{figure}

The redshifted SW lobe is distributed in two main components, a low-velocity wide-angle wind and a faster inner flow of apparent conical shape, 
with an opening angle decreasing with velocity.  
To quantify the variation of the aperture, we determined, for selected channels from 9 to 28 \kms and separately for the SE and NW sides, the angle between the system axis and a line connecting the source and the position of the peak emission at the flow borders (identified by the contour profiles), at an offset of 4\farcs5 and 7\farcs5 from the source. The results of this analysis are shown in Figure \ref{fig:opening_angle}. The half-opening angle  
decreases on average with velocity, 
from about 40$\degr \pm$ 2$\degr$ at 10 \kms to 27$\degr \pm$ 2$\degr$ at 26 \kms. The difference in value displayed in a given channel by the various curves reflects the deviations in the shape of the flow from a perfect cone.

\begin{figure}
    \centering
    \includegraphics[width=0.9\linewidth]{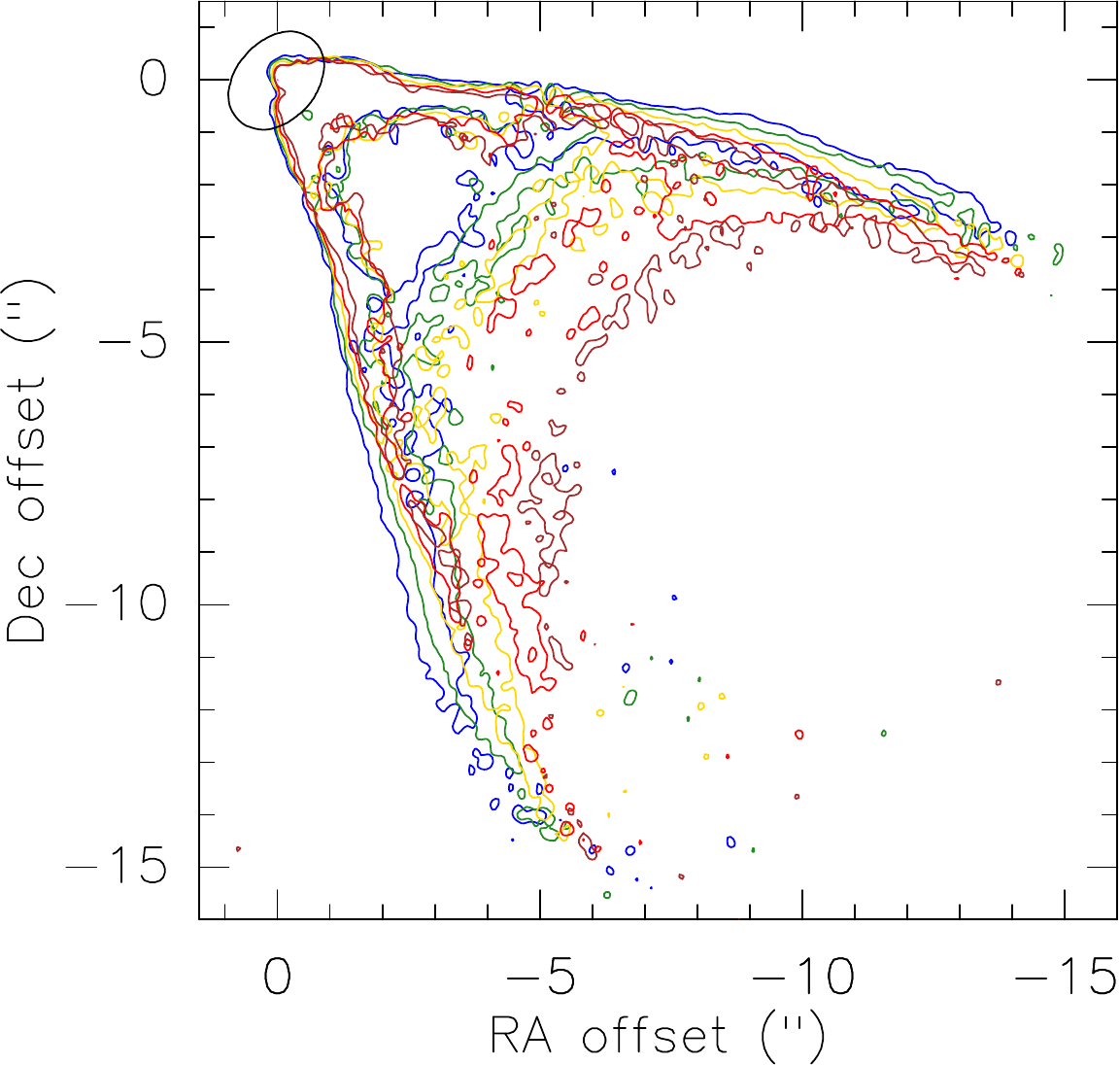}
    \caption{
    Contours of the CO emission at 4$\sigma_{\rm CO}$ 
    in the channels at 15.4 (blue), 16.2 (green), 17.0 (yellow), 17.8 (red), and 18.6 (brown) \kms. The disk is indicated by a black contour at 10$\sigma_c$. The structures are seen at progressively larger separations from the source in contiguous channel maps.
    }
    \label{fig:ell_move}
\end{figure}

Inscribed in the apparent cones one finds a series of arcs and bubble-like features. Each of these features can be followed over a number of adjacent channels, and it maintains the same shape and orientation with respect to the axis. The distance from the source and the size of each feature increase monotonically with \vLSR. As an example,
\cref{fig:ell_move} illustrates the shift in position of the arc-shaped structures of family B with increasing velocity.

The base of the redshifted wind (with reference to the SE side with respect to the axis, as the NW side hosts the emission from the receding part of the disk) appears resolved in the channel maps at low velocities between \vLSR = 7.4 and 9.4 \kms. 
Although the emission pattern is not clear, this velocity interval is consistent with the one expected under the DW hypothesis.
In this scenario, the flow starts with $V_p =0$ at the footpoint (corresponding to \vLSR = \vsys), and it is gradually faster as it leaves the disk. However, due to the absorption of the ambient medium, no material with \vLSR\,$\leq$ 7.3 \kms\, is visible (see, e.g., \cref{fig:PVparB&W}).
On the other hand, the maximum \vLSR\, at which the wind base will be imaged is \vLSR = \vsys + $V_p^* \cos{i}$, where $V_p^*$ is the poloidal velocity
achieved at an altitude $h = s / \cos(i) \sim$ 60 au above the disk, where   $s \sim 40$ au is the spatial resolution element.  Compared with the curves of the DW model discussed in  \cref{su:comparison},  $V_p^*$  is expected to be between 2.0 and 3.5 \kms from B2 to B1a (see \cref{fig:rz-Vz-Jz.vs.model}, where the observed quantities are normalized by the corresponding $r_0$ and $V_K (r_0)$). Projecting along the l.o.s. and adding \vsys, one finds that the base of the flow should be visible in channel maps up to \vLSR $\sim $  8.4 -- 9.5 \kms, as we find, with the highest \vLSR\, corresponding to inner B1a. As predicted by the model, in the
channel maps at \vLSR$>9.5 $ \kms the wind base is not resolved.

\FloatBarrier

\section{Atlas of transverse position-velocity diagrams and application of the ellipse fitting}
\label{app:PVperp}

\begin{figure}
    \centering
    \includegraphics[width=\linewidth]{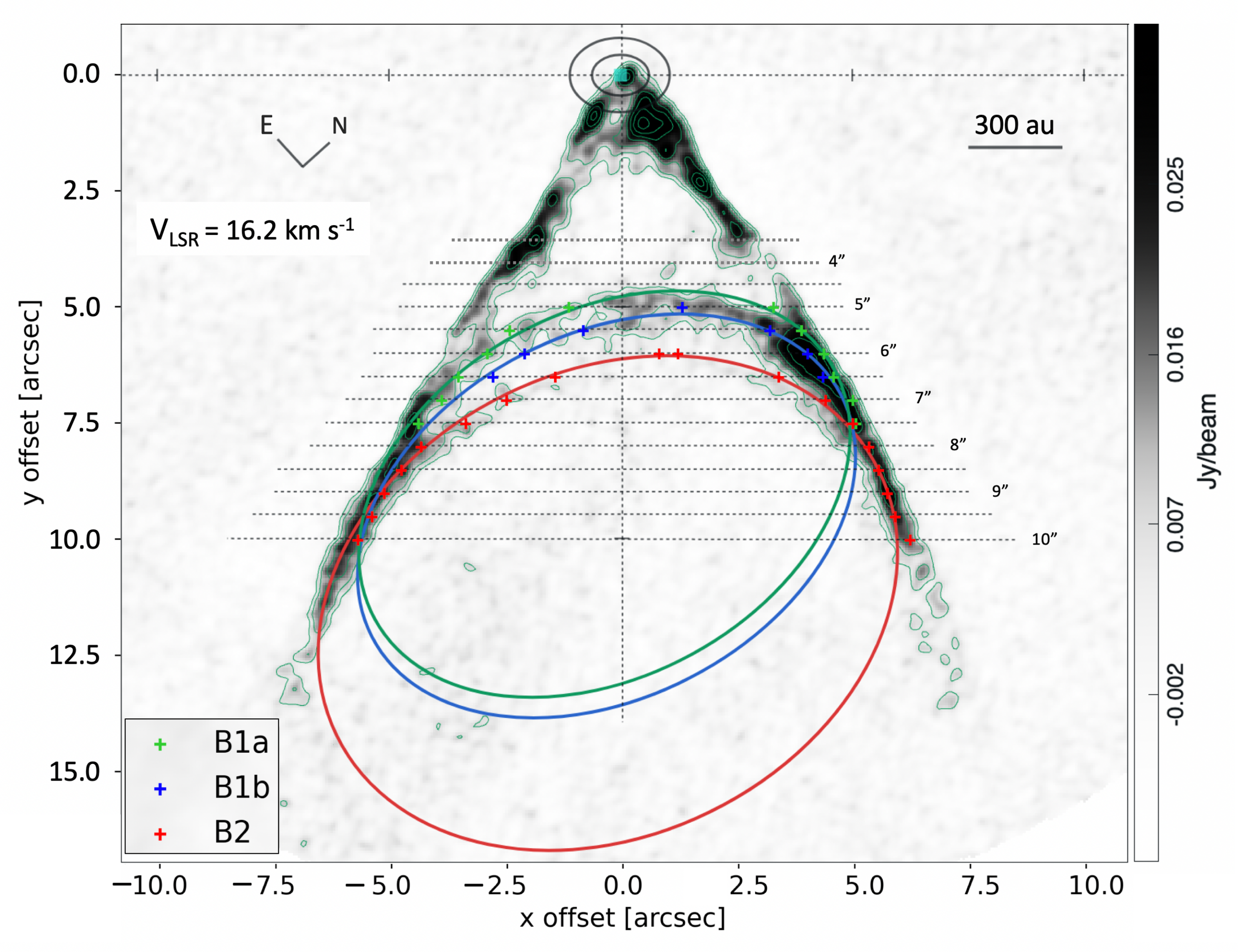}
    \caption{
    Example of cross-check between the positions of the traces B1a, B1b, and B2 in the transverse \pvperp\, diagrams and in the velocity channel maps. In the map at \vLSR=16.2 \kms  (rotated clockwise by 48\degr as in \cref{fig:xoffsets})  
    the horizontal dashed lines indicate the offset from the source of the diagrams in \cref{fig:perpred2_1,fig:perpred2_2,fig:perpred2_3}. 
    The crosses indicate the $(x,y)$ positions of the traces B1a, B1b, and B2 in the \pvperp\, diagrams formed at each offset $y$ and at the selected velocity. The three ellipses are drawn  to show the good correspondence with the observed arcs in the channel maps. 
    }
    \label{fig:ell_chmap16.2}
\end{figure}

\begin{figure*}[h!]
\centering
 \includegraphics[width=16cm]{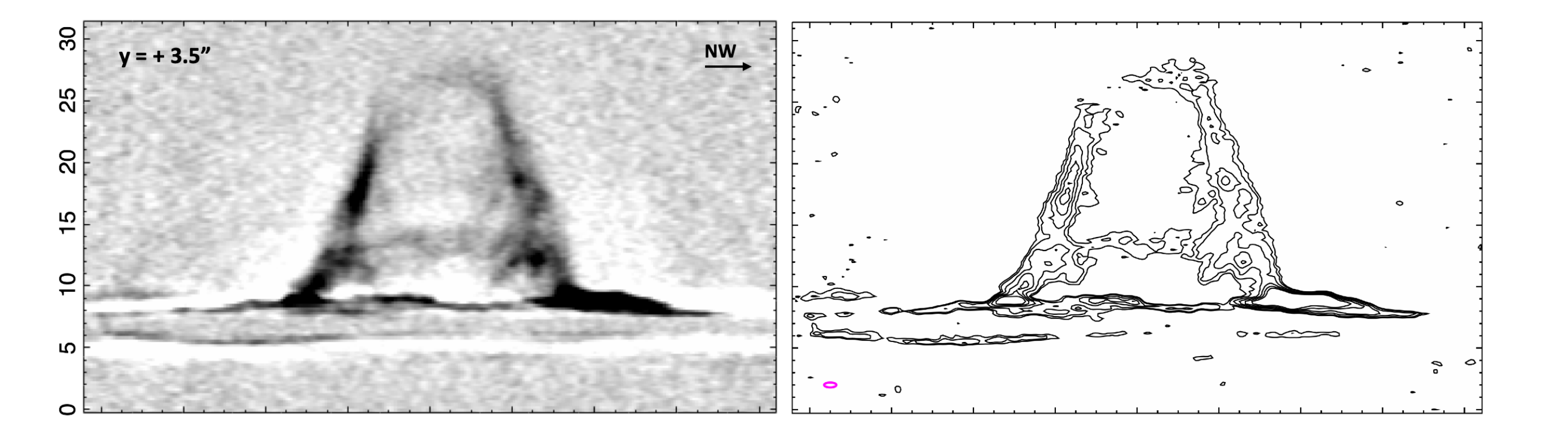}
 \includegraphics[width=16cm]{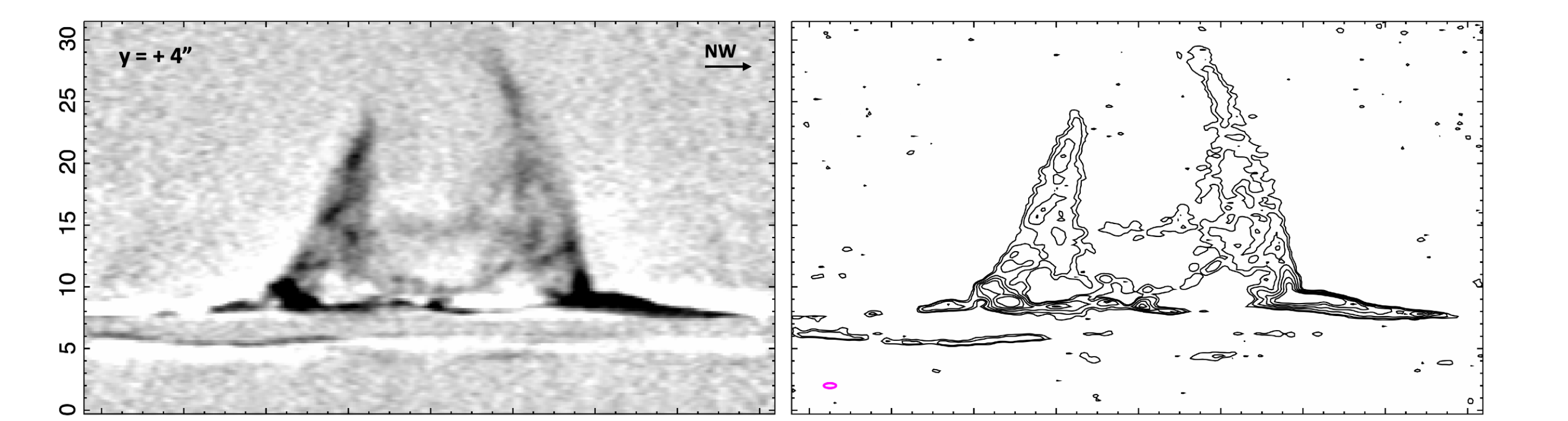}
 \includegraphics[width=16cm]{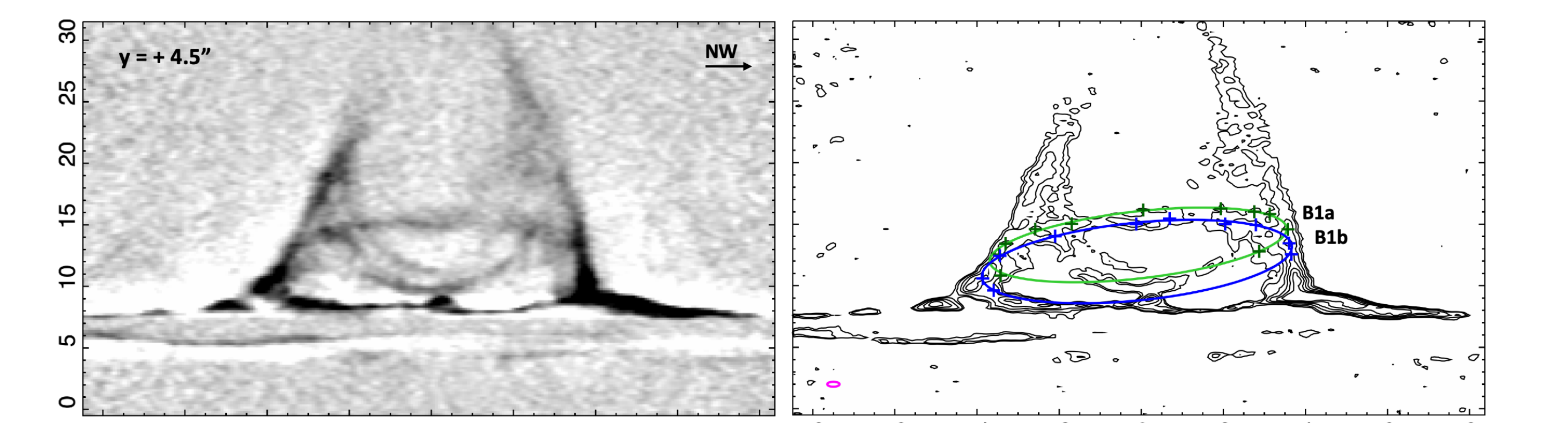}
 \includegraphics[width=16cm]{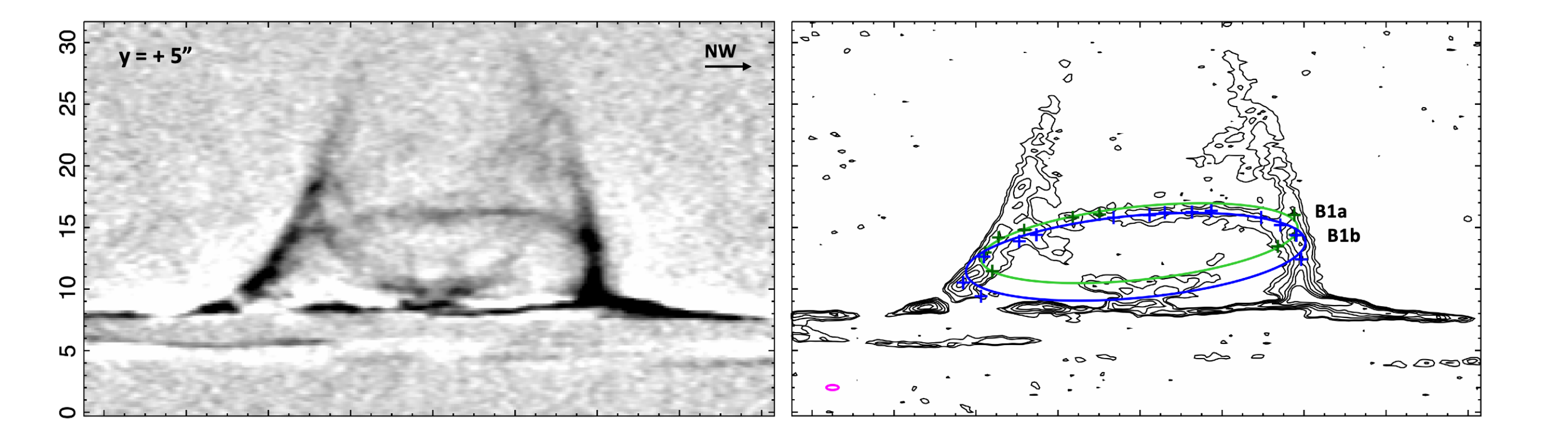}
 \includegraphics[width=16cm]{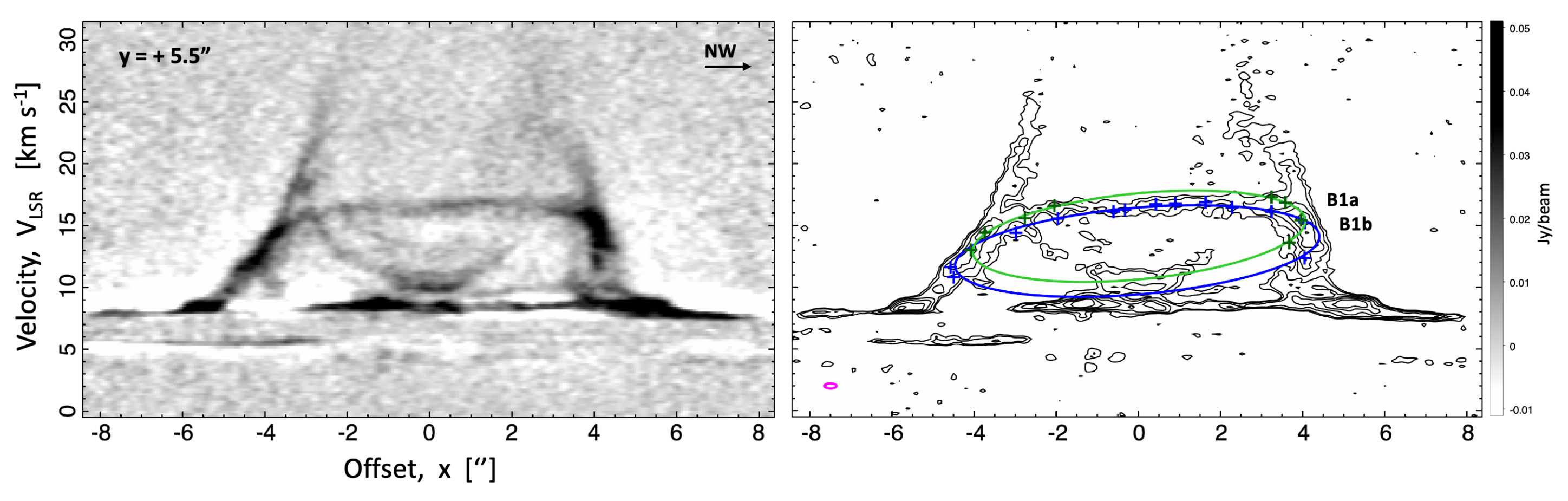}
       \caption{
       {\em Left panels:}Position-velocity diagrams taken across the redshifted outflow lobe (\pvperp\, diagrams), with a 0\farcs3 wide virtual slit centered on the flow axis  and perpendicular to it (PA = 318\degr), with positive $x$ offsets toward NW. The separation of the virtual slit  from the source in arcseconds is given in the upper left corner. {\em Right panels:} Ellipse fits to the features in the corresponding PV diagram, where an identification was possible. The contour levels are at 5, 10, 20, 30, 40, 50 mJy beam$^{-1}$. The crosses indicate the  sample points, all chosen above an emission threshold at the 3$\sigma_{\rm PV}$ level, with $\sigma_{\rm PV}= 2$ mJy beam$^{-1}$ (see text for details). Color code is green, blue, and red for features B1a, B1b, and B2, respectively. The small magenta ellipse at the bottom left corner indicates the uncertainty (0\farcs3$\times$0.2 \kms).
       } 
          \label{fig:perpred2_1}
    \end{figure*}

\begin{figure*}[h!]
\centering
 \includegraphics[width=16cm]{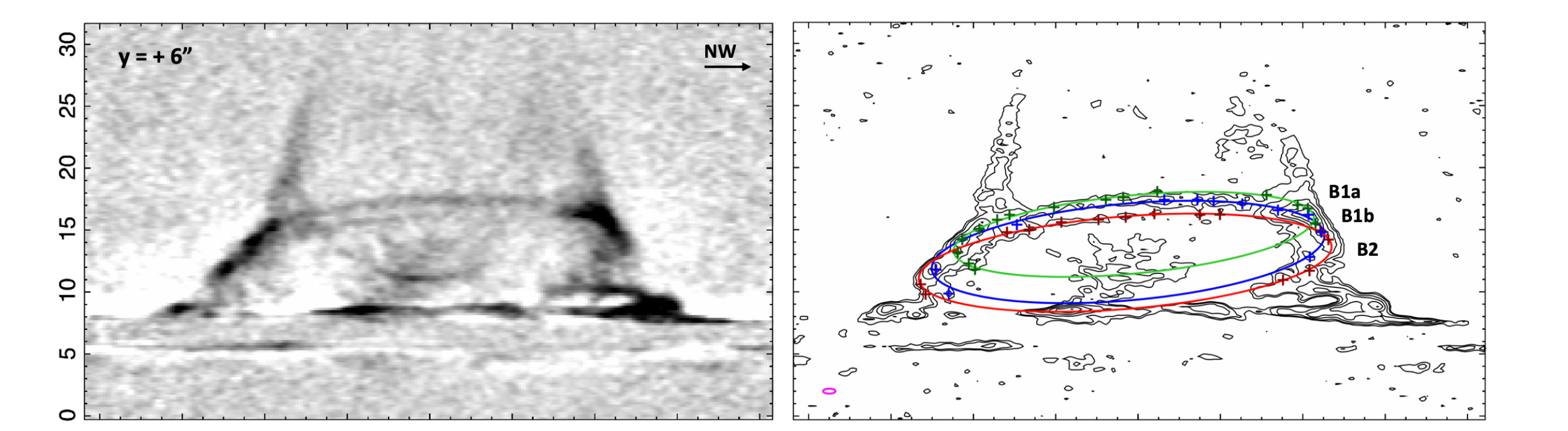}
 \includegraphics[width=16cm]{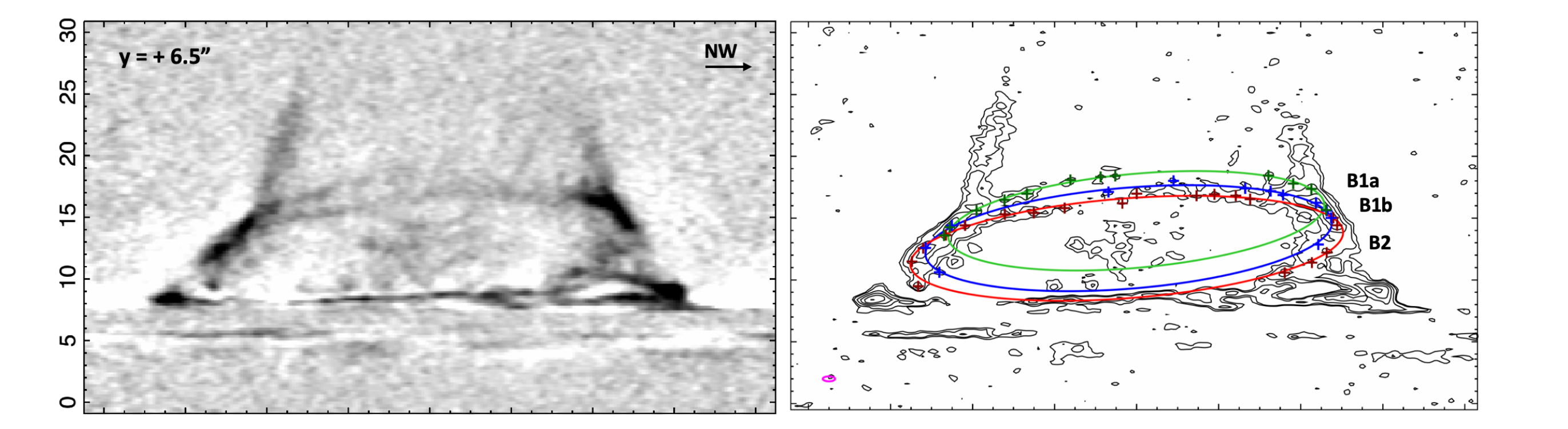}
 \includegraphics[width=16cm]{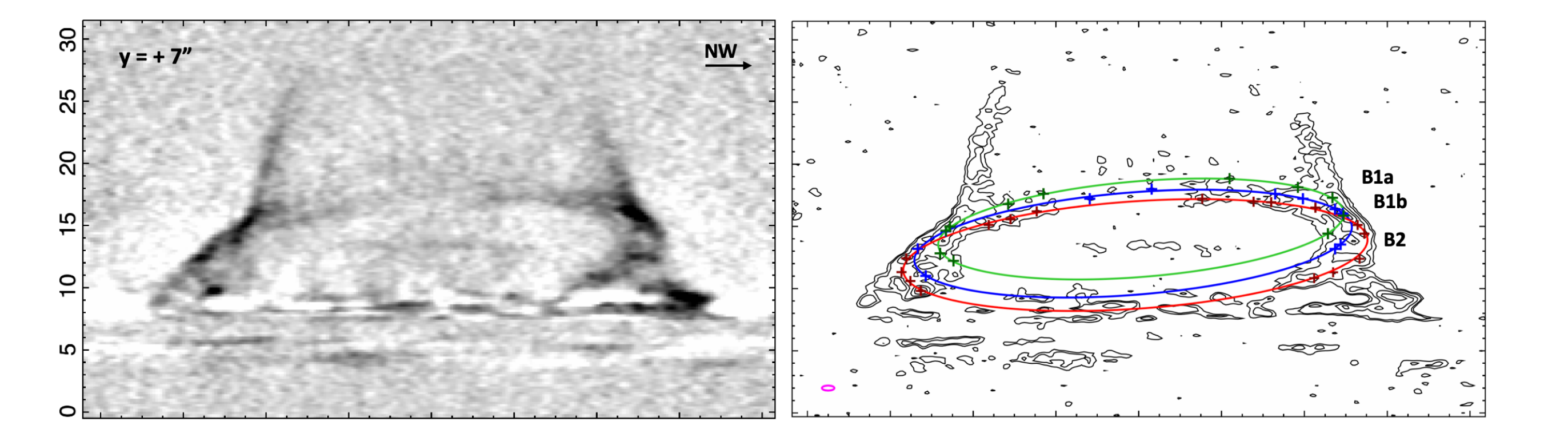}
 \includegraphics[width=16cm]{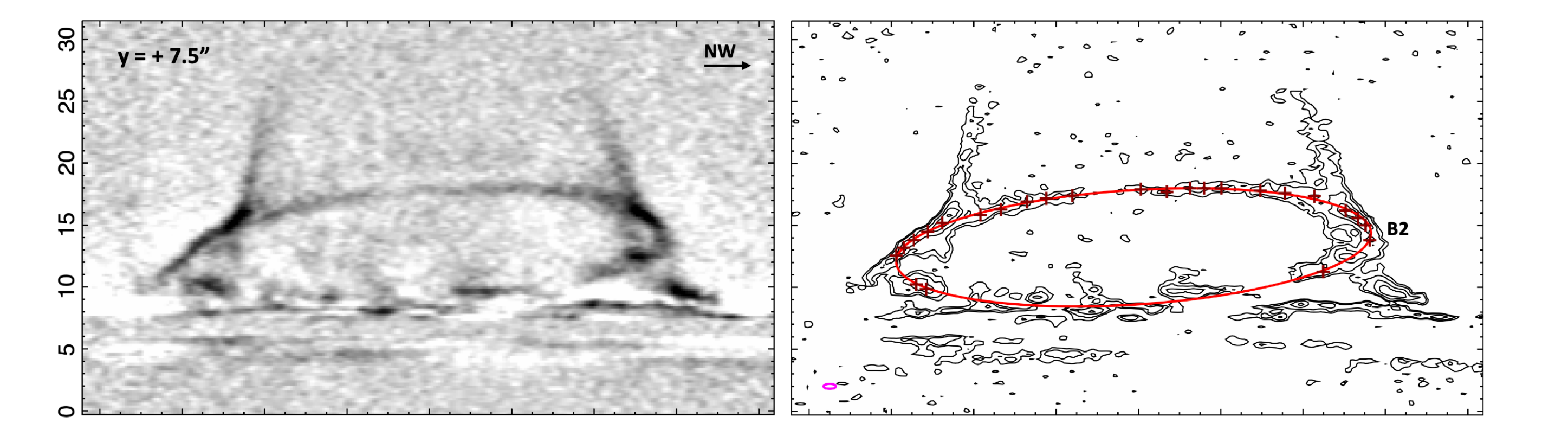}
 \includegraphics[width=16cm]{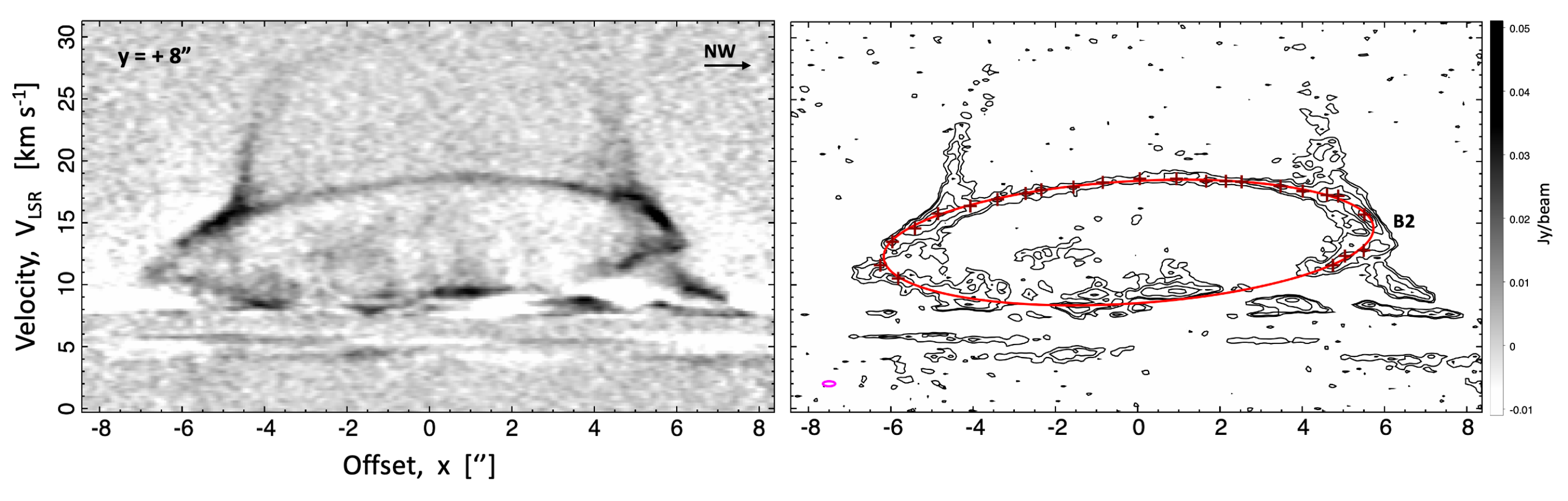}
   \caption{
   Same as \cref{fig:perpred2_1}, but for the transverse PV cuts at separations from the source  from 6 to 8 arcseconds.
   }
          \label{fig:perpred2_2}
    \end{figure*}

\begin{figure*}[h!]
\centering
 \includegraphics[width=16cm]{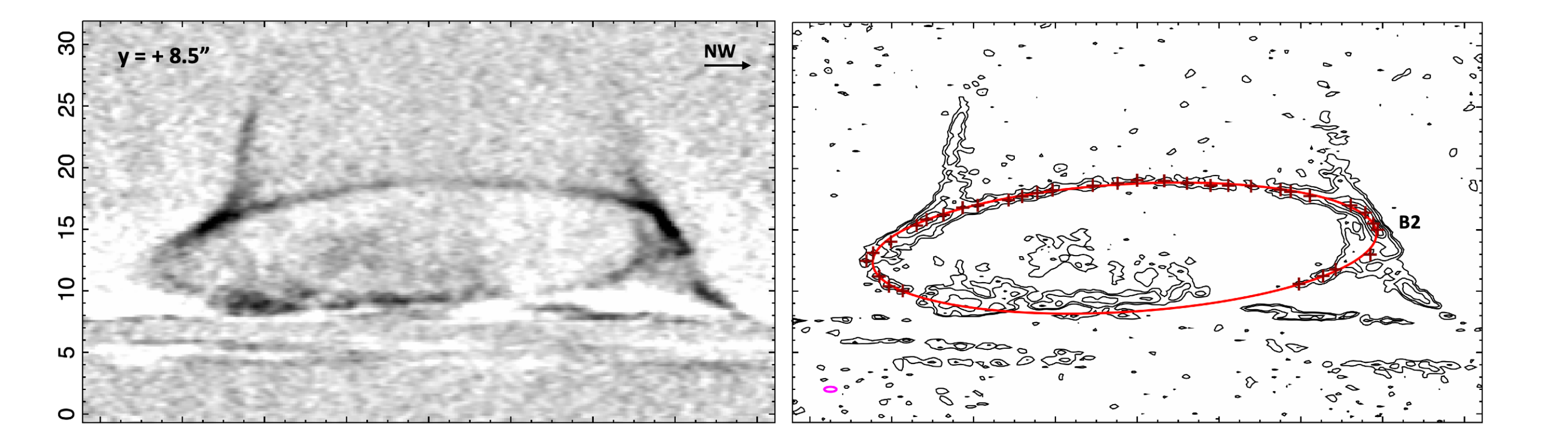}
 \includegraphics[width=16cm]{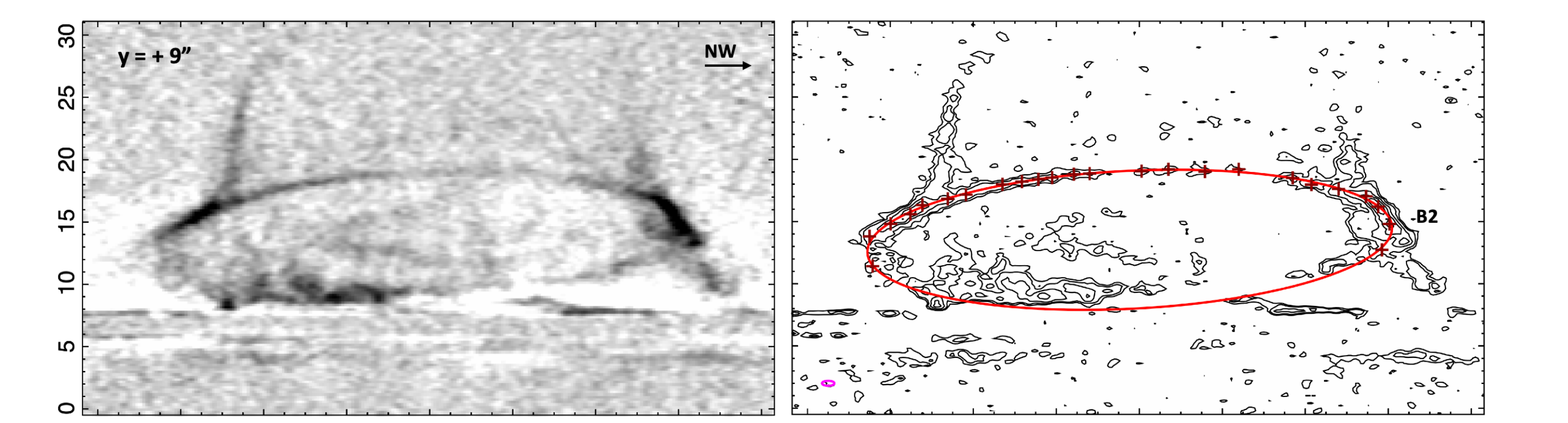}
 \includegraphics[width=16cm]{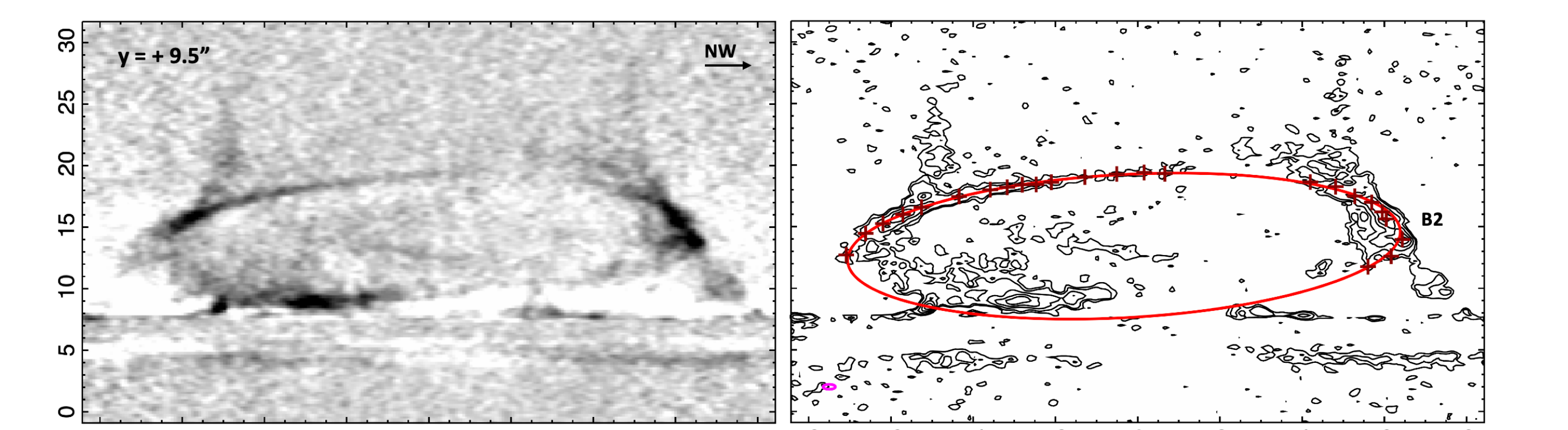}
 \includegraphics[width=16cm]{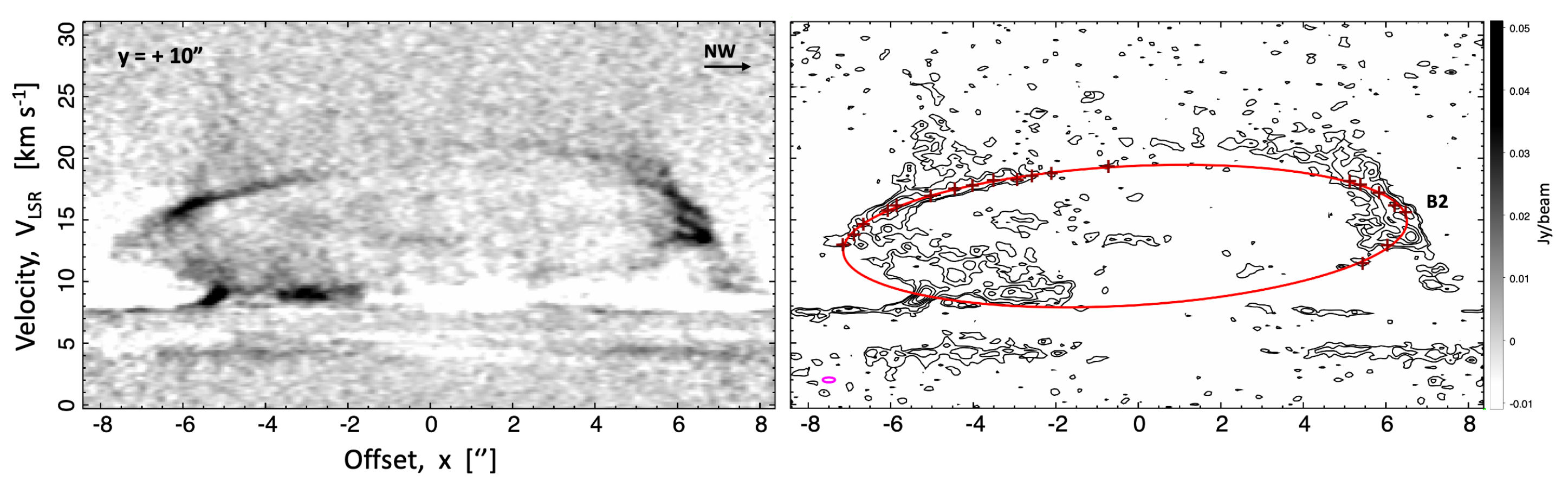}
   \caption{
   Same as \cref{fig:perpred2_1}, but for the transverse PV cuts from 8.5 to 10 arcseconds from the source. At these distances only feature B2 is identified.
   }
          \label{fig:perpred2_3}
    \end{figure*}

Figures \ref{fig:perpred2_1}, \ref{fig:perpred2_2} and \ref{fig:perpred2_3} present an atlas of the transverse Position-Velocity diagrams (\pvperp) obtained for the red lobe with a 
0\farcs3-wide virtual slit centered on the axis and oriented at PA=318${^\circ}$  (i.e., perpendicular to the axis and with positive offsets toward NW), at separations $y$ from the source from 3\farcs5 to 10\arcsec\, in steps of 0\farcs5, 
as indicated by the horizontal dashed lines in \cref{fig:ell_chmap16.2}. 

The left panels show the observed emission in grayscale, while the right panels illustrate the same emission with a contour map. Superposed to the latter, for the PV cuts from 4\farcs5 to 10\arcsec, mathematical ellipse fits of the features are reported. The fit is performed with the purpose of identifying the lateral edges of the observed features. The coordinates of these points are used in a tomographic reconstruction of the morphology and kinematics of the flow shell (see \cref{su:tomography}).

The critical aspect of the procedure is the difficulty in identifying unambiguously the traces corresponding to each shells B1a, B1b, and B2 in the faint or crowded regions of the transverse PV diagrams. To give an identification as reliable as possible, a cross-check on the positions of the emission peaks common to different projections of the datacube was performed. 
Such a correspondence is expected if the observed features originate from cuts of coherent structures in the datacube. 
To this aim, we first found sequences of relative intensity peaks in each \pvperp\, diagram that could trace a feature, all above an emission threshold of a 3 $\sigma_{\rm PV}$ level, where $\sigma_{\rm PV}=2$ mJy beam$^{-1}$ is the average over all panels estimated in a region along the $x=0$ axis at \vLSR $>$ 20 \kms, where the signal of the traces is almost absent.  We then looked for the signature of these sequences in the other two projections of the datacube. 
In particular, the position of the traces in the fan of the \pvpar\, diagram  (\cref{fig:PVparB&W}) identifies the expected velocity of each feature along the $x=0$ axis.
On the other hand, the position of the traces in the channel maps at each offset $y$ gives information on the expected offset $x$ of the emission in the \pvperp\, diagram formed at the distance $y$ from the source, and at the velocity of the selected channel map. We verified this correspondence by reporting on the channel map at a given velocity the $(x,y)$  positions of the traces in the \pvperp\, diagrams at each offset $y$, at the selected velocity. An example is given in \cref{fig:ell_chmap16.2} for the map at \vLSR = 16.2 \kms.
The quantitative comparison of the coordinates in the three projections allowed the selection of the sequence of peaks belonging to a single feature in each diagram. 

Then, a mathematical fitting routine based on a least-squares algorithm was applied to each identified sequence of relative intensity peaks. The fitting is anchored to points in the upper arc and side edges, but not in the lower arc, as the latter is difficult to identify due to overlap of low-velocity traces, signal absorption, and deformation due to flow inclination.
The appearance of the traces in the \pvperp\ projections suggests an ellipse as the geometrical form closest to the actual shape of the features, so the fit requires a number of input points greater than five. 
Sampling sets that met the above criteria could be found
for shells B1a and B1b at offsets from 4\farcs5 to 7\arcsec from the star, and for B2 from 6\arcsec to 10\arcsec. The input points are reported as crosses in Figures \ref{fig:perpred2_1} -  \ref{fig:perpred2_3}, with the fitted ellipses drawn in the corresponding color. The routine provides the coordinates of the center, the length of the semi-major and semi-minor axes, and the inclination with respect to the horizontal direction (i.e., \vLSR = constant). 

The SE and NW edges of the fitted ellipses, that is, 
the points on the curve farthest from the $x=0$ axis, 
were considered the vertices of each feature at the borders of the shell
(in a couple of cases where the geometric extreme of the ellipse had no emission associated, the closest emission peak was selected). 
The coordinates of these points provided the quantities necessary for the tomographic reconstruction.

With respect to the error estimate, statistically, the applied fitting procedure is expected to reduce the uncertainty with respect to that of the input points. In practice, however, the advantage is contrasted by the difficulty in identifying the traces, the lack of reliable sampling points in the lower part of the curve, the difference of the actual trace shape with respect to a geometrical ellipse, and the absence of emission at the extremes in a few cases. Therefore, we considered the uncertainty in the final coordinate values to be comparable to that of the input points in the spectral images, that is, $\pm 0\farcs15$ in position and $\pm 0.1$ \kms\, in velocity, reported as a small magenta ellipse in the right panels.

\FloatBarrier

\section{Influence of wiggling and precession}
\label{app:wigggle}

We tested the influence of wiggling and precession on the derivations in \cref{su:param_tomography} by analyzing the deviation of the centers of the fitted ellipses of features B with respect to the axis $x=0$ of the \pvperp\, diagrams (the perpendicular to the virtual slit of the diagram), corresponding in the sky to PA = 228\degr. The axis of propagation of the shells is found to deviate slightly from this direction, by 0.5, 2 and 3 degrees toward SE, for the B1a, B1b, and B2 shells, respectively. The scatter of the centers with respect to these axes shows a hint of wiggling pattern consistent among the shells, but the shift is anywhere lower than 3/5 of the observational uncertainty on the shell radius. In any case, our derivation of $r$, V$_z$ and V$_\phi$ relies on the relative positions of the apexes of the ellipses at the borders of the flow and does not use the information on the position of the ellipse center. Therefore, the determination of these quantities is not affected by wiggling, as long as the shell retains its coherence. In contrast, the different PA of the flow with respect to the perpendicular to the virtual slit can introduce a bias. We checked the importance of this effect by evaluating the expected variation in the worst case, that is, for shell B2 at a large offset $y=9\farcs5$ from the star. The difference between the measured and real values of $r$ and $z$ is negligible, while the measured $V_z$ turns out to be larger by 1\% of the real one, and $V_{\phi}$ smaller by 8\%. In any case, these values are lower than the observational uncertainty
5\% and 10\% in the poloidal and toroidal velocity, respectively, 
which justifies the adoption of \cref{eq:rvpvphi} for our analysis.
\vspace{0.5cm}

\section{Estimate of the radial velocity V$r$}
\label{app:vr}

\begin{figure}[h!]
\centering
 \includegraphics[width=9cm]{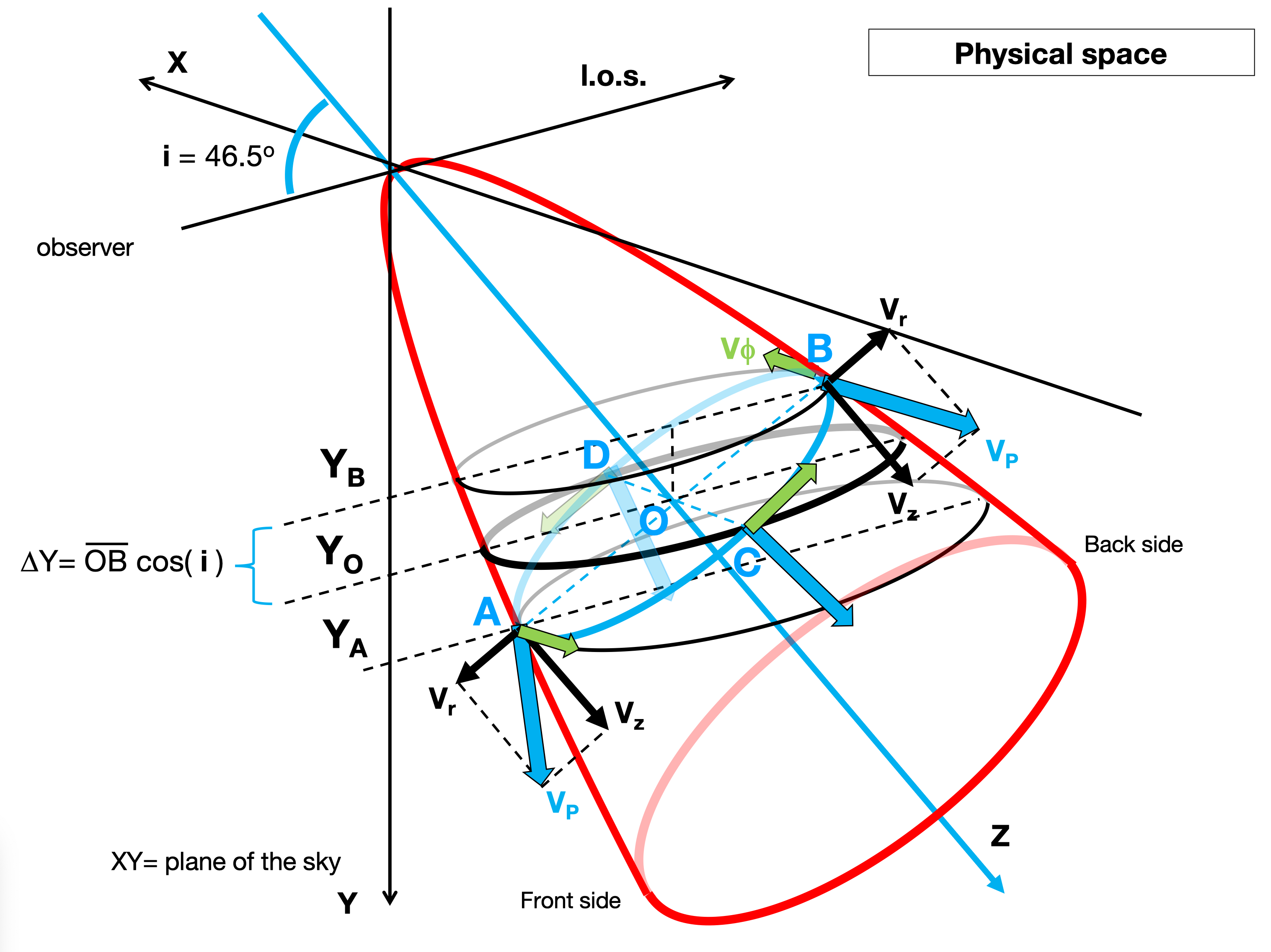}
   \caption{Sketch showing the geometry of the flow in physical space relative to  the determination of  the radial component of the velocity $V_r$, in the general case in which the poloidal velocity $V_p$ does not lie along the shell surface (see text). }
          \label{fig.sketchvr}
    \end{figure}

In \cref{su:tomography} the assumption is made that
the wind flows along the detected shells. Formally, this translates into the fact that in the cylindrical coordinate system aligned with the redshifted flow axis $z$ and centered on the star,
the poloidal velocity $V_p =  V_r \hat{e}_r + V_z \hat{e}_z$ is tangent to the shell surface. 
To test the validity of this assumption, one should verify that $V_r/V_z = \tan(\alpha$), where $\alpha$ is the local tangent angle to the shell, determined from the derivative of the measured $r=r(z)$ and reported in \cref{tab:PVperp_ell}.  While $V_z$ and $\alpha$ at a given offset from the source are found directly from the measurements at the lateral apexes of the fitted ellipses in one single diagram, the search for $V_r$ is complicated by
the combination of the detected variation of $V_z$ along the shell and the inclination of the flow axis. With reference to \cref{fig.sketchvr}, and assuming axisymmetry, to determine the radial velocity $V_r$ corresponding to the axial velocity $V_z$ evaluated at offset $Y_O$, one has to measure \vLSR\, at either point A or point B of the same shell section (the blue circle in the figure), that is, at the points where the projection of $V_\phi$ along the l.o.s. vanishes:
\begin{equation}
V_r = {{V_{\rm LSR}(B) - V_z \cos{i} - V_{\rm sys}}\over{\sin{i}}} =
{{V_z \cos{i} + V_{\rm sys} -V_{\rm LSR}(A)}\over{\sin{i}}}
\label{eq:vr}
\end{equation}
However, points A and B fall
at offsets $Y_{\rm A} = Y_O + R_O (Y_O) \cos{i}$ and $Y_{\rm B} = Y_O - R_O (Y_O) \cos{i}$.
Unfortunately, with the given geometry, these points correspond in almost all cases to regions of the diagrams with faint or confused emission,  out of the field of view,  hidden by the brightness of the gas close to the source, or lost because of the CO auto-absorption around \vsys. These factors make the determination of $V_r$ extremely difficult.

In practice, this quantity can be reasonably determined only for shell B2 in the two positions at 9\farcs5 and 10\arcsec, using the cleaner diagram \pvpar\, and limiting the search to the upper point of type B.
Taking into account the radius of the shell in these positions, the searched point B is at offsets 4\farcs9 and 5\farcs3, where one finds \vLSR(B)= 15.5$\pm$ 0.3 \kms and 15.7 $\pm$ 0.3 \kms, respectively. Equation \ref{eq:vr} then leads to
V$_r$ = 2.6 $\pm$ 0.4 \kms and 2.8 $\pm$ 0.4 \kms. In turn, the orientation angle of V$_p$ with respect to the $z$ axis turns out to be
15.2\degr $\pm$ 3.0\degr\, at offset 9\farcs5 and 16.3\degr $\pm$ 3.2\degr\, at offset 10\arcsec. According to \cref{tab:PVperp_ell}, the tangent angle $\alpha$ at these positions is 17.5\degr, which implies that here V$_p$ is compatible with being tangential to the shell surface within the errors.

\FloatBarrier

\section{Comparison with the tomographic procedure in \citet{DeValon22}}
\label{app:devalon}

A tomographic analysis similar to the one presented in \cref{su:tomography} was applied in \cite{DeValon22} to derive the morphology and the kinematic properties of the molecular outflow from the young star DG Tau B. In this source the flow presents a complex system of internal substructures with characteristics analogous to those of the flow from HL Tau. However, the adopted technique is not identical and leads to different outcomes.

The system of \cref{eq:rvpvphi} is equivalent to the one in Eqs. (4)-(7) in Sect. 3 of \cite{DeValon22}.\footnote{the inclination angle of the flow with respect to the line of sight differs by 180\degr with respect to ours, which introduces a minus sign in $V_z$}  In both cases, $V_z$ and $V_{\phi}$ are derived from the values of \vLSR\,  extracted along the bell profile in the \pvperp\, diagrams.  The difference resides in the selection of the relevant points, a choice that derives from a different assumption regarding the distribution of the gas in the flow. 
In \cite{DeValon22} the wind is considered homogeneous in the transverse direction to the axis. Thus, the points are selected with continuity throughout the bell profile. Under the assumption of a perfectly symmetric flow with respect to the axis, the procedure gives at each distance $z$ from the source two continuous functions $f_-$=\vLSR$(-\delta x)$ and $f_+$=\vLSR$(+\delta x)$ for varying $x$ offset $\delta x$. From these, maps of $V_z$ and $R V_{\phi}$ in the $(R,z)$ plane are derived over a transversely homogeneous wind (see Fig. 3 in \cite{DeValon22}). The presence of the observed arcs and cusps in the channel maps and of the ellipse-shaped traces in the PV diagrams does not enter the analysis at this stage. In fact, the features are interpreted as increases in density moving in a homogeneous flow. The justification of such increases invokes the presence of perturbations in the flow caused by a non-stationary mechanism, influencing either the accretion flow through the disk or the ejection of the axial jet (see Fig. 19 in \cite{DeValon22}).

In our work, the substructures are interpreted as the signature of a wind that is transversely inhomogeneous and structured in a finite number of distinct nested shells. In this way, the observed features naturally arise from the cuts of the wind distribution in the projections of the PPV datacube (cf. \cref{fig:sketch}).  As a consequence, non-stationary mechanisms are not needed to justify the observations. Technically, in our analysis we search for the edges of each ellipse-shaped trace in subsequent \pvperp\, diagrams, obtaining for each shell a set of pairs of correlated points ($x_-$,\vLSR$_-$; $x_+$,\vLSR$_+$) as a function of the distance from the source. From this set, we derive the properties of the flow along that particular shell, and the bell profile is interpreted as the superposition of the limb-brightened traces of the subsequent shells. Part of these traces are bright enough to be clearly identified also inside the bell profile (as the B family), part of them are too faint for a direct analysis and will be examined in a forthcoming work.

\section{Evaluation of the dynamical time of the substructures}
\label{app:tdyn}

\begin{figure}
    \centering
    \includegraphics[width=0.9\linewidth]{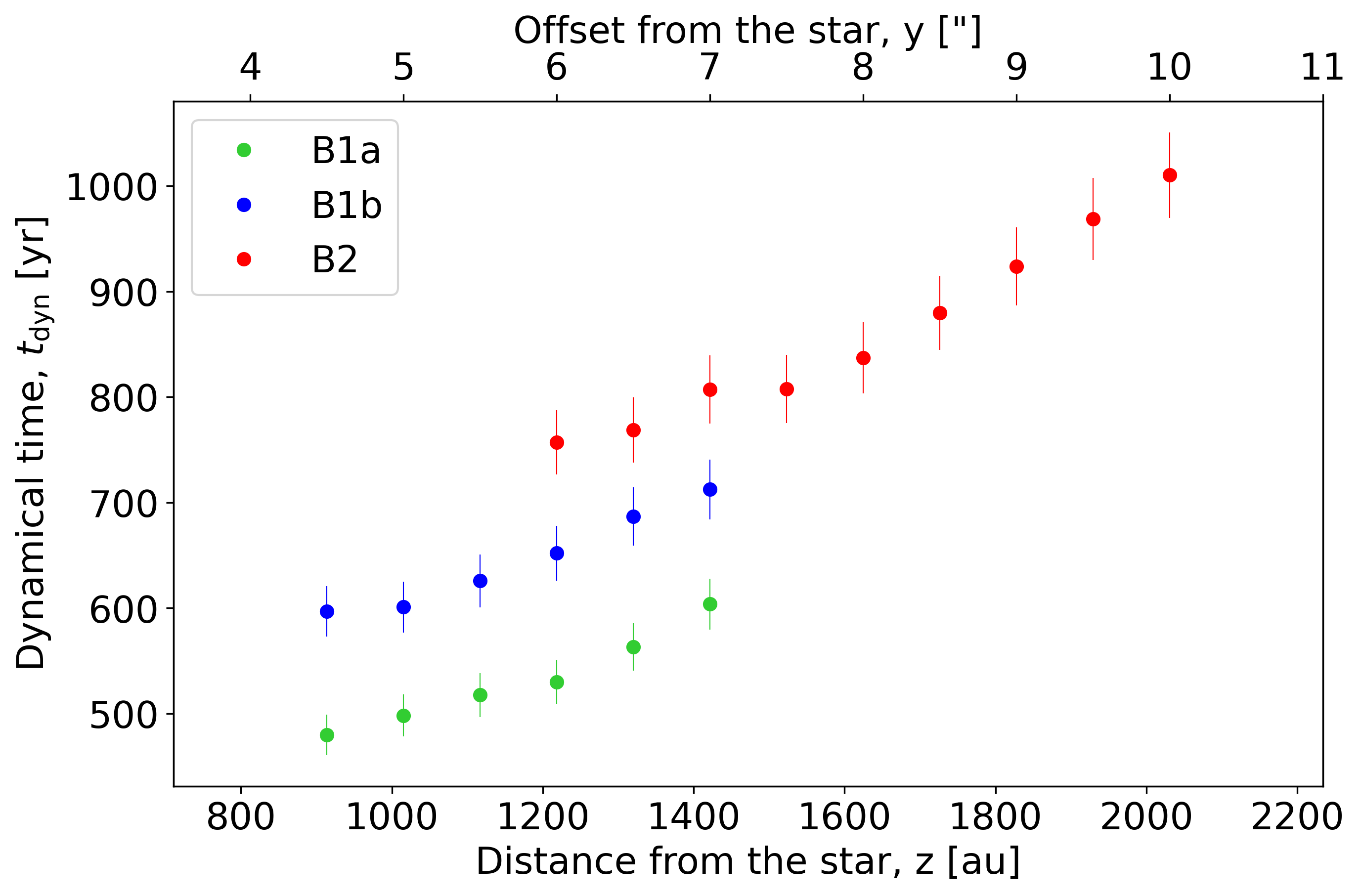}
        \includegraphics[width=0.9\linewidth]{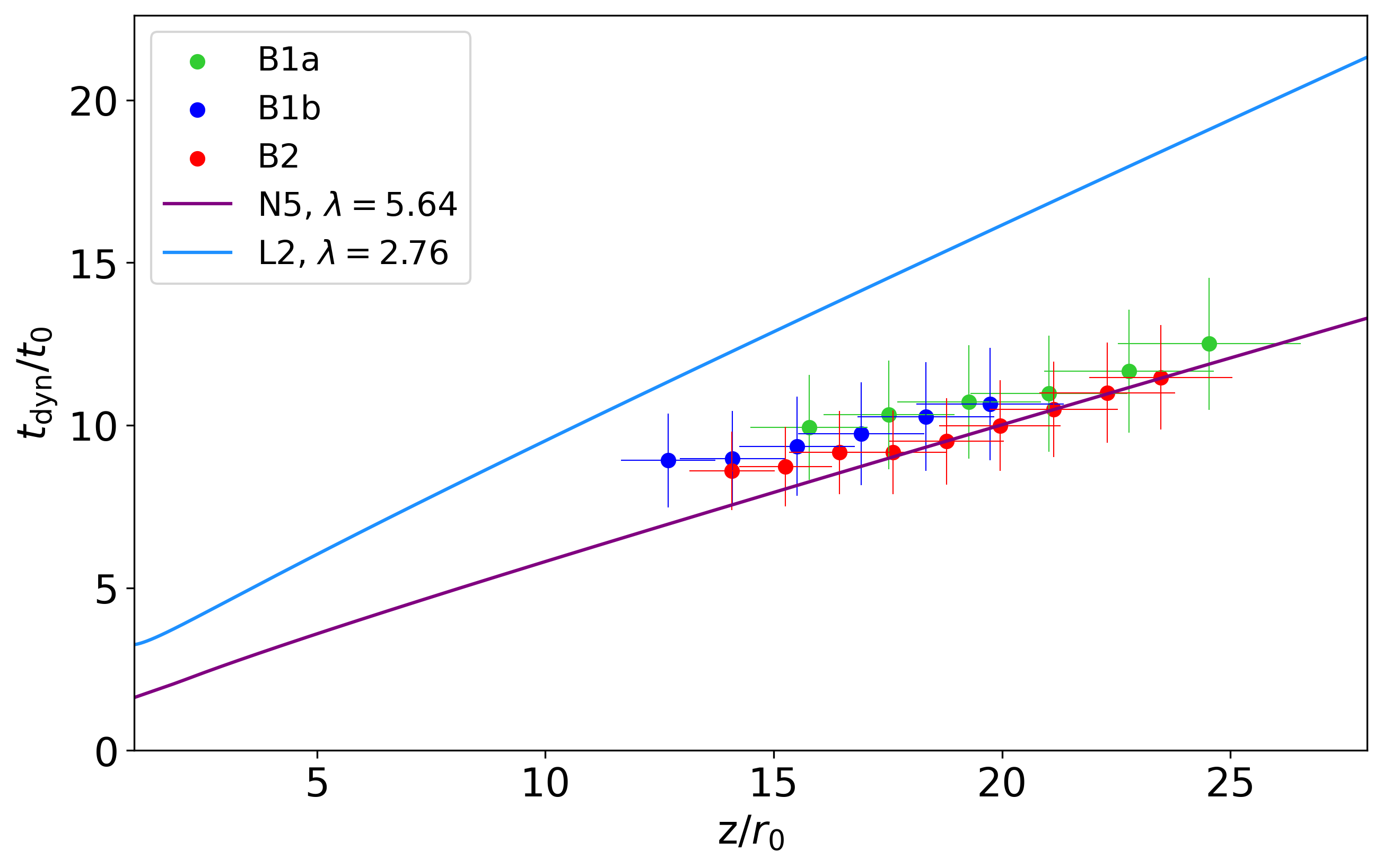}
    \caption{
    {\em Top}: 
    Variation of the dynamical time $t_{\rm dyn} = z /V_z$ with vertical distance from the source $z$  of the flow substructures B1a, B1b, and B2 as derived from the tomographic analysis. 
    {\em Bottom}:
Comparison between the normalized dynamical time  $t_{\rm dyn}/t_0 = (z/V_z)/(\bar{r}_0/V_{\rm K}(\bar{r}_0)$ derived for each of the three shells and the analogous quantity estimated for the MHD disk wind models L2 (blue) and N5 (purple) illustrated in \cref{su:comparison}.
    }
   \label{fig:tdyn}
\end{figure}

In the upper panel of \cref{fig:tdyn} we report the dynamical time estimated as $t_{\rm dyn} = z /V_z$, as a function of the distance from the source $z$, for the three substructures analyzed.
The dynamical time is higher for the slower outer shells, and for each shell increases monotonically with distance from the source,
from 4.8 to 6$\times$ 10$^2$ years for B1a and
from 6.0 to 7.1$\times$ 10$^2$ years for B1b, over a distance from 0.9 to 1.4$\times$ 10$^3$ au from the source. For the outer shell B2 $t_{\rm dyn}$ is measured to vary between 7.6 and 10$\times$ 10$^2$ years over a distance of 1.2 to 1.9$\times$ 10$^3$ au from the source.
The observed trends are compatible with a steady ejection of material and steady propagation along the nested shells as postulated in a magnetohydrodynamic disk wind. The applicability of the DW scenario is quantitatively confirmed by \cref{fig:tdyn}, bottom panel, in which the dynamical times derived for the three shells, normalized through their footpoint radii and corresponding Keplerian velocities, are compared with the analogous quantity estimated for the DW models L2 and N5 (cf. \cref{su:comparison}). The points show the same trend as the curves, closely following the one for model N5.  

Alternatively, assuming that the observed features are swept-up shells caused by subsequent pulsed ejections, from the comparison of $t_{\rm dyn}$ at the same altitude one could estimate the time separation between subsequent ejections, obtaining intervals of about one hundred years. However, in this scenario one would rather expect for each shell a constant dynamical time along $z$, as the bulk of the material ejected in each event would reach the current altitude at the same time. However, it cannot be excluded that the interaction with the environment and/or with the material ejected previously causes a slowdown of the flow in the forward part of the shell, leading to an apparent increase in dynamical time at high $z$.   

\FloatBarrier

\section{Comparison with models: Variation of $\lambda_\phi$ along the streamline}
\label{app:trendsz}

\begin{figure}[h!]
\centering
 \includegraphics[width=9cm]{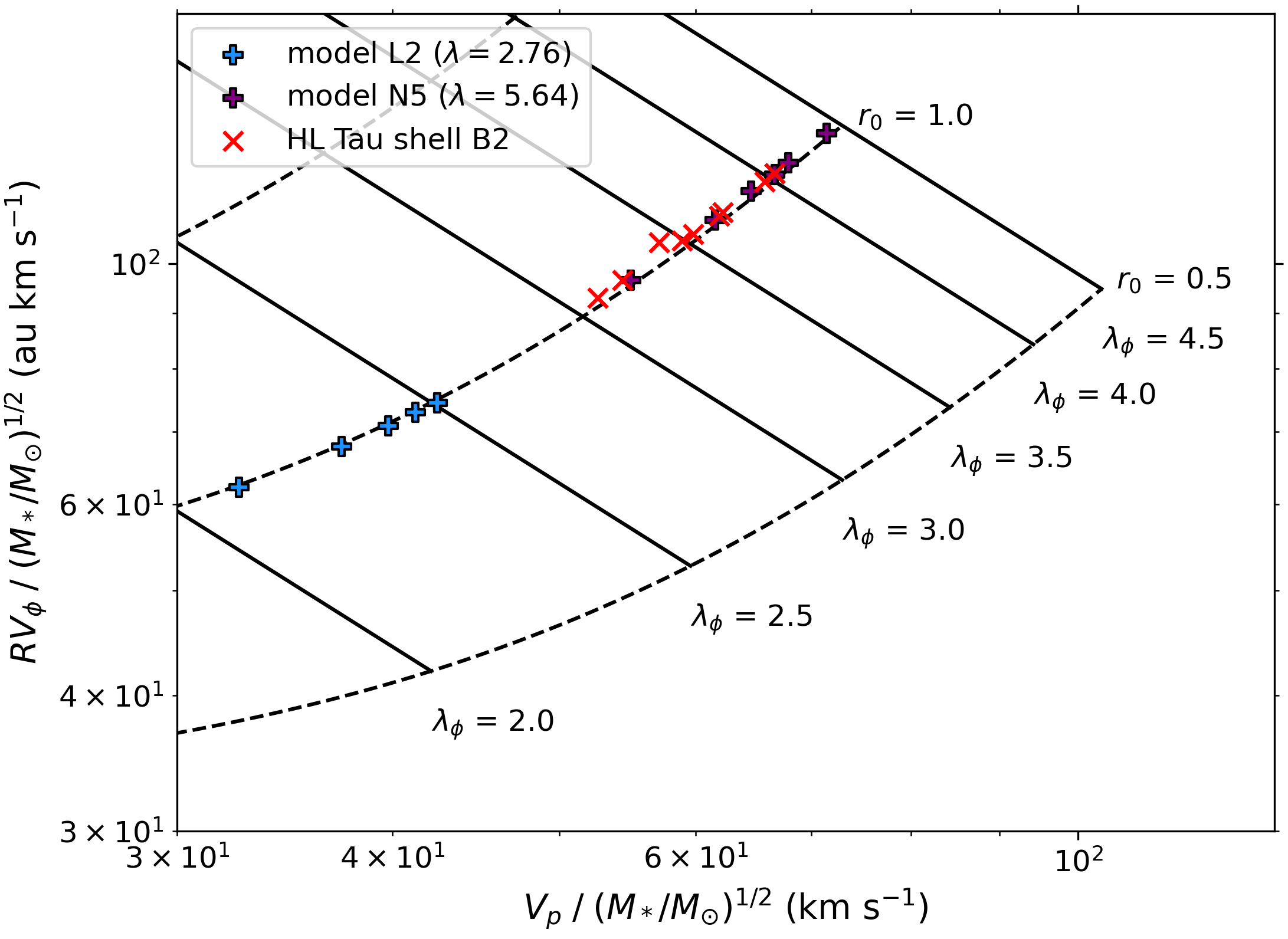}
   \caption{Trend of the normalized specific angular momentum $\lambda_{\phi}$ along a streamline originating from 1 au in a disk around a 1 $M_{\odot}$ star following models L2 (blue) and N5 (purple), for  separations from the source $z/r_0$ = 10, 20, 30, 40, 50 from left to right, with an additional point at  $z/r_0$ = 100 for model N5. The superposed red crosses illustrate the trend for  $\lambda_{\phi}$ along shell B2 as in \cref{tab:PVperp_ell}, after normalization of $r$, V$_p$, and V$_\phi$ by the derived  average footpoint radius $\bar{r}_0 $ and corresponding Keplerian velocity, for $M_\star$ = 2.1 M$_\odot$ (see \cref{su:comparison}).}
          \label{fig.j_vpo-models}
    \end{figure}

Figure \ref{fig.j_vpo-models} illustrates the similarity of the trend of variation of the specific angular momentum $\lambda_{\phi}$ along a streamline in the DW models examined and in the observed wind. Models L2 \citep{Lesur21} and N5 \citep{Zimniak24} are calculated for $r_0$=1 au and a star of one solar mass. The points derived for shell B2 from the data are normalized by the radius of the retrieved footpoint and by the corresponding Keplerian velocity, for a stellar mass of 2.1 M$_\odot$.  In this way the wind streamline is considered to originate at 1 au, which is usually assumed to be the upper limit of the extension of the turbulent inner disk. In this way we have the most effective comparison between the turbulence- and AD-dominated disk wind models (N and L types of solutions, respectively) and the observed wind. 
In all cases the values are seen to follow closely the curve at constant $r_0$=1, with a progression toward higher $\lambda_\phi$ with distance, as expected in steady MHD disk winds.

\end{appendix}

\end{document}